\documentclass[]{aa}  

\usepackage{float}
\usepackage{graphicx}    
\usepackage{color}  
\usepackage{txfonts}
\usepackage{inputenc}
\usepackage{multicol}
\usepackage{multirow}
\usepackage{siunitx}
\DeclareSIUnit{\solarmass}{\text{M}_{\odot}}
\DeclareSIUnit{\parsec}{pc}
\DeclareSIUnit{\year}{yr}
\DeclareSIUnit{\mag}{mag}
\DeclareSIUnit{\arcsec}{arcsec}
\usepackage{amsmath}
\usepackage{xcolor}
\usepackage{booktabs}
\usepackage{color} 

\usepackage{algorithm} 
\usepackage{algpseudocode} 
\usepackage{hyperref}
\hypersetup{colorlinks, allcolors=blue}

\begin{document}

   \title{Significance Mode Analysis (\texttt{SigMA}) for hierarchical structures}
   \subtitle{An application to the Sco-Cen OB association}
   
   \author{Sebastian Ratzenb\"ock\inst{1,2,3}
          \and
          Josefa~E.~Gro\ss schedl\inst{1}
          \and
          Torsten M\"oller\inst{2,3}
          \and
          Jo\~ao Alves\inst{1,2} 
          \and
          Immanuel Bomze\inst{2,5}
          \and
          Stefan Meingast\inst{1}
          }

   \institute{University of Vienna, Department of Astrophysics,
              T\"urkenschanzstra{\ss}e 17, 1180 Vienna, Austria\\
              \email{sebastian.ratzenboeck@univie.ac.at}
         \and
             University of Vienna, Research Network Data Science at Uni Vienna, Austria
        \and 
            University of Vienna, Faculty of Computer Science, W\"ahringer Straße 29/S6, A-1090 Vienna
         \and
            University of Vienna, ISOR/VCOR, Oskar-Morgenstern-Platz 1, A-1090 Vienna
             }

   \date{Submitted March 31, 2022; Resubmitted November, 2022; ... }
   
  \abstract
    {We present a new clustering method, Significance Mode Analysis (\texttt{SigMA}), to extract co-spatial and co-moving stellar populations from large-scale surveys such as ESA Gaia. The method studies the topological properties of the density field in the multidimensional phase space. We validate \texttt{SigMA} on simulated clusters and find that it outperforms competing methods, especially in cases where many clusters are closely spaced. We apply the new method to Gaia DR3 data of the closest OB association to Earth, Scorpio-Centaurus (Sco-Cen), and find more than 13,000 co-moving young objects, with about 19\% of these having a sub-stellar mass. \texttt{SigMA} finds 37 co-moving clusters in Sco-Cen. These clusters are independently validated by their narrow HRD sequences and, to a certain extent, by their association with massive stars too bright for Gaia, hence unknown to \texttt{SigMA}.
    We compare our results with similar recent work and find that the \texttt{SigMA} algorithm recovers richer populations, is able to distinguish clusters with velocity differences down to about 0.5 km\,s$^{-1}$, and reaches cluster volume densities as low as 0.01 sources/pc$^3$.
    The 3D distribution of these 37 coeval clusters implies a larger extent and volume for the Sco-Cen OB association than typically assumed in the literature. Additionally, we find the association to be more actively star-forming and dynamically more complex than previously thought.
    We confirm that the star-forming molecular clouds in the Sco-Cen region, namely, Ophiuchus, L134/L183, Pipe Nebula, Corona Australis, Lupus, and Chamaeleon, are part of the Sco-Cen association. 
    The application of \texttt{SigMA} to Sco-Cen demonstrates that advanced machine learning tools applied to the superb Gaia data allows to construct an accurate census of the young populations, to quantify their dynamics, and to reconstruct the recent star formation history of the local Milky Way.
    }


   \maketitle
\defcitealias{Kerr2021}{KRK21}
\defcitealias{Schmitt2022}{SCF22}
\defcitealias{Squicciarini2021}{SGB21}
\defcitealias{Damiani2019}{DPP19}
\defcitealias{Luhman2022a}{L22A}
\defcitealias{MiretRoig2022b}{MR22}
\defcitealias{Zerjal2021arxiv}{ZIC21}
\defcitealias{BricenoMorals2022}{BMC22}


\section{Introduction}\label{sec:intro}

The ESA Gaia mission \citep{Brown2016, Brown2018, Brown2021, Vallenari2022arXiv} is transforming our knowledge of the local Milky Way, particularly considering the distribution of young stellar populations. However, disentangling and extracting coeval populations remains notoriously difficult. This is reflected in the wide variety of methods applied to the Gaia data  \citep[e.g.,][]{Oh2017, Kushniruk2017, Zari2017, Castro-Ginard2018, Cantat-Gaudin2018b, Galli2018, Zari2019, Damiani2019, Meingast2019, Kounkel2019, Chen2020, Hunt2021, Olivares2021, Meingast2021}. The wide range of approaches in the literature reflects the rather complex feature space\footnote{Stars in the data set are represented as points in a five- or six-dimensional space with three positional axes and two or three kinematic axes. In a machine-learning context, this space is referred to as feature space. The term ``feature'' is synonymous with dimension or coordinate axis.} from where the stellar populations are extracted. Firstly, these initially compact objects are stretched into elongated, sometimes non-convex structures in position space as a consequence of interactions with the Milky Way potential, spiral arms, and giant molecular clouds \citep[e.g.,][]{Kamdar2021}. This ``galactic-stretching'' leads to a variety of cluster\footnote{In this paper,  we use the word ``cluster'' in the statistical sense, namely, an enhancement over a background. This avoids creating a new word for the spatial/kinematical coherent structures we find in Sco-Cen. None of the Sco-Cen clusters is expected to be gravitationally bound.} shapes from compact (when young) to low-contrast, spread-out, sometimes S-shaped clusters dominated by Milky Way tidal forces \citep[e.g.,][]{MeingastAlves2019, Roeser2019, Meingast2019, Beccari2020, Kounkel2019, Jerabkova2019, Ratzenboeck2020, Meingast2021, Jerabkova2021, Kerr2021, Kamdar2021}. Secondly, due to the low number of available radial velocities, about 2\% in the Gaia DR3 database \citep{Vallenari2022arXiv, Katz2022arXiv}, one is, for the most part, restricted to two tangential velocity axes plus the spatial three-coordinate axes derived from Gaia positions, parallaxes, and proper motions (5D phase space). Thus, even under the assumption of perfectly Gaussian-distributed 3D velocities within clusters, the projection on the sky distorts the multivariate Gaussian (5D space) into arbitrary shapes depending on the orientation, distance, and size of the stellar cluster. To make matters worse, stellar cluster members constitute a minute subset of the Gaia data, with unrelated field stars creating background noise that is not easily removable in the 5D space. Thus, the feature space consists of stellar clusters of various shapes and densities embedded in a sea of noise. 

To tackle the challenge of identifying sub-populations in a star-forming region, we have developed a method that analyses the topological structure of the 5D density field spanned by 3D positions and 2D tangential velocities. We apply a fast modality test procedure, which introduces a measure of significance to peaks in the density distribution, thus, providing an interpretable cluster definition. This clustering method is called Significance Mode Analysis, or \texttt{SigMA}, and it is designed to extract co-spatial and co-moving stellar populations from large-scale surveys such as ESA Gaia. 

The goal of this paper is to present the \texttt{SigMA} method, and apply it to the Scorpius-Centaurus OB association \citep[Sco-Cen,][]{Kapteyn1914-tt,Blaauw1946, Blaauw1952, Blaauw1964a, Blaauw1964b} to identify the different sub-populations, and compare results with recent papers attempting similar goals. Sco-Cen is the closest and best studied OB stellar association \citep[e.g.,][]{deGeus1989, deGeus1992, DeBruijne1999, Preibisch1999, deZeeuw1999, Lepine2003, Preibisch2008, Makarov2007a, Makarov2007b, Diehl2010, Poeppel2010, Rizzuto2011, Pecaut2012, Pecaut2016, Forbes2021}, with an age\,$\lesssim$\,20\,Myr \citep*{Pecaut2012}. These and many other papers in the literature have established Sco-Cen as an important laboratory for star formation, for the characterization of stellar associations, and understanding the impact of massive stars on the ISM and planet formation. Since the advent of large-scale astrometric data from the ESA Gaia mission that started in 2016 \citep{Brown2016}, there has been a renewed interest in this benchmark region focusing on the kinematics and 3D structure of the association
\citep{VillaVelez2018, Wright2018, Goldman2018, Damiani2019, LuhmanEsplin2020, Grasser2021, Squicciarini2021, Kerr2021, Luhman2022a, Schmitt2022, MiretRoig2022b, BricenoMorals2022}. 

In this paper, we present the method \texttt{SigMA} in Sect.~\ref{sec:methods}, using Gaia DR3 data (Sect.~\ref{sec:data}), with an application on Sco-Cen discussed in Sect.~\ref{sec:results}, including comparisons to previous work (Sect.~\ref{sec:compare}). In Sect.~\ref{sec:conclusion} we summarize our findings.

\section{Data}\label{sec:data}

In this work, we apply the newly developed method presented in this paper, \texttt{SigMA}, to Gaia DR3 data at and around the Sco-Cen OB association. To this end, we select a box of about $1.5 \times 10^7$\,pc$^3$ from the Gaia DR3 Archive \citep{Vallenari2022arXiv}, which extends well beyond the traditional and well-studied Sco-Cen regions. Several hints in the literature suggest that the Sco-Cen OB association is a larger complex than traditionally defined by \citet{Blaauw1964a}. It includes several star-forming regions that have originally not been assigned to Sco-Cen \citep[e.g.,][]{Lepine2003, Sartori2003, Bouy2015, Kerr2021, Zucker2022}. The box is defined in a Heliocentric Galactic Cartesian coordinate frame (XYZ) within: 
\begin{equation}
    \label{eq:box}
    \begin{aligned}
        -50\,\mathrm{pc} & <\mathrm{X}<250\,\mathrm{pc} \\
        -200\,\mathrm{pc} & <\mathrm{Y}<50\,\mathrm{pc} \\
        -95\,\mathrm{pc} & <\mathrm{Z}<100\,\mathrm{pc}
    \end{aligned}
\end{equation}
The 3D space positions (XYZ) are derived from the Gaia DR3 positions right ascension ($\alpha$, deg) and declination ($\delta$, deg), and the parallax ($\varpi$, mas). The distance ($d$, pc) is estimated from the inverse of the parallax, which is a fairly good approximation of the distance for sources within 200\,pc and with small errors (see also Appendix~\ref{app:box}).

To reduce the influence of spurious measurements, we apply the following quality criteria to the Gaia DR3 data within the selected box:
\begin{equation}
    \label{eq:gaia-quality}
    \begin{aligned}
        &\mathrm{fidelity\_v2} > 0.5 \\
        &\varpi/\sigma_\varpi = \mathrm{S/N}_\varpi > 4.5 \\
    \end{aligned}
\end{equation}
The parameter fidelity\_v2 is a classifier to identify spurious sources in the Gaia DR3 and eDR3 \citep{Brown2021} catalogs, developed by~\citet{Rybizki2022}, which can be used to select high fidelity astrometry. The \texttt{parallax\_over\_error} cut (similar to a S/N threshold) reduces additional uncertainties in distance. This leaves 980,348 sources inside the box to which we apply the \texttt{SigMA} clustering algorithm, which we describe in Sect.~\ref{sec:methods}. In Appendix~\ref{app:box} we give more details on data retrieval and the choice of the quality criteria. 

The clustering is done in a 5D phase space, using the 3D spatial coordinates XYZ in pc, and the 2D tangential velocities $v_{\alpha}$ and $v_{\delta}$ in km\,s$^{-1}$, as derived from the observed proper motions ($\mu_\alpha^* = \mu_\alpha \cos(\delta)$, $\mu_\delta$) and parallaxes (see Appendix~\ref{app:box}).
The tangential velocities $v_{\alpha}$,$v_{\delta}$ are strongly influenced by the Sun's reflex motion.
If this is not accounted for, the distribution of Sco-Cen members in tangential velocity space is strongly correlated and depends on a cluster's position and apparent size (see Fig.~\ref{fig:vtan_theoretical}). Such non-linear relationships contradict a central underlying assumption of many clustering algorithms. These clustering methods assume a universally valid metric, which implies a global correlation behavior (e.g., the commonly used Euclidean norm assumes no correlation between input features). To avoid formulating a locally adaptive metric, we transform tangential velocities to velocities relative to the local standard of rest (LSR), written as $v_{\alpha,\mathrm{LSR}}$ and $v_{\delta,\mathrm{LSR}}$ in km\,s$^{-1}$. This transformation reduces the influence of the reflex motion of the Sun. 

We use the barycentric standard solar motion relative to the LSR from \citet{Schoenrich2010}, while there are different values in the literature, which would give slightly different resulting motions \citep[e.g.,][see also Appendix~C in \citealp{Grossschedl2021}]{KerrLyndenBell1986, FrancisAnderson2009}. However, the differences are only marginal and they are not relevant for our purposes, since the main goal is to reduce the strong positional correlation, which is applied consistently to all stars when deciding for one standard Solar motion correction. 
The effect of the transformation on tangential velocities is highlighted in Fig.~\ref{fig:vtan} and Appendix~\ref{app:vtan}. The LSR conversion of the proper motions is accomplished with Astropy, as outlined in Appendix~\ref{app:box}. Finally, the different dimensions are scaled to each other, as described in the methods in Sect.~\ref{scaling-factors}.

For validation purposes, in Sect.~\ref{noise-removal} we use Gaia DR3 radial velocities \citep{Katz2022arXiv} to remove noise. However, for our clustering, we do not use the third velocity dimension (radial velocities, $v_r$), since Gaia DR3 only includes $v_r$ for about 2\% of the sources with parallaxes (or about 20\% in the selected Sco-Cen box if considering sources with $\sigma_{v_r}<2$\,km\,s$^{-1}$). Adding auxiliary $v_r$ data from other surveys would improve the statistics, but would lead to a very inhomogeneous data sample with 6D phase space information. Therefore, we restrict our clustering procedure to the 5D phase space, as provided by Gaia, allowing us to create a homogeneous and more complete overview of the existing clusters in regions like Sco-Cen. Moreover, by focusing on the 5D phase space, we can create a method that does not strongly rely on radial velocity information, which can then be used more widely on larger data samples. 


\section{Methods}\label{sec:methods}

In this section, we first give a brief overview of the basic definitions of several widely used clustering algorithms, which leads to detailed explanations on the buildup of the Significance Mode Analysis clustering algorithm (\texttt{SigMA}) in Sect.\,\ref{sec:sigma}, as developed in this work. An in-depth description of related work underlying \texttt{SigMA} can be found in Appendix~\ref{app:methods}.

\subsection{Clustering algorithms: a brief review}
Understanding the Milky Way, or any object in the Universe, is directly linked to the quantity and quality of the available data. Nowadays, the biggest effort is no longer data collection, but its large size and high-dimensionality greatly impact all parts of the analysis pipeline --  storage, processing, modeling, and interpretation. 
``Big data'' usually contain extensive information, diversity, and complexity, and thus, we require more complex methods to model its observations. However, many traditional analysis techniques have time and memory complexities that fail to perform under millions or even billions of data samples~\citep{Kumar:2020}. Consequently, many studies start with an exhaustive pre-filter step to improve downstream analyses~\citep[e.g.,][]{Zari2019, Kerr2021}.

To deal with large complex data, new interpretive methods need to be tailored to the particular scientific question, in our case, the identification of co-moving and coeval groups of stars inside the 1+ billion stars in the Gaia archive.
Clustering analysis, or unsupervised machine learning, has recently become essential to the identification of coeval stellar structures. Clustering aims to obtain an organization of data points into meaningful groups. However, due to the lack of labeled data, partitioning into ``meaningful'' clusters is in general an ill-posed problem \citep{CORNUEJOLS:2018}. The choice of the algorithm and its parameters have to match the problem at hand. 
Clustering methods can generally be split into space partitioning algorithms \citep[e.g., $K-$Means,][]{MacQueen:1967}, hierarchical algorithms \citep[e.g., single linkage clustering,][]{Johnson:1967}, density based techniques~\citep[e.g., DBSCAN,][]{Ester:1996}, and model-based or parametric clustering algorithms \citep[e.g., Expectation Maximization,][]{Dempster1977}. For an introduction into cluster analysis and classical methods, see, for example, \citet{Jain:1999}.  

From this list of clustering methods, parametric clustering algorithms and density-based methods are commonly used on astronomical data sets \citep{Kuhn2017, Hunt2021}. In the following, we give an overview of model-based clustering algorithms in Sect.~\ref{sec:parametric_clustering} with an extension to Bayesian formulation in Sect.~\ref{sec:bayesian}. In Sect.~\ref{sect:nonparam_clustering} we present density-based clustering methods, which build the foundation for \texttt{SigMA}, discussed in Sect.~\ref{sec:sigma}.

\subsubsection{Parametric clustering}\label{sec:parametric_clustering}
Parametric clustering algorithms are appealing because of the probabilistic interpretation of the clusters these algorithms generate. The model-based approach introduces a finite mixture of density functions of a given parametric class. The clustering problem reduces to the parameter estimation of the mixture components, which is typically done using the expectation-maximization (EM) algorithm \citep{Dempster1977}. The EM algorithm tries to find maximum likelihood estimates of given parameters iteratively. A popular approach is to model the mixture components as multivariate Gaussian density \citep[e.g.,][]{Gagne2018c, Cantat-Gaudin2019b, Kuhn2017}. 

A considerable downside that limits the versatility of parametric clustering algorithms is their dependence on the unknown number $k$ of mixture components. Depending on data characteristics such as dimensionality, the number of samples per group, and group separation, determining $k$ is a difficult problem. Addressing this problem is paramount, as the resulting model is very sensitive to the choice of $k$~\citep{Celeux:2018}. 

Although model selection methods such as Akaike Information Criterion \citep[AIC,][]{Akaike:1974} and Bayesian Information Criterion \citep[BIC,][]{Schwarz:1978} aim to provide a principled approach on selecting $k$ they make the rather restricting assumption that the data is sampled from a model within the collection to be tested. This assumption can result in overestimating the number of $k$ in practical situations~\citep{Celeux:2018}. Further, BIC and AIC only work well in cases with plenty of data samples, well-separated clusters, and a well-behaved background distribution~\citep{Hu:2003}. These circumstances make extracting clusters with a low signal-to-noise ratio difficult, especially in the low-density regime. 
Many of the above-presented problems can be mitigated by framing them in a Bayesian analysis approach.

\begin{figure*}[!t]
    \centering
    \includegraphics[width=0.99\textwidth]{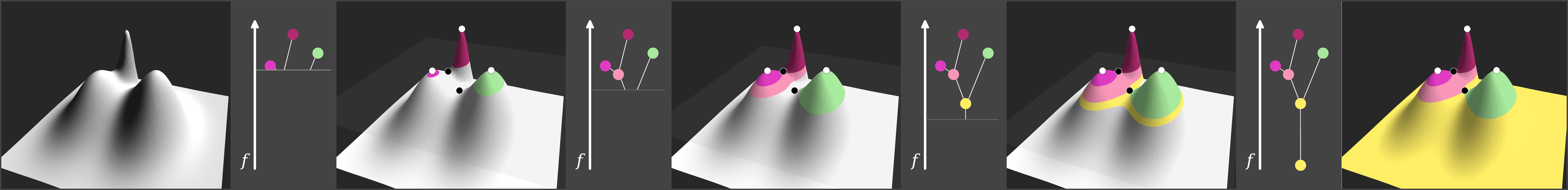}
    \caption{Level set method generating a hierachical merge tree. Via a continuous change of $\lambda$ from $\infty$ to $-\infty$ a new component is created at each maximum (white points). At each saddle point (black points) components are merged. The merge tree is fully computed when $\lambda$ reaches the global minimum.} 
    \label{fig:level_set}
\end{figure*}

\subsubsection{Bayesian clustering approach}\label{sec:bayesian}
The Deviance Information Criteria for missing data models~\citep[DIC,][]{Celeux:2006} is commonly mentioned as a Bayesian model selection criteria. However, in a Bayesian setting, the unknown number $k$ of mixture components can naturally be treated as a random variable that is estimated jointly with the component-specific parameters.

Notably, two Bayesian formulations of selecting $k$ exist, finite and infinite mixtures models. Finite mixture models often rely on reversible jump Markov Chain Monte Carlo~\citep[RJMCMC,][]{richardson1997bayesian} which can navigate between finite mixtures densities with variable $k$. Similarly, sparse finite Gaussian mixtures~\citep{Malsiner:2016}  involve specifying sparse priors on the mixture parameters and can be performed using classical MCMC methods. 
In contrast, nonparametric Bayesian approaches are based on mixture models with countably infinite number of components. In this case, the prior over the mixing distribution takes typically the form of a Dirichlet process~\citep{Mueller:2013}.

Although Bayesian analysis methods provide well-established methods for identifying the number of clusters $k$, fitting models to data requires a statistical model that can generate the data set reasonably well~\citep{Hogg2010}. The observed morphological structure of co-moving stellar systems in position space is significantly more intricate than simple multivariate Gaussians. In recent years many new cluster shapes such as tidal tails~\citep{MeingastAlves2019, Roeser2019, Jerabkova2021}, streams~\citep{Meingast2019}, strings~\citep{Kounkel2019}, rings~\citep{Cantat-Gaudin2019a}, snakes~\citep{Wang2021}, and pearls~\citep{Coronado2022} have been identified. Additionally, the projection of 3D space velocity onto the celestial sphere provides another degree of complexity that requires a flexible clustering scheme.

The variety of observed cluster shapes and the possibility of yet undetected morphological structures make parametric clustering impracticable and error-prone. Instead, we focus on non-parametric clustering methods that use an alternative cluster definition that does not depend on probabilistic models.

\subsubsection{Non-parametric, density-based clustering}\label{sect:nonparam_clustering}
The premise of non-parametric density-based methods states that the observed data points\footnote{In the following, bold, lower-case variables denote $d$-dimensional vectors.} $X = \{\Vec{x}_1, \dots, \Vec{x}_N\}$ with $\Vec{x}_i \in \mathbb{R}^d$ are drawn from an unknown density function $f$.
The goal of non-parametric cluster analysis is then to understand the structure of the underlying density function, which is estimated from data. In one of the earliest formulations,~\citet{Wishart:1969} argues that clusters are data samples associated with modes in $f$. The work proposed by~\citet{Koontz:1976} and the widely used Mean-Shift algorithm and its variants~\citep{Cheng:1995, Comaniciu:2002, Vedaldi:2008} are examples of this \emph{mode-seeking} category.

Mode-seeking methods proceed to group the data by locating local peaks in $f$ and their corresponding attraction basins. Attraction basins are regions in which all gradient trajectories converge into one single peak. However, the gradients and modes are highly dependent on the density function approximation $\hat{f}$. To increase the robustness of the result, Mean-Shift, for example, seeks to reduce random fluctuations by employing a smoothing kernel to $\hat{f}$. The introduction of an extra parameter shifts the issue to the user, who is tasked to carefully select the non-intuitive smoothing factor in order to obtain a satisfying clustering result. Moreover, the time complexity of at least $\mathcal{O}(N^2)$ makes them not great candidates for application to astronomical data sets.

\citet{Hartigan:1975} proposed a similar definition of clustering in which a cluster is defined as the connected components of the level-sets\footnote{Often also referred to as superlevel-sets.} of $f$. Given a data set $X$ drawn from an unknown density function $f$ which has compact support $\mathcal{X}$ we can formally write the resulting level-sets for the threshold $\lambda$ as:
\begin{equation}\label{eq:level_set}
    L(\lambda) := \{ \Vec{x} \in \mathcal{X} ~ : ~ f(\Vec{x}) \ge \lambda \}
\end{equation}
Thus, $L(\lambda)$ constitutes a set of connected components which we identify as clusters. Varying the parameter $\lambda$ from $\infty$ to $-\infty$ gives a hierarchical data summary, called the merge tree. Figure~\ref{fig:level_set} highlights the generation of such a merge tree which builds the basis for hierarchical density based clustering. For more details see Appendix~\ref{app:methods}.

In the level-set framework, popular clustering algorithms such as DBSCAN can be simply thought of as a single level which is obtained by fixing $\lambda$. DBSCAN avoids estimating the data density explicitly, by employing a radius parameter, usually called $\epsilon$, along with a minimum number of points parameter, \texttt{min\_points}. Clusters are defined as connected regions of points that contain at least \texttt{min\_points} within $\epsilon$-sized shells around them.

The connected components of the level-set $L(c)$ are the resulting clusters while the remaining data is treated as noise. However, the choice of the parameter $\lambda$ which is related to DBSCAN's $\epsilon$ parameter, is ambiguous, a task which gets especially challenging when the number of clusters varies greatly between levels. We find a reflection of this difficulty in choosing the right parameters in the astronomical literature, which employs a variety of different heuristics to select the parameter $\epsilon$ \citep[e.g.][]{Castro-Ginard2018, Zari2019, Fuernkranz2019, Hunt2021}. 

For many data sets containing clusters with variable densities, employing a single threshold $\lambda$ cannot reveal all peaks in $f$. A hierarchy of clustering solutions can be obtained by considering all possible threshold values at once, see Fig.~\ref{fig:level_set}.

\begin{figure*}[t!]
    \centering
    \includegraphics[width=0.99\textwidth]{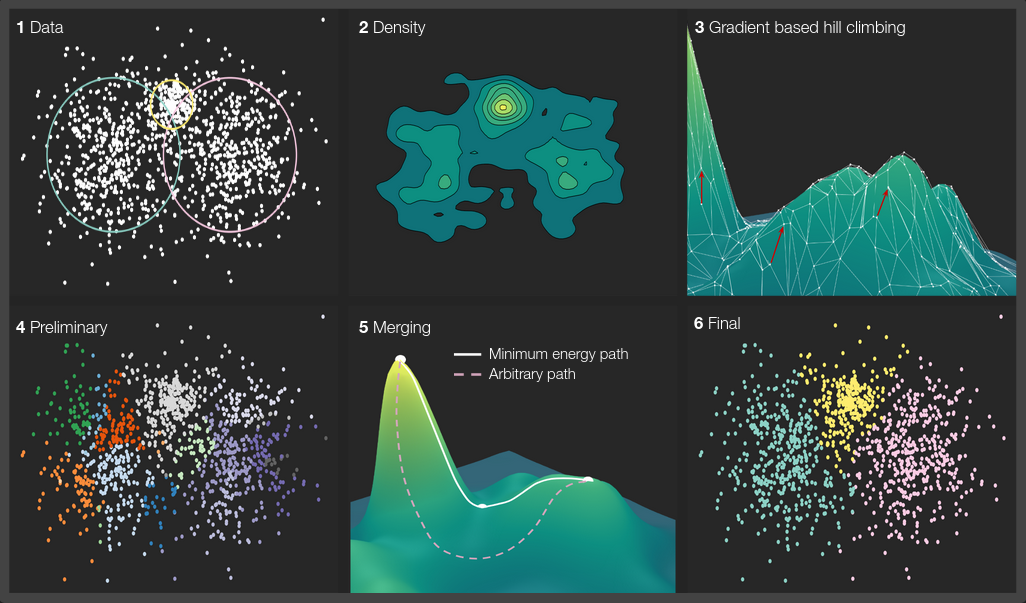}
    \caption{The proposed clustering process \texttt{SigMA}, highlighted on a 2D toy data set of three Gaussians with variable covariance matrices and means. \textbf{(1)} The generated toy data set consisting of three bivariate Gaussians is shown in white alongside $2\sigma$ confidence ellipses in color. \textbf{(2)} The clustering procedure starts off by estimating the density of the input data. \textbf{(3)} Next, a graph-based hill climbing step is performed in which points are propagated along gradient lines towards local peaks. 
    \textbf{(4)} This gradient propagation results in a preliminary segmentation of input samples which typically is far too fine-grained. \textbf{(5)} These segmented regions are iteratively merged with a parent mode if a modality test along the ``minimum energy path'' detects no significant density dip.
    \textbf{(6)} The final segmentation retains all three clusters.
    } 
    \label{fig:algorithmic_schematic}
\end{figure*}

\subsection{\texttt{SigMA}: Significance Mode Analysis}\label{sec:sigma}
This section describes our clustering pipeline, \texttt{SigMA}, which builds on several established methods from data mining and statistics, that we discuss in further detail in Appendix~\ref{app:methods}. 

\texttt{SigMA} is tuned to astrometric data provided by Gaia, and aims at producing astrophysically meaningful clustering results.
Our technique seeks to identify modal regions in the data set (5D phase space) which are separated by dips. 
By applying a modality test for each pair of neighboring modes, we obtain a clustering result with measures of significance. The workflow is schematically highlighted in Fig.~\ref{fig:algorithmic_schematic}. 
A modal region is defined as the set of points that all end in a particular mode when following the path tangent to the gradient field at each point. It is important to note that modal regions fully segment the data set, as seen in Fig.~\ref{fig:algorithmic_schematic} (panel 6). Thus, modal regions are a mixture of cluster members and field stars, while the field stars will be removed as noise as outlined in Sect.~\ref{noise-removal} and shown in Fig.~\ref{fig:cluster-noise-classifier}.

\subsubsection{A fast modality test procedure}\label{sect:modality_test}
We consider the hypothesis test introduced by~\citet{Burman:2008} which examines the modality structure of a path between two peaks in the density. Conceptually two neighboring peaks are ``true'' clusters in the data if there exists no path between them that does not undergo a significant dip in density. 

Given the $d$-dimensional data $X = \{\Vec{x}_1, \dots, \Vec{x}_N\}$ drawn from $f$ and any point $\Vec{r}$ on a path connecting two modes $\Vec{c}_i$, $\Vec{c}_j$ in $f$, ~\citet{Burman:2008} show that 
\begin{equation}
    \label{eq:4_text}
    \widehat{\text{SB}}(\Vec{r}) = d \sqrt{k/2} \left[ \text{log}~d_k(\Vec{r}) - \text{max} ( \text{log}~d_k(\Vec{c}_i),~\text{log}~d_k(\Vec{c}_j) \right]
\end{equation}
is asymptotically standard normal distributed. Here $d_k(\Vec{z})$ denotes the distance to the $k$'th nearest neighbor of the point $\Vec{z}$. The null hypothesis of uni-modality is rejected at significance level $\alpha$ if
\begin{equation}
    \label{eq:5_reject_cond_text}
    \widehat{\text{SB}}(\Vec{r}) \ge \Phi^{-1}(1-\alpha)
\end{equation}
where $\Phi$ is the standard normal cumulative distribution function (cdf). For a more thorough derivation of Eq.~(\ref{eq:4_text}) and Eq.~(\ref{eq:5_reject_cond_text}) see Appendix~\ref{app:modality_test}.

Since Eq.~(\ref{eq:4_text}) processes a single point rather than a complete path, the modality test in Eq.~(\ref{eq:5_reject_cond_text}) describes a pointwise procedure. \citet{Burman:2008} employ the test with samples generated along the straight line connecting two modal candidates to determine the modality for an entire path. The null hypothesis is rejected if any single test fulfills Eq.~(\ref{eq:5_reject_cond_text}). However, this procedure only applies to convex clusters and does not scale well as tens to hundreds of distance computations along each path increase the run-time drastically. 

Instead of computing the test statistic $\widehat{\text{SB}}(\Vec{r})$ for multiple values of \Vec{r}, we propose to limit the calculation to only a single realization. Importantly, reducing the number of pointwise evaluations of pointwise tests does not interfere with the distributional assumption of the test statistic itself. \citet{Burman:2008} show that the test statistic $\widehat{\text{SB}}$ along the entire path $p$ with $p = [\Vec{r}_1, \dots \Vec{r}_N]$ follows an $N$-dimensional multivariate normal distribution with zero mean and identity covariance matrix under null hypothesis. Thus, the null hypothesis is independent of the number $N$ of pointwise tests performed.

Modifying the modality test procedure to a single evaluation of the test statistics reduces the overall statistical power of the test. Because we undersample the path between two modes, we decrease the chance of sampling in places where a significant drop in density occurs. Therefore, the probability of type II errors increases, i.e., failing to reject the null hypothesis even though it is false. To maintain statistical power while also extending the test procedure to non-convex cluster shapes, we analyze the nature of possible connections between modal candidates in the data.

Of all possible paths between two peaks, only the ``minimum energy path'' (MEP) needs to be considered. The MEP is the optimal solution for the problem of finding the continuous path from one peak to another through input space $\mathcal{X}$ with highest minimal density. Thus, the density dip along the MEP is the minimal possible dip that can exist between two neighboring peaks. 

Given a set of initial modal candidate regions in $\hat{f}$ the MEP leads over the connecting saddle point when moving from one mode to another. At the saddle point position, the path reaches its global density minimum. Figure~\ref{fig:algorithmic_schematic} (panel 5) schematically illustrates two possible paths, the MEP and a second arbitrary path.

To effectively reduce the number of pointwise tests without losing all statistical power we need to evaluate Eq.~(\ref{eq:4_text}) in areas close to the maximum density dip while ignoring other areas irrelevant for the rejection decision. This maximum density dip at point $\Vec{s}$ maximizes the test statistic and, thus, dominates the test procedure. Due to the test statistics proportionality to the distance $d_k(\Vec{s})$, its value is maximal when the density is minimal. 

For two neighboring modal regions the modality test procedure can, therefore, be reduced to a single pointwise test at the saddle point $\Vec{s}$ connecting the two peaks. As the saddle point governs the modality test, we can assign a $p$-value which takes the following form:
\begin{equation}
    \label{eq:6_pvalue_text}
    p = 1 - \Phi \left(d \sqrt{k/2} \left[ \text{log}~d_k(\Vec{s}) - \text{max} ( \text{log}~d_k(\Vec{c}_i),~\text{log}~d_k(\Vec{c}_j) \right] \right) 
\end{equation}
At the end of Appendix~\ref{app:modality_test}, we empirically show that these assumptions hold and introduce a small correction factor to the variance of the standard normally distributed test statistic under $H_0$ that is valid for Gaia phase space data.

Determining the saddle point is discussed in the following section. If all density minima lie on the boundary of modal regions, the saddle point of two neighboring modes lies at their common border. Using this monotonous property assumption, we aim to provide a fast and yet accurate test procedure to examine the modality structure of the data.

\subsubsection{Identifying and pruning modal candidates}\label{sec:model_candidates}
To identify modal regions from the data set $X$ we implement a graph-based, hill-climbing algorithm analogous to~\citet{Koontz:1976} where the vertex set of the graph $G$ represents the data $X$. The initial modal search is performed in one pass over the vertices of $G$ sorted in descending $\hat{f}$-order. 

A data point is defined as a local mode of $\hat{f}$ if all its neighbor connections have lower densities. Alternatively, points are propagated according to their slope in $\hat{f}$. Each point is iteratively assigned to neighbors with maximum $\hat{f}$-value (see Fig.~\ref{fig:algorithmic_schematic}, panel 3, for a schematic illustration). After this pass the data is separated into $m$ disjoint modal sets $\Vec{M} = \{ M_1,\dots,M_m \}$.

Since graph-based hill-climbing procedures are susceptible to perturbations in $\hat{f}$, a second pass is needed to merge insignificant modal regions into their stable parent mode. To determine the merge order we compute the cluster tree of $\Vec{M}$. As described in Sect.~\ref{sect:nonparam_clustering}, the cluster tree is obtained by varying the density threshold $\lambda$ from $\infty\to -\infty$ and registering modal regions when $\lambda$ passes through a peak in $\hat{f}$ and their unification when $\lambda$ passes through the respective saddle point. To finalize the cluster tree we need to identify the saddle points between modal regions of $\Vec{M}$. 

We determine the saddle point between two modes via an edge search in $G$. Specifically, we consider edges which connect vertices that lie in different modal sets. We assume extracted modal regions are proper ascending manifolds. Thus, the modal regions are devoid of local minima on the inside, which only lie on the border; consequently, saddle points are found at the common boundary of both regions. The ``saddle edge'' represents the bridge between two modal regions where the density is maximal. We define edge density as the minimum density along the connecting line segment. To account for density dips along the edge path while limiting the number of distance computations, the edge density is set to be the minimum density between its two vertices and the density at the geometric mean of the vertex positions. The corresponding saddle point density between two adjacent modal regions is approximated by this edge density.

The merging of spurious modes then proceeds by iterating over the set of predetermined saddle points sorted in descending $\hat{f}$-value order. At each step, the uni-modality test in Eq.~(\ref{eq:6_pvalue_text}) is evaluated and neighboring modal regions are merged if the respective $p$-value exceeds the significance level $\alpha$. Therefore, the significance level $\alpha$ provides an immediate and meaningful way to simplify the initial cluster tree.

\subsection{Parameter selection}\label{sect:parameter_selection}
In the following, we discuss various parameter choices which affect the final clustering result. The presented mode seeking methodology is agnostic to the choice of the (1) graph used in the hill-climbing step, (2) density estimator, and (3) scaling factors between positional and velocity features. In the following, we will explain our decisions on these three algorithmic aspects.

\subsubsection{Graph}\label{sec:graph}
The choice of the graph directly affects the gradient approximation. For instance, in a complete graph where every pair of vertices are connected via an edge, the graph-based gradient approximation loses its locality-meaning entirely. In this case, the hill-climbing algorithm merges each vertex with the densest point in the data set on the first pass. Thus, over-connected graphs lead to clusters that falsely merge numerous distinct modes in the data set. 

Conversely, under-connected graphs such as minimum spanning trees restrict the gradient estimation too much, producing vast amounts of spurious clusters. Furthermore, the low number of neighboring vertices greatly restricts the possible paths between two initially formed modes. Thus, under-connected graphs introduce significant errors in determining saddle points, which drastically compromises the validity of extracted modal regions.
We consider \emph{empty region graphs} (ERG) to strike a balance between over and under connecting points in the data set $X$. In an ERG, a vertex between two points is created if a given region around them does not contain any other point \citep[see][for a review]{Jaromczyk:1992}. 

The $\beta$-skeleton~\citep{Kirkpatrick:1985} is a one-parameter generalization of an ERG where $\beta$ determines the size of the empty region. For $\beta = 1$ the graph becomes the Gabriel graph~\citep{Gabriel:1969}, while for $\beta < 1$ and $\beta > 1$ edges are added or removed from it, respectively. \citet{Correa:2011} find that critical point searches (important for topological decomposition, clustering, and gradient estimation) are more accurate with $\beta$-skeletons, with $\beta < 1$ compared to $k$-nearest neighbor graphs and the Gabriel graph. Since the number of vertices grows very fast in size as $\beta$ gets smaller we choose a value of $\beta = 0.99$.

Adopting a $\beta$-skeleton on our 5D data we find that points have on average approximately $50$ neighbors. To reduce the chance of separate modal regions being connected via vertices and, thus, erroneously merging in the first hill-climbing step, we prune the initially computed graph in a post-processing step. We remove vertices that show a significant density dip as one moves from one vertex to another. For simplicity, we assume that the saddle point lies at the arithmetic mean of the two vertex points.

\subsubsection{Density estimation}\label{sec:density_est}
As the graph choice, density estimation is a core part of the algorithmic pipeline that affects gradient propagation and, consequently, the initial mode finding step (see panels 2 and 3 in Fig.~\ref{fig:algorithmic_schematic}). Since we cannot describe the complex stellar distribution via parametric models, we employ a model-agnostic, non-parametric estimator for the underlying density.

The most popular non-parametric density descriptors are kernel density estimation (KDE) and k-nearest-neighbor ($k$-NN). The KDE technique estimates the density $f$ by convolving the data with a symmetric kernel function. The bandwidth parameter can be thought of as the standard deviation of the kernel, which determines the smoothing effect of convolution. A gradual increase in bandwidth and its impact on the density is shown in Fig.~\ref{fig:scalespace}.
The $k$-NN method takes a more naive approach to estimate the underlying density. The density value at any given point in the phase space is inversely proportional to the distance to its $k$-th nearest neighbor. 

The KDE inherits the smoothness properties of the kernel. Thus, the density becomes infinitely differentiable for a Gaussian kernel. Conversely, the $k$-NN density estimate is not smooth and, in fact, not even continuous. Despite its non-continuous nature, the $k$-NN density estimation method has several advantages for modal clustering. Notably, \citet{Dasgupta:2014} show that point modes of a $k$-NN density estimate approximate the true modes of the underlying density function. The nearest neighbor method is also able to provide a more accurate estimate of high density regions compared to the kernel method \citep{Burman:1992}. 

In contrast to KDE, the computational cost of nearest neighbor methods is highly efficient due to the use of kd-tree\footnote{Short for $k$-dimensional tree.} queries that provide desirable memory complexity~\citep{Bentley:1975}. Further, choosing the number of neighbors $k$ is more straightforward than the bandwidth parameter for KDE. Finally, the locality of the $k$-NN approach provides a versatile method to determine densities when structures exist at different densities scales. Since KDE employs a constant bandwidth, it can only adapt to a single characteristic density scale. A fixed, ``intermediate'' bandwidth may adequately resolve medium-density clusters when structures are present at various scales. However, fine-grained and large-scale patterns will be over-smoothed or under-smoothed, respectively. 

We employ a $k$-NN estimator to approximate the density function considering these advantages. Specifically, we use a density estimator based on the Distance To an empirical Measure (DTM) described by \citet{Biau:2011}. It is a weighted $k$-nearest neighbor estimate, which incorporates distances $d_1, \dots, d_k$ to all nearest neighbors up to $k$. The DTM is a distance-like function robust to the addition of noise and is used to recover geometric and topological features such as level sets. It is defined as follows,
\begin{equation}\label{eq:d_dtm}
    d_{m}(\Vec{x}) = \sqrt{ \frac{1}{k} \sum_{\Vec{y}_i \in N_k(\Vec{x})} ||\Vec{y}_i-\Vec{x}||^2 }
\end{equation}
where $N_k(\Vec{x})$ is the neighborhood point set of $\Vec{x}$ of size $k$. In other words, the distance to empirical measure takes the form of a mean distance from the point $\Vec{x}$ to its $k$ nearest neighbors. The density estimator is defined via the inverse of this quantity,
\begin{equation}\label{eq:f_dtm}
    \hat f_{m}(\Vec{x}) = \frac{1}{n V_d} \left( \frac{\sum_{j=1}^{k} j^{2/d}}{k d_{m}^2(\Vec{x})} \right)^{d/2}
\end{equation}
where $V_d$ denotes the volume of the $d$-dimensional unit ball and $n$ is the number of data points. 
Since in our use case the order of density values is important, we can ignore constant normalization terms in Eq.~(\ref{eq:f_dtm}).

The $k$-NN algorithm is not only used to estimate the density but also during the modality test procedure (see Sect.~\ref{sect:modality_test}). Since classical $k$-NN, as employed in the modality test, automatically ignores points within its k-distance, \texttt{SigMA} has a built-in limit to the size of structures it can resolve. This allows us to determine a lower bound on the velocity dispersion of a population that \texttt{SigMA} can identify. We find the minimally resolvable tangential velocity dispersion to be $0.5$\,km\,s$^{-1}$ by analyzing the distribution of k-distances with a lower bound on $k=15$, which we also assume to be the minimum cluster size. Clusters with lower velocity dispersion get smoothed to at least this minimum dispersion. This value increases as $k$ gets larger.

\subsubsection{Scaling factors} \label{scaling-factors}
The clustering analysis of co-moving populations in position and velocity occurs in a combined positional and kinematic phase space. Distance relationships among stars are needed to express densities and build a graph from the input data. Since tangential velocities are measured in km\,s$^{-1}$ and galactic coordinates in pc, both sub-spaces have different ranges. Significant range discrepancies between dimensions influence the clustering process as it directly impacts the distance function. Individual 1D distance contributions along feature axes with narrow ranges can be ignored when features with large standard deviations are present. Hence, we consider scaling factors between positional and kinematic feature sub-spaces.

Scaling factors $c_i$ put weight on specific sub-spaces to in- or decrease their importance in the clustering process. The multiplicative factor affects the range of feature axes impacting the distance function. Thus, scaling factors $c_i > 1$ increase the distance to objects in a given dimension $i$, increasing their importance in the process. We apply the same scaling $c_{v}$ to both tangential velocity axes while leaving the positional axes unchanged with $c_{x} = 1$. \texttt{SigMA} is applied to the following set of dimensions $\mathcal{D}$:
\begin{equation}
    \mathcal{D} = \{X, Y, Z, c_{v} \times v_{\alpha,\mathrm{LSR}}, c_{v} \times v_{\delta,\mathrm{LSR}} \}
\end{equation}

Theoretical considerations of the scaling relationship $c_x/c_v$ depend on various initial cloud and cluster configurations and interactions. However, the estimation of these influences is plagued by substantial uncertainties. Instead, we aim to determine a suitable scaling factor empirically by considering successful past extractions. Since the tangential velocity is inverse proportional to parallax, our goal is to extract a relationship between a stellar cluster's distance and its scaling factor. 

The Sco-Cen association is at a distance of about 100--200\,pc from us. To model the empirical distance-scaling relationship and subsequently apply it to Sco-Cen, we need data on stellar clusters within at least $300$\,pc. \citet{Cantat-Gaudin2020a} have compiled a list of open clusters in the Milky Way disk. However, using a single cluster census can introduce a bias in the resulting scaling factor as only a single member selection function was used to obtain the sample. Thus, we substitute and add groups covered by \citet{Gagne2018a}, who have used a multivariate Bayesian model to identify members of young associations within $150$\,pc.\footnote{We cross match the Gaia DR1 sources identified by \citet{Gagne2018a} and DR2 sources from \citet{Cantat-Gaudin2020a} with DR3 for more precise astrometry. If a cluster appears in both surveys, we opted for the \citet{Gagne2018a} census. Compared to density-based methods, the mixture of Gaussian densities deals naturally with scaling factors (provided the Gaussian assumption holds). The scaling factors can be compensated (to some extent) in the covariance matrix of the individual Gaussian components.}

\begin{figure}[t]
    \centering
    \includegraphics[width=0.98\columnwidth]{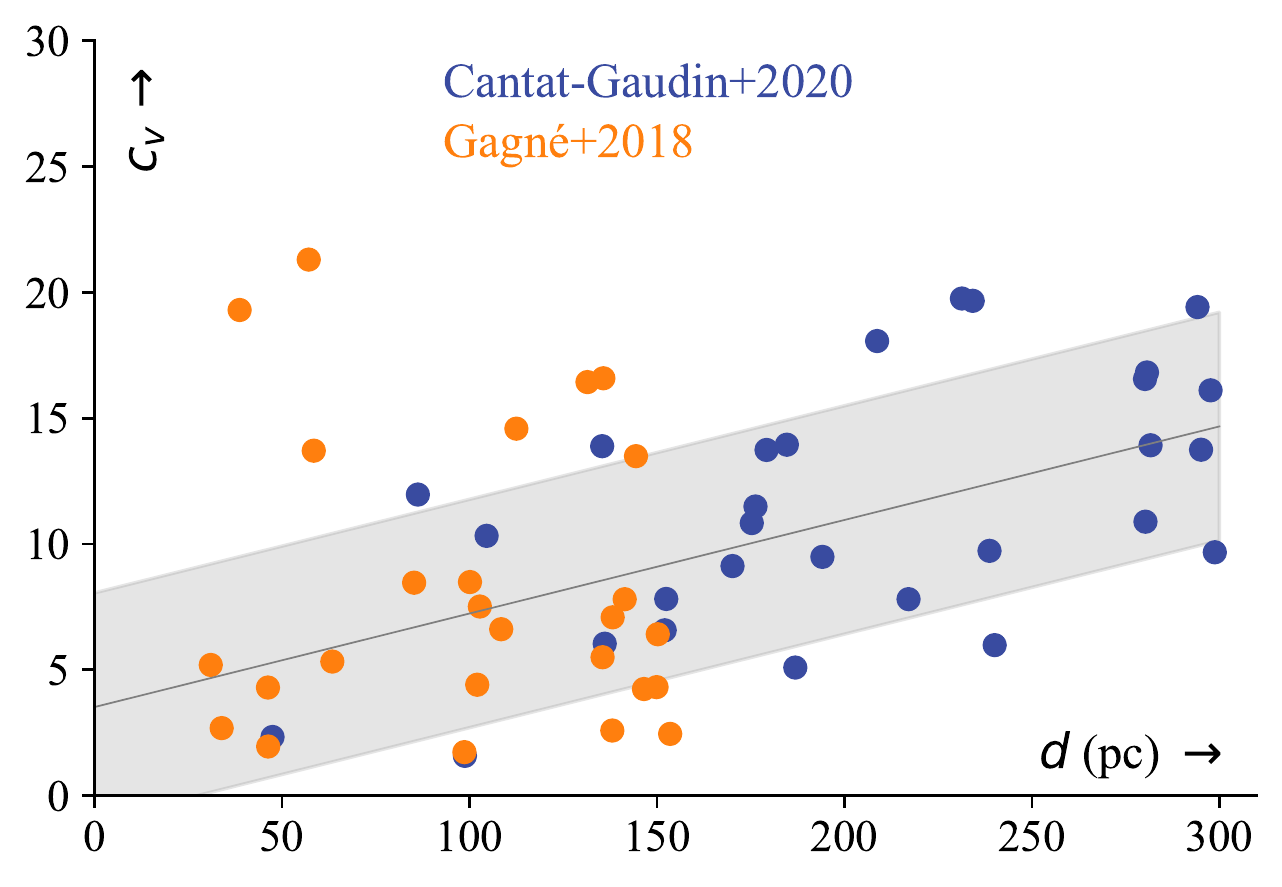}
    \caption{Empirical distance-scaling relationship using data from \citet{Gagne2018a} and \citet{Cantat-Gaudin2020a}. The x-axis represents the distance to stellar groups; the y-axis shows the dispersion ratio of positional over kinematic sub-spaces. This dispersion directly corresponds to the velocity scaling factor $c_v$ as discussed in Sect.~\ref{scaling-factors}.}
    \label{fig:linear_regression}
\end{figure}

The scaling fraction should account for the distance differences between positional and kinematic sub-spaces. To quantify this idea, we consider the distance distribution of sources to the cluster's center in each sub-space. Specifically, we compare the median absolute deviation of sources from their centers in position and velocity space, providing a robust statistic for statistical dispersion. We refer to this ratio of observed dispersion in the respective sub-spaces as the x-v dispersion ratio. The dispersion of cluster members in a given feature provides a measure of the scale of stellar populations in that dimension. Since the x-v dispersion ratio is not one (see Fig.~\ref{fig:linear_regression}), we have to prevent an unequal emphasis of one subspace against the other. To compensate for the bias towards positional axes during clustering, the velocity features must be scaled by a factor $c_v$ equal to the observed x-v dispersion ratio.

Figure~\ref{fig:linear_regression} shows the relation between a cluster's distance and its x-v dispersion ratio, which equivalently is our choice of $c_v$. We identify a linear trend and fit a linear model to the data, the gray band indicates a deviation of one standard deviation away from the mean assuming constant Gaussian model uncertainty. Since we observe several outliers, we use the Huber loss~\citep{Huber:1964}, which is less sensitive to anomalies. 

Using this empirical model, we find mean suitable scaling factors $c_v$ between approximately 5--15, assuming the groups of Sco-Cen are at a distance of about 100--200\,pc. Without the LSR correction, these values translate to a range of 4--9, which are comparable the correction factors by~\citet[][using $v_\alpha$,$v_\delta$]{Kerr2021} who used the values 5 and 6 in their clustering approach (see also Fig.~\ref{fig:vtan} for a comparison of these two velocity spaces).
At first glance, the model suggests sampling values in the range of 5--15 or using the mean 10. However, we also observe a significant scatter around the model that we need to consider. Instead of a single mean scaling factor, we aim to obtain a distribution of values from a given range of distances to the groups we aim to find. 

As discussed in Appendix~\ref{app:scaling_factors}, possible scaling factors can be expressed by the conditional probability integrated over a range of distance values. Given the linear model and associated Gaussian model uncertainties, we find a resulting distribution of scaling factors within distances of 100--200\,pc. Keeping the number as small as possible is essential since we need to perform a separate clustering run for each sample that we draw from the distribution. We generate ten samples, which try to cover the sample space while keeping the underlying probability distribution in mind. The resulting samples can be seen in Fig.~\ref{fig:pdf_cdf}.\footnote{We want to point out that the distance notation in the appendix changes from $d$ to $r$ to minimize confusion in the derivation of the final pdf.} 

We run the clustering pipeline for each scaling fraction sample, creating an ensemble of 10 clustering solutions. By summarizing the (potentially conflicting) results, we obtain a single consensus clustering solution. The consensus result is more robust against noisy data by aggregating multiple clustering solutions. This aggregation technique creates a meta-solution that usually provides better accuracy than any single clustering result can~\citep{Strehl:2002, Vega:2011}. 

A consensus function aims to produce a result which, shares as much information as possible with individual clustering results among the ensemble. In particular, we are interested in robust cluster solutions that exist through multiple velocity scales while ignoring unstable solutions where clusters randomly break apart or merge into others. We outline our consensus clustering approach in Appendix~\ref{app:consensus_clustering} where Fig.~\ref{fig:consensus_pipeline} shows a schematic of the proposed pipeline.

\subsection{The role of measurement uncertainties}\label{sec:uncertainty}
Rigorous integration of measurement uncertainties into the modality testing procedure of \citet{Burman:2008} is a highly complex task, primarily due to the heteroscedastic nature of the uncertainties. Instead, we use a Monte Carlo approach that attempts to approximate the sensitivity of the modality structure under statistical uncertainty. We do this by resampling the data using a Gaussian distribution centered on each point with an appropriate covariance matrix obtained from Gaia data.

Re-computing the modal structure on each resampled data set individually is computationally expensive. Therefore, we aim to study the effect of deviations on the initially computed modal layout instead. Since every merge decision impacts the final modal structure, we must evaluate the impact of uncertainty at each saddle point. 
While looping through all saddle points, we re-evaluate the hypothesis test for each resampled modal and saddle point density. However, testing each hypothesis multiple times increases the likelihood of rejecting an individual null hypothesis. Instead of focusing on single tests, we need to combine these individual tests and simultaneously test the global null hypothesis that no $p$-value is significant. As a result, a global hypothesis test can ``borrow'' information from the other test statistics to gain significance.

A popular method of computing the global $p$-value is  Fisher's method~\citep{fisher:1934}. Fisher's method assumes that individual $p$-values are uniformly distributed in the interval $\left[ 0,1\right]$. Consequently, the negative logarithm follows an exponential distribution: $-\text{log}\,p_i \sim \text{Exp}(1)$. The test statistic $t$ then becomes the sum of the negative log-sum of $n$ $p$-values which follows a $\chi^{2}$ distribution with $2n$ degrees of freedom. 
\begin{equation}\label{eq:fishers_method}
    t = - 2 \sum_{i=1}^{n} \text{log}\,p_i \sim \chi^{2}(2n)
\end{equation}

Fisher’s method is especially attractive in the case of densely packed cluster agglomerates such as Sco-Cen. If clusters have only marginally different velocities and positions, our point-wise test might produce a p-value slightly larger than the rejection threshold. In such cases finding the precise saddle point position is difficult leading to a type II error. Although no single test (or very few) can reject the null hypothesis, many small effects add up in Eq.~\ref{eq:fishers_method} enabling us to reject the global null hypothesis. However, Fisher's method assumes statistical independence between individual tests. Since the resampled data sets are not independent of each other, this assumption is violated to some extent (for a more detailed discussion, see Appendix~\ref{app:resampling}). 

Instead, we employ the Cauchy Combination Test \citep[CCT,][]{Liu:2020} which is similar to Fisher's method in a sense that it is also able to combine multiple individual $p$-values that aggregate multiple small effects. Compared to Fisher's method, the authors show that CCT is still powerful under arbitrary dependency structures among $p$-values. The test statistic $t$ is defined in the following:
\begin{equation}\label{eq:cct}
    t = \sum_{i=1}^{n} w_i \left[ (0.5 - p_i) \pi \right]
\end{equation}
The weights $w_i$ must sum to one and can reflect the power of respective hypothesis tests. Since all tests are performed equally, we distribute the weights evenly by choosing $w_i = 1/n$. 

In practice, we compute distances to the $k$`th neighbor of each initial modal candidate and corresponding saddle points across resampled data sets. The number of resampled data sets limits the proposed procedure, as data generation is costly. Thus, we restrict the number of samples to $n = 50$, rejecting the global null hypothesis at a $5\%$ significance level, hence $t < 0.05$.

\subsection{Noise removal} \label{noise-removal}
Following the procedures described above, we obtain a data set segmentation into prominent peaks by iteratively merging modal regions separated by insignificant dips in density. This segmentation yields a list of non-overlapping areas in the data set without a noise characterization in mind. In principle, each modal region contains a dense core and background population corresponding to the stellar group and field content. In this section, we aim to remove the field star component from the modal region to obtain a final clustering result. 

\begin{figure}[!t]
    \centering
    \includegraphics[width=0.71\columnwidth]{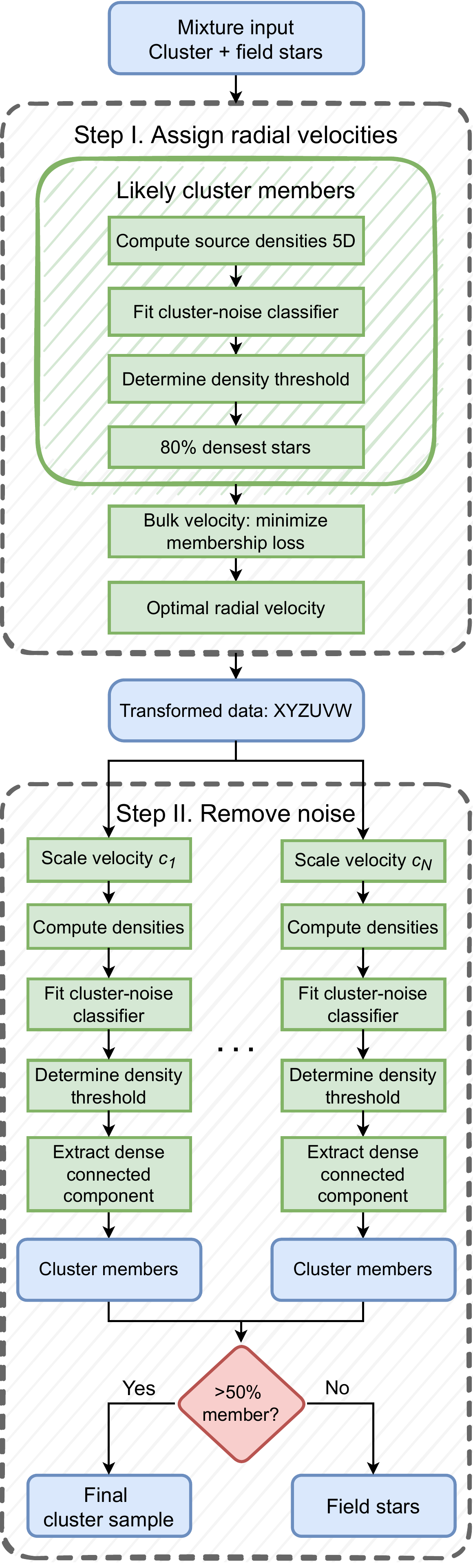}
    \caption{Noise removal pipeline separating cluster members from field stars. The pipeline is split into two main parts. In the first part we aim to assign a radial velocity to each source to transform the data into Galactic Cartesian coordinates. In the second step we use the transformed data to fit several cluster-noise classifiers to separate signal from background. Blue colors represent data products at various steps, green colors denote processes, and red shows decisions. For more details see Sect.~\ref{noise-removal}.}
    \label{fig:noise_removal_pipeline}
\end{figure}

In the following, we discuss the noise removal pipeline, which is schematically highlighted in Fig.~\ref{fig:noise_removal_pipeline}. The noise removal scheme is based on a density-based member selection technique, which we motivate in Sect.~\ref{sec:db_memberselection}. The pipeline is roughly split into two main parts. In the first part, we aim to assign a radial velocity to each source to transform the data into 6D Galactic Cartesian coordinates (XYZUVW). To estimate missing radial velocity information, we determine the cluster's 3D space motion, discussed in Sect.~\ref{sec:bulkvelocity}. In the second part, we describe the automated cluster member selection using so-called \textit{cluster-noise classifiers}, see Sect.~\ref{sec:remove_fieldstars}. Finally, we discuss contamination and completeness estimates of our member selection procedure in Sect.~\ref{sect:contamination_completeness}.

\subsubsection{Density based member selection}\label{sec:db_memberselection}
The identification of signal and background sources can be formulated using a mixture model approach in which cluster and field star populations are modeled directly in phase space. However, due to complex cluster shapes found in the literature (see Sect.~\ref{sec:bayesian}), we cannot create a generative model of the data set at hand. Instead, we return to the density-based formalism of clustering, where we treat clusters as enhancement of density over the background. To select cluster members as over-densities in phase space, we reduce the 5D phase space information to the univariate density information. 
A single density threshold in this univariate space corresponds to an isosurface in the original phase space. 

To automatically obtain a suitable isosurface threshold to separate signal from the background we aim to describe the univariate density distribution as a mixture model. This model should be able to capture the point-wise density distribution of field stars and cluster members. Before fitting a mixture model to data we first need to define the number of mixture components and distributions we use.

In a first approximation, a mixture of two components seems plausible as the algorithm divides the input space into regions containing a signal and background component. By design, each region consists of a single-density peak in phase space. This distributional condition forces the field star component to lie locally around the cluster while exhibiting no extra peaks. To concur to uni-modality, the background distribution is restricted to being uniform or featuring a single-density peak that coincides with the signal mode. As the former is more likely, we assume the field component to be approximately uniform in phase space around a cluster. Uniform distributions in N-dimensional feature spaces translate to a single Gaussian in the univariate density space.

The distribution of cluster star densities is harder to model. Cluster members are commonly modeled as multivariate Gaussians \citep[e.g.][]{Gagne2014, Sarro2014, Crundall2019, Riedel2017}. As discussed in Appendix~\ref{app:dist_density}, given a $k$-NN density estimator, a multivariate Gaussian in phase space approaches a Gaussian distribution in univariate density space as the dimensionality grows. However, observational findings point to more complex morphologies \citep[e.g.,][]{MeingastAlves2019, Roeser2019,  Meingast2019, Kounkel2019, Cantat-Gaudin2019a, Jerabkova2021, Wang2021, Coronado2022} and significant mass being contained in the low-density region outside the cluster core~\citep{Meingast2021}. As discussed in Sect.~\ref{sec:bayesian}, we lack critical information to formulate a generative model for signal distribution in phase space and consequently in univariate density space. Instead of explicitly building a univariate signal model, we employ multiple Gaussian mixture components to describe the point-wise $k$-NN density distribution. Thus, during the fitting procedure, we do not restrict the number of Gaussians components in an effort to provide flexibility to capture more complex distributions. 

To decide on a proper density threshold $\rho_0$, we determine the number of mixture components by minimizing the BIC. The background is automatically identified as the Gaussian component with a low relative mean (i.e., lower point-wise densities), small variance (uniform background component has less relative variance around its mean density than the signal), and large weight (the number of field stars exceeds cluster members by about 100:1). This procedure can be seen in Fig.~\ref{fig:cluster-noise-classifier} where we show an example of two Gaussians fitted to the univariate density data (denoted by $\rho$) of one modal region.

\subsubsection{Bulk velocity estimation}\label{sec:bulkvelocity}
The Gaussianity assumption of density components is appropriate only in the original Cartesian coordinate system. Densities computed from tangential velocities suffer from perspective effects leading to deviations from normality due to the nonlinearity of projections onto the celestial sphere. We find such distortions also empirically when analyzing distributions of various modal regions in projected 2D (see the tangential velocity space in Fig.~\ref{fig:vtan_theoretical}) compared to Cartesian 3D velocities. This effect is reduced by correcting the observed tangential velocities for the Sun's motion, yielding motions relative to the LSR (see Fig.~\ref{fig:vtan}). However, very nearby clusters, which cover large areas in the sky, are still affected by the observer's point of view from Earth. To eliminate these observational effects, we move to the Galactic Cartesian coordinate system. Thus, we transform the data into the six-dimensional space (XYZUVW) to facilitate efficient signal and background models. 

A transformation from proper motion space to a 3D Cartesian velocity space is only possible if radial velocity information is available. However, the majority of radial velocity measurements of sources are missing ($\sim 62\%$ in our box, $80\%$ if sources with $\sigma_{v_r} > 2$\,km\,s$^{-1}$ are removed). Nevertheless, we can exploit the co-moving property of stellar populations. We aim to adopt a similar strategy to~\citet{Meingast2021}, inspired by convergence point ideas \citep[e.g.,][]{vanLeeuwen2009}. The expected radial velocity value can be determined when the 3D bulk motion of stars alongside their positions is known. However, some groups lack enough statistics to compute their bulk motion. Before we can determine individual radial velocities of member stars, we have to estimate the space motion of individual populations. In the following, we describe bulk motion estimation, which provides a mean to estimate an optimal radial velocity. We summarize the process in the first part of Fig.~\ref{fig:noise_removal_pipeline}.

We determine the space motion $\Tilde{\Vec{v}}$ of individual populations of size $n$ by minimizing the following loss function, henceforth called \textit{membership loss}:
\begin{align}
    \label{eq:vel_loss}
    L(\Tilde{\Vec{v}}) &= \sum_{i=1}^{n} \left( 
      \frac{\Delta v_{\alpha, i}^{2}}{\sigma_{v_{\alpha, i}}^{2}} 
    + \frac{\Delta v_{\delta, i}^{2}}{\sigma_{v_{\delta, i}}^{2}}
    + \frac{\Delta v_{r, i}^{2}}{\sigma_{v_{r, i}}^{2}} 
    \right)        \\
    \Delta v_{x, i} &= v_{x, i}^{\text{obs.}} - \Tilde{v}_{x, i}
\end{align}
The minimization is done over the tangential ($v_\alpha$, $v_\delta$) and radial ($v_r$) velocities\footnote{Compared to the clustering analysis, which assumes a universally valid metric implying a global correlation behavior (see Sect.~\ref{sec:data}), the optimization procedure is not affected by non-linear relationships between input features. Thus, to avoid propagating errors through the LSR conversion, we stick to observed tangential and radial velocities.}. 

The delta terms in the membership loss describe the offset between observed and computed values at the specified velocity $\Tilde{\Vec{v}}$. Although we introduce an additional observational error via the parallax uncertainty, we choose the tangential velocities to match the unit of radial velocities, the essential component in the sum in Eq.(\ref{eq:vel_loss}). Each term in the sum is weighted by its respective uncertainty, which decreases the influence of observations with large measurement errors on the membership loss. If all observations lack radial velocities, then the last term is set to zero; if only a subset of $v_r$'s is missing, their values are imputed with the average of its complement. 

For a perfectly co-moving population, the membership loss has a global minimum with a value of 0 at the group motion. Observational uncertainties, contamination from field stars, and a non-zero velocity dispersion will increase the minimum value accordingly. To search the 3D bulk motion that minimizes the membership loss, we use the quasi-Newton method of Broyden, Fletcher, Goldfarb, and Shanno (BFGS)~\citep{Nocedal:1999} with an initial guess of the mean 3D velocity\footnote{If no radial velocities are available, our initial guess is the null vector. We empirically find that the optimization converges to the same results for different initial velocities.}. We denote the velocity, which minimizes the membership loss (Equ.~(\ref{eq:vel_loss})) as the optimal bulk motion ($\Vec{v}_\mathrm{OBM}$).

\begin{figure}[!t]
    \centering
    \includegraphics[width=0.98\columnwidth]{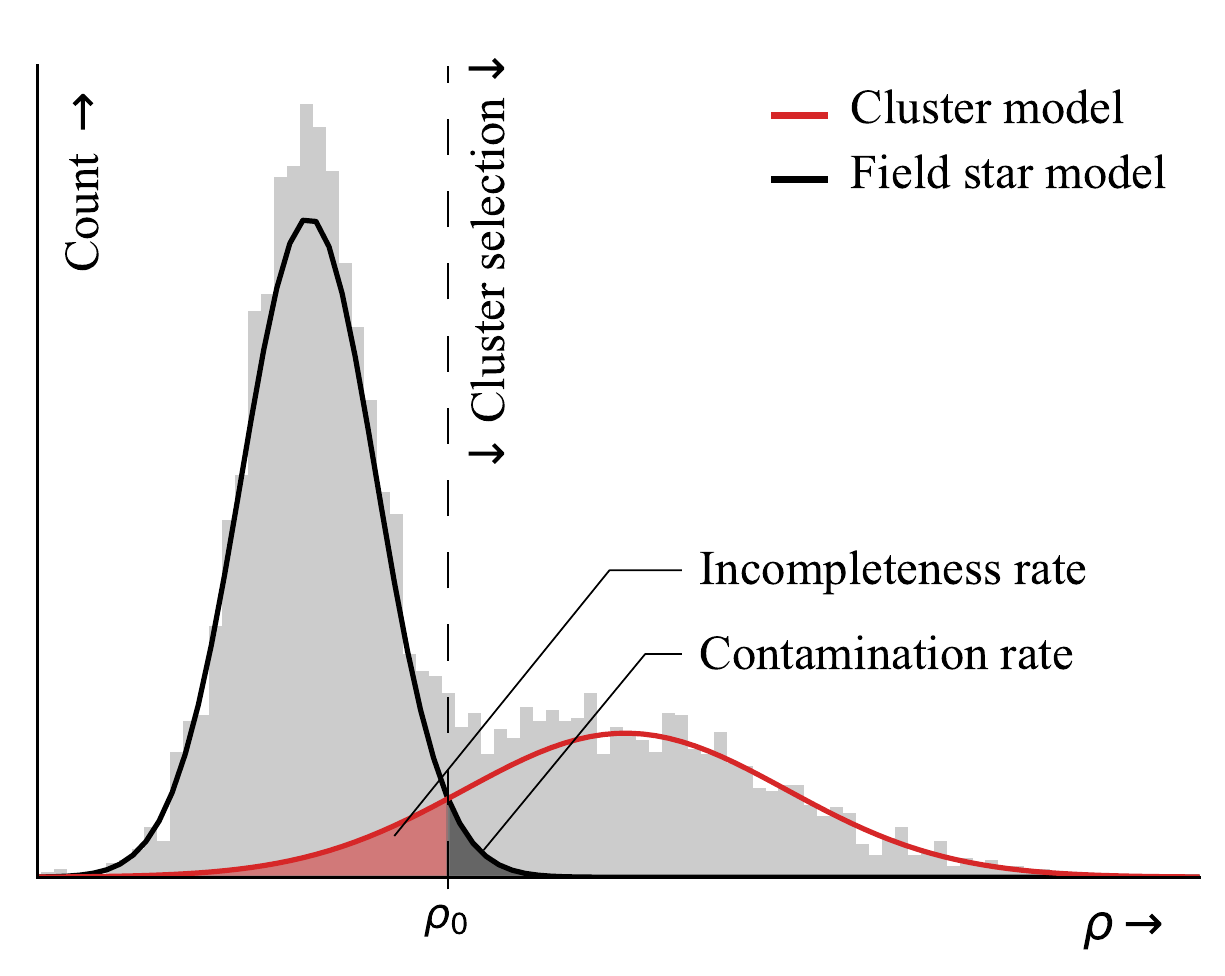}
    \caption{Noise reduction schematic. We fit the observed uni-variate density distribution $\rho$ with a mixture of two Gaussians modeling the cluster (red line) and field star (black line) population, respectively. We obtain an approximation to the field star contamination, and incompleteness rate in the cluster sample by considering the cluster-noise classifier's confusion matrix entries; in particular, false positives, false negatives, and true positives.}
    \label{fig:cluster-noise-classifier}
\end{figure}

To determine the group motion of the co-moving population via our minimization approach, finding $\Vec{v}_\mathrm{OBM}$ needs a large and pure selection of cluster sources; meaning truly co-moving stars. We attempt to obtain a rather clean sample of cluster stars via the aforementioned mixture model approach (see Fig.~\ref{fig:cluster-noise-classifier}, and ``Likely cluster members'' in Fig.~\ref{fig:noise_removal_pipeline}). By fitting a mixture of univariate Gaussians to the density distribution of a modal region we get a classifier that separates cluster from field stars\footnote{We use a simple threshold classifier where both mixture components have equal class (posterior) probability. The likelihoods and class fractions are estimated using a univariate GMM.}. Since the input density is one-dimensional, the classifier -- also referred to as cluster-noise classifier -- becomes a simple threshold classifier. 

Sources with a density greater than the threshold $\rho_0$ are likely cluster members. As the classifier is trained on densities determined in the 5D space, which experiences projection distortion, we only use the $80\%$ most dense stars in the cluster sample to determine $\Vec{v}_\mathrm{OBM}$. This density filter is designed to remove likely field star contaminants (false positives) which are typically expected to be less dense than cluster members. Figure~\ref{fig:cluster-noise-classifier} shows an example of the contamination estimation.

The optimal bulk motion $\Vec{v}_\mathrm{OBM}$ is used to infer an ``ideal'' radial velocity. The ideal radial velocity minimizes the Euclidean distance between $\Vec{v}_\mathrm{OBM}$ and the velocity vector constrained by the measured proper motions. We refer to the computed 3D space motion, which is a combination of measured proper motions and the inferred radial velocity, as the minimally different velocity ($\Vec{v}_\mathrm{MDV}$). On the one hand, genuine cluster members should receive an estimated space velocity that is very close to their true motion (assuming low intra-cluster velocity dispersion of a few km\,s$^{-1}$). Field stars (if not an interloper in phase space) show incompatible observations with the co-moving population and are on average assigned a different space velocity. Together with sparseness in positional space, field stars consequently show lower densities in phase space.

We infer $\Vec{v}_\mathrm{MDV}$ for sources without $v_r$ measurements as well as for sources with large uncertainties of $\sigma_{v_r} > 2$\,km\,s$^{-1}$. After this step, all sources have an associated radial velocity, either measured or inferred. We provide these inferred radial velocities ($\hat{v}_r$) in our final catalog\footnote{Given as \texttt{v\_ERV} (estimated radial velocity) in Table~\ref{tab:member-catalog}.}. In comparison to the \texttt{SigMA} pipeline, we explicitly determine radial velocity estimates for all sources regardless of uncertainty.

The bulk velocity estimate is directly correlated with observed proper motions and radial velocities. Therefore, systematic errors as in the case of binary or multiple stellar systems introduce corrupted measurements that potentially bias the final result. However, directly flagging binary stars and removing them from the inference process does not significantly alter the results as only a tiny fraction (0.05\%) of Gaia sources is identified as multiples~\citep{Arenou:2022}. Since the work by \citet{Arenou:2022} does not represent the full binary population, we investigated another indicator for multiples in the Gaia catalog. For example, the RUWE parameter \citep{Lindegren2018, Lindegren2021} can be used as a discriminator, which is a measure of how well the astrometric solution is fitted to a single star model, as also discussed in  \citet{Penoyre2022a, Penoyre2022b}. These authors also show that binaries, which have been observed with longer time baselines (e.g., comparing \textsc{Hipparcos}, DR2, and DR3) could still deliver parallaxes and proper motions that are close to the true values \citep[see also][]{Kervella2022}. As a consequence, binaries can still be selected as true-positive members of a cluster if selected with 5D Gaia astrometry, as can be seen, for example, by the clear binary sequences in HRDs \citep[e.g.,][]{Meingast2021}. However, even in these cases, the multiple systems do not comprise a large fraction of the cluster selection. As a consequence, we argue that binaries inherently contribute only marginally to the bulk velocity computation.

\begin{figure}
    \centering
    \includegraphics[width=\columnwidth]{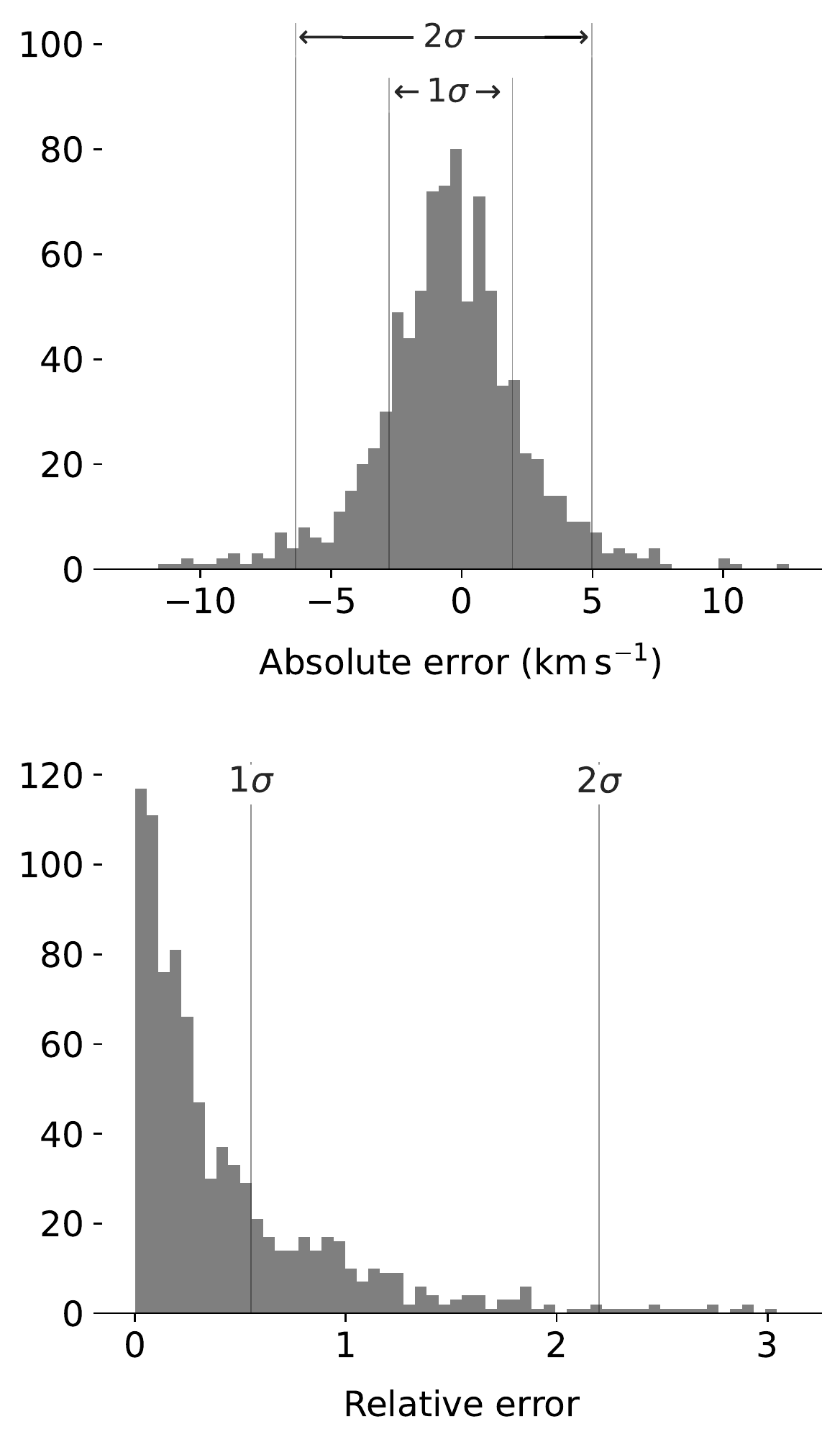}
    \caption{Absolute and relative error of inferred radial velocities compared to observed values in Gaia DR3 with radial velocity errors below 2 km\,s$^{-1}$. During inference, we randomly removed $95\%$ of available radial velocities to facilitate this comparison. Only inferred values are shown where the Gaia observable has been removed. We highlight the $1\sigma$ and $2\sigma$ percentiles and find that the majority ($68\%$) of absolute errors are within $\pm 2.35$\,km\,s$^{-1}$ and have relative errors below $0.55$.}
    \label{fig:vr_inferred}
\end{figure}

To validate the bulk velocity and optimal 3D velocity estimation, we apply the presented method to clusters found in the Sco-Cen OB association (see Sect.~\ref{sec:results}). During inference, we randomly remove $95\%$ of radial velocities to facilitate a comparison with observed values. The absolute $\Delta v_r$ and relative errors $\delta v_r$ to Gaia measurements with $\sigma_{v_r} <2$\,km\,s$^{-1}$ are shown in Fig.~\ref{fig:vr_inferred}. The absolute error is defined as the difference between the estimated radial velocity $\hat{v}_r$ and the observed value $v_r$:
\begin{equation}\label{eq:abs_errors}
    \Delta v_r = \hat{v}_r - v_r
\end{equation}
The relative error expresses the magnitude of the absolute error compared with its measured magnitude:
\begin{equation}\label{eq:rel_errors}
    \delta v_r = \left| \frac{\hat{v}_r - v_r}{v_r} \right|
\end{equation}
We find that $68\%$ of sources ($1 \sigma$) have absolute errors of less then $\pm 2.35$\,km\,s$^{-1}$ and $95\%$ ($2 \sigma$) of absolute errors are within $\pm 5.66$\,km\,s$^{-1}$. Thus, the average error is close to the large statistical uncertainties (2\,km\,s$^{-1}$) in the sample, which constitutes an approximation for the lower bound for the mean estimation error. The majority ($1 \sigma$) of relative errors are below $0.55$. Thus, inferred radial velocities are in good agreement with observations, validating our method and highlighting its robustness to (not yet fully understood) binary effects and contamination.

The space velocity information is used to determine cluster membership in the following steps. We pre-filter unlikely members via a kinematic selection before applying the cluster-noise classifier (see Sect.~\ref{sec:remove_fieldstars}) to a complete 6D phase space including the computed $\hat{v}_r$ estimates. The pre-filter removes possible contaminant stars that have vastly different 3D motion, namely sources that differ more than 10\,km\,s$^{-1}$ from the determined bulk motion, hence $||\Vec{v}_\mathrm{MDV}  - \Vec{v}_\mathrm{OBM}|| < 10$\,km\,s$^{-1}$. 

\begin{figure*}
    \centering
    \includegraphics[width=0.99\textwidth]{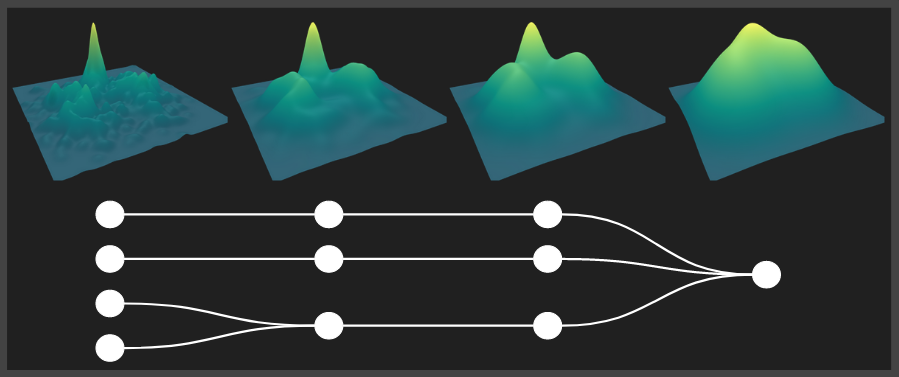}
    \caption{Schematic figure linking the cluster number to the density estimation process. Applying a smoothing operator generates a family of density fields. This hierarchical family of functions is called a scale space.} 
    \label{fig:scalespace}
\end{figure*}

\subsubsection{Removing field star contamination}\label{sec:remove_fieldstars}
Figure~\ref{fig:noise_removal_pipeline} shows the noise removal pipeline, which consists of two main parts. In the previous section, we have discussed the first part in which we impute missing radial velocities by assuming the presence of a single dense co-moving population in the input data. In the following, we discuss the second part, in which we aim to fit cluster-noise classifiers to remove the field star content in the combined space of Heliocentric Galactic Cartesian position and determined $v_\mathrm{MDV}$'s.

Combining positional and kinematic spaces (here XYZUVW) directly puts an emphasis on one of the sub-spaces (position or velocity) due to different value ranges (see Sect.~\ref{scaling-factors} for more details). Large axis ranges automatically dominate the extraction as distances along these dimensions are penalized, more drastically impacting the density estimation. Instead of selecting a single scaling factor we choose multiple plausible scaling factors and compute a univariate density distribution $\rho$ for each one. We obtain scaling factors $c_1, \dots c_N$ by repeating the procedure discussed in Sect.~\ref{scaling-factors} in Galactic Cartesian phase space (XYZUVW), setting $N$ to 10 (see forking into $N$ subprocesses in Step II in Fig.~\ref{fig:noise_removal_pipeline}). 

We separate the stellar population from the field star component using a cluster-noise classifier (the classifier is motivated and discussed in detail in Sect.~\ref{sec:bulkvelocity}). This classifier is applied to the 1D density estimation $\rho$ determined in 6D phase space, using measured and estimated radial velocities (see the x-axis in Fig.~\ref{fig:cluster-noise-classifier}). This thresholding method results in a global isosurface selection that is independent of positional information of sources in the original feature space (see Fig.~\ref{fig:level_set}). To reduce the contamination of random field star components, we employ the $\beta$-skeleton as a locality-aware neighborhood graph from which we delete vertices that fall below the computed density threshold $\rho_0$, as shown in Fig.~\ref{fig:cluster-noise-classifier}. For more details on this graph based approach, see the related work discussion in Appendix~\ref{app:methods}. 

Field stars account for the majority of sources in the given samples. Thus, the number of vertices that fall below the density threshold makes up most of the graph. Removing them disconnects the graph and splits it into multiple connected components. We define sources within the densest (and typically the largest) connected component as cluster members. To extract cluster members more robustly we compute one extraction for a range of scaling parameters (see Sect.~\ref{scaling-factors}). We obtain a final cluster catalog by removing unlikely members that appear in less than half of the $N$ extractions when using different scaling factors $c_i$ (see Fig.~\ref{fig:noise_removal_pipeline}).

\subsubsection{Contamination and completeness estimate} \label{sect:contamination_completeness}
The cluster-noise classifier is a discriminative model, whose conditional densities (or mixture components) describe the cluster and field star distributions. In combination with the decision threshold $\rho_0$, we can internally compute estimates for the field star contamination fraction $f_\mathit{cont}$ and the incompleteness fraction $f_\mathit{inc}$ for each cluster sample. 

In the fitting procedure, the number of mixture components $k$ is a free parameter determined by minimizing the BIC. Thus, the discriminative model can have a different number $k$ of Gaussian components for each cluster. To formalize a consistent definition across a different number of components, we introduce the following notations. 
We denote the set variables specifying the identity of the mixture component by $Z$; its $k$ components are then $Z = \{ z_1, \dots, z_k \}$. Individual Gaussian components can then be formulated as densities conditioned on the mixture component:
\begin{equation}
    p(x \mid z_i) = \mathcal{N}(\mu_i, \sigma_i)
\end{equation}
The mixture density can then be written as:
\begin{equation}
    p(x) = \sum_{z \in Z} p(z) p(x \mid z) 
\end{equation}
where $p(z)$ is the prior probability of the mixture component $z$, also called mixture weight. The $k$ mixture weights must sum to one. 

The subset of components that describe the distribution of cluster members is denoted with $S$. It encompasses all mixture components whose mean (or expected value) exceeds the threshold $\rho_0$, thus, $S = \{ z \in Z : \mathbb{E}[p(x \mid z)] > \rho_0 \}$. The relative complement of $S$ with respect to $Z$ then contains all components describing the field star distribution; we denote this set as $B$, defined by $B = Z \setminus{S}$.

Finally, with this formulation in mind we can express both the incompleteness fraction $f_\mathit{inc}$ and contamination fraction $f_\mathit{cont}$. The incompleteness fraction, as shown in Fig.~\ref{fig:cluster-noise-classifier}, is the probability of observing a sample from the cluster distribution with a value less than $\rho_0$. 
\begin{equation}
   f_\mathit{inc} = \frac{\sum_{s \in S} \int_{-\infty}^{\rho_0} p(s) p(x \mid s)\,dx}{\sum_{s \in S} p(s)}
\end{equation}
The contamination fraction is defined as the fraction of false positive samples in the overall cluster sample. Thus, $f_\mathit{cont}$ can be expressed as the probability of observing a sample from the field star distribution among all samples with a value larger than the threshold $\rho_0$.
\begin{equation}
    f_\mathit{cont} = \frac{\sum_{b \in B} \int_{\rho_0}^{\infty} p(b) p(x \mid b)\,dx}{\sum_{z \in Z} \int_{\rho_0}^{\infty} p(z) p(x \mid z)\,dx}
\end{equation}
Both $f_\mathit{inc}$ and $f_\mathit{cont}$ are schematically shown in Fig.~\ref{fig:cluster-noise-classifier} for the example of a two-component mixture.

The contamination and incompleteness are computed for each group. We obtain a mean contamination estimate across all groups in Sco-Cen of $5.3 \%$ with a standard deviation of $3.1 \%$ across groups. This value agrees well with photometric contamination estimates via the Gaia HRD, as shown in Fig.~\ref{fig:hrd} and described in Sect.~\ref{sec:results} and Appendix~\ref{ap:old-star-cont}. Although we find good agreement, we do not completely trust the stated values due to a lack of knowledge on systematic uncertainties. 

The major source of systematic uncertainty is a deviation from Gaussianity of any of the mixture components. Especially, in the case of more than two mixture components, we expect that these internal estimations have increased error rates. By departing from the paradigm of ``one mixture component per signal and background'' we increase the accuracy of the model (and ideally the obtained cluster members) at the cost of direct model interpretability (and all of its consequences). 
Further uncertainty is added via the density estimation to which the mixture model is fit. Since we do not have access to the true underlying density $f$, we inevitably make mistakes by substituting it with our estimate $\hat{f}$.

We find a mean completeness across groups of approximately $89.2 \%$ with a standard deviation of $8.3\%$. Similarly to the contamination fraction, determining the incompleteness depends on the mixture components and density approximation. Still, compared to the contamination fraction, the incompleteness estimate is relatively high. A caveat of our noise reduction procedure is that we reduce high-dimensional phase space information into a univariate variable that is used to filter the data. This univariate formulation lacks descriptions of local positional and kinematic relationships that might help to increase the completeness of our catalog. 
Further, we estimate the actual value even lower, as we find multiple connected components in the neighborhood graph of which we only extract the main component. We also only admit stars that pass a threshold of $50\%$ across different scaling fractions. All these decisions increase the precision of our sample at the cost of a reduced recall. 
Additionally, we evaluate the estimated completeness fraction by comparing our sample to past extractions in the literature in Sect.~\ref{sec:compare}. These comparisons suggest sample completeness of about $90\%$ (e.g., when compared to \citealp{Damiani2019} or \citealp{Luhman2022a}) which agrees with our estimate. However, these surveys can also not be considered complete. On the contrary, a direct comparison shows (see Sect.~\ref{sec:results}) that others applications on Sco-Cen are missing sources that \texttt{SigMA} is able to uncover. 

Instead of comparing estimated values to past extractions, we aim to evaluate the accuracy of internal contamination and completeness estimates, see Sect.~\ref{sec:validation_simulations}. Using simulations, we find that on average the contamination from field stars can be approximated quite well using our univariate mixture model approach. However, especially in dense cluster environments, our approach seems to overestimate completeness. If a large portion of the cluster exists in the low-density environment outside the cluster core the density approach fails to adequately capture the true number of missing cluster members. see Sect.~\ref{sec:dense_clusters} for a detailed discussion.

\subsection{Multi-scale clustering}\label{sec:scale-space}
The density field is the main parameter of the proposed clustering method. Its topology is affected by the estimation process, which impacts the final result. Especially the smoothing parameter can create, on the one hand, a very rough and, on the other hand, an over-simplified density field. The schematic Fig.~\ref{fig:scalespace} illustrates the dependence of the cluster number to the density estimation process. Applying a smoothing operator generates a family of density fields, called a scale space~\citep{Witkin:1987}. We use this scale-space concept to study the dependence of extracted clusters on density estimation. Clusters with a long lifetime in the scale space are preferred over, for example, ``short-lived'' children.

We approximate the scale space by running \texttt{SigMA} $N$ times obtained by progressively smoothing the initial density field. Given an ensemble of $N$ density estimates $\{\hat{f}\}_i, i \in [0,N]$, we track clusters through various density filters. To track clusters through different levels of scale space we use three cluster connection rules based on cluster modes, which we approximate by the densest point in a modal region. The connections we define are the following: \emph{direct link}, \emph{merge}, and \emph{split}. 

\begin{figure}[t!]
    \centering
    \includegraphics[width=\columnwidth]{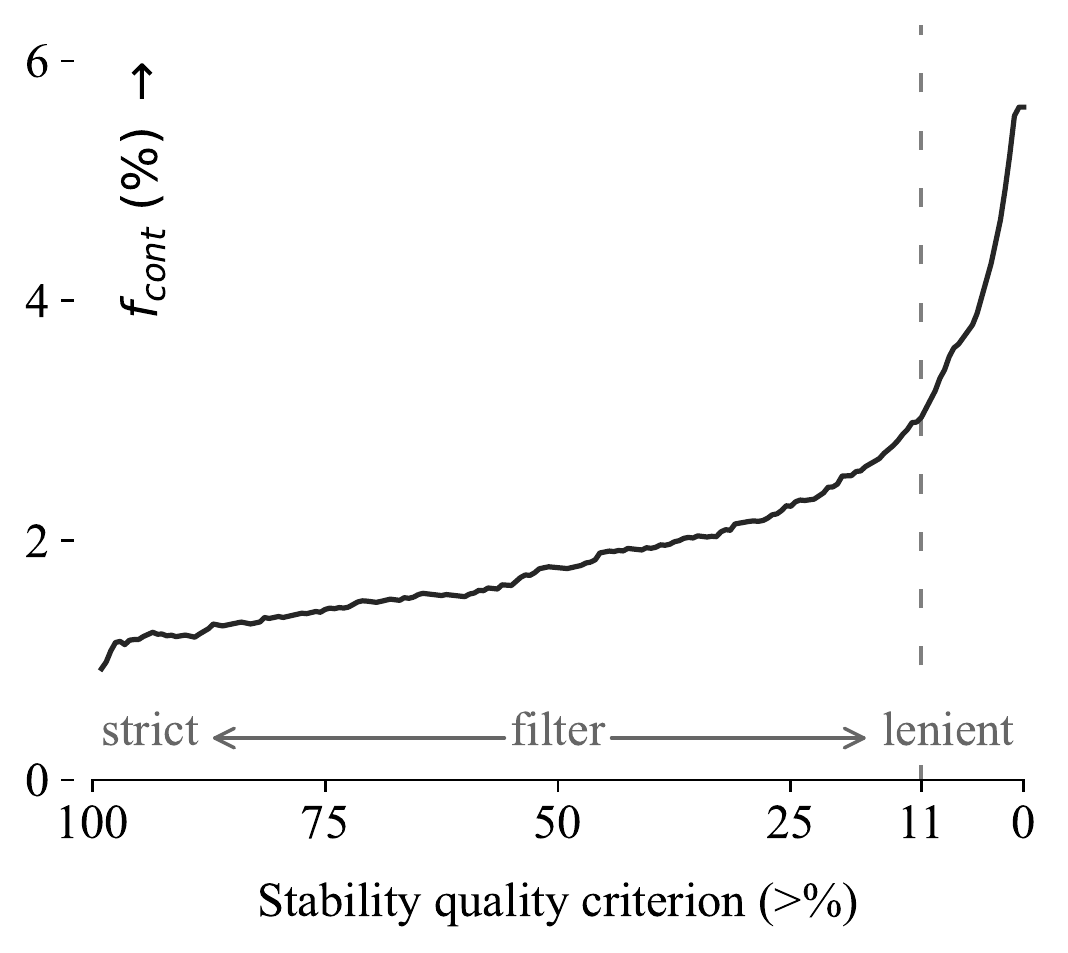}
    \caption{Stability over estimated contamination rate. The contamination estimate was determined via source positions in the HRD relative to the $25\,$Myr isochrone as shown in Fig.\ref{fig:hrd}, selecting potential contaminants from older populations. We identify a sharp drop in contamination for low stability values which levels off at around $11$\%.}
    \label{fig:stability}
\end{figure}

A direct link connection denotes a connection between two modal regions whose Jaccard similarity is larger than $50\%$ and both cluster modes lie in the intersection set. A merge connection is a weaker condition and is only placed if no direct link can be established. A merge link is made when a parent cluster\footnote{The parent cluster resides in the $i+1$'th level, whereas the child cluster is from level $i$.} contains the cluster mode of its child. If both conditions for direct and merge links are not satisfied, a split connection is placed between a parent and child cluster if the child contains the cluster mode from its parent.

The emergence of critical points, or additional clusters, in smoother versions of the scale space, is a result of the non-exact nature of our density estimation~\citep{Reininghaus:2011, Lifshitz:1990} as well as due to randomness introduced by our Monte Carlo strategy. In the absence of noise, smoother density filters result in a simplified topology. Thus, we apply the pruning strategy introduced by~\citet{Reininghaus:2011} to the resulting merge-split graph, which generates a simplified merge tree. The merge tree for our running toy data set is schematically illustrated in Fig.~\ref{fig:scalespace}. 

We extract stable components from the resulting merge tree via a consensus clustering approach, discussed in Sect.~\ref{scaling-factors}. In total, we find 60 stable clusters in the search box, while 37 are discussed in more detail in Sect.~\ref{sec:results} as being part of the Sco-Cen association. We find the 23 remaining groups to be unrelated to the young Sco-Cen association (see Sect.~\ref{sec:results}). Often these appear as incomplete (or truncated) cluster extractions; in particular, the shape and position of the majority of these groups suggest that they extend beyond the defined search box. 

The consensus approach also lets us characterize how often individual sources appear throughout the cluster ensemble\footnote{The cluster ensemble is the collection of $N$ clustering solutions corresponding to the $N$ density estimates.}. We report this value as ``stability'' in our cluster catalog. The stability criterion can be used as an effective measure to remove spurious sources from the catalog. Figure~\ref{fig:stability} highlights the effect of the stability criterion on sources when empirically estimating the contamination from older sources (hence likely unrelated sources) via an HRD. The $x$-axis shows a given stability criterion where we filter sources with stability\,$>\,x$. The $y$-axis shows the empirical contamination estimate, which was determined via positions of filtered (older) sources in the HRD. The fraction of sources to the left of the $25\,$Myr isochrone are used to estimate the false positive rate while sources to the right account for true positives (see Fig.~\ref{fig:hrd} and further details in Sect.~\ref{sec:results} and Appendix~\ref{app:hrd}). Based on this result, we recommend a quality criterion of stability\,$>$\,$11\%$.

However, due to the density-based nature of \texttt{SigMA}, the stability criterion is strongly correlated with density. Especially clusters with minor density enhancement over the field background are short-lived in scale space. Thus, although \texttt{SigMA} detects them clearly, some clusters contain members with overall disproportionately small stability values causing them to fall out of the sample for relatively low stability values (e.g., the cluster Oph-North-Far, see Sect.\,\ref{sec:L134}).
Therefore, we do not recommend generally using the stability quality criterion, but to investigate its behavior per cluster, to get potentially cleaner cluster samples.

\begin{figure*}[!t]
    \centering
    \includegraphics[width=0.9\textwidth]{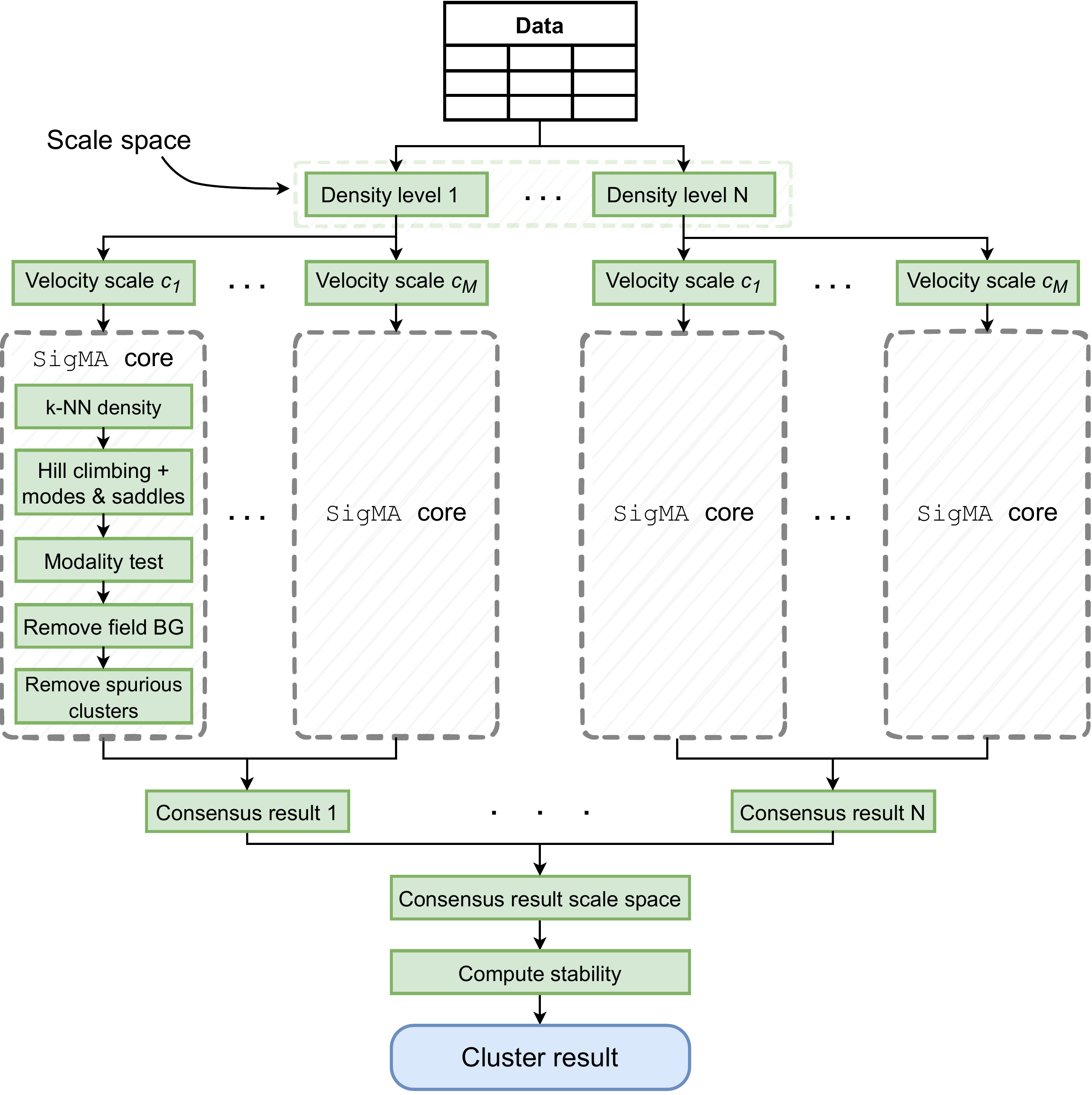}
    \caption{Schematic illustration of the \texttt{SigMA} pipeline. The pipeline consists of two main parts, the \texttt{SigMA} core and two consensus clustering steps. For a detailed explanation, see the main text and the references therein in Sect.~\ref{sec:sigma_pipeline}.}
    \label{fig:sigma_pipeline}
\end{figure*}

\subsection{Removing spurious cluster solutions}\label{sec:spurious_removal}
Unstable cluster solutions are automatically filtered out in our scale space approach in Sect.~\ref{sec:scale-space}. However, the distribution of sources in the Milky Way in phase space is far from uniform. \texttt{SigMA}'s job is to separate the input data space into uni-modal regions. This segmentation does not distinguish between compact, clustered over-densities and long-range, low-density modes inherent to the Milky Way distribution. We aim to remove the latter from our cluster sample. 

We assume a natural clustering of ``real'' and ``spurious'' groups in 1D density space. That means if the density of all members across the $N$ extracted groups are plotted we expect a bi-modal distribution. The modes of these distributions then correspond to the real and spurious groups. 

To classify each group into any one category we employ an iterative approach that starts off by assigning all groups to the category ``real''. Subsequently, we loop through all clusters sorted by their median member density (computed in 5D phase space) in ascending order. In each iteration $i$, with $i \in \{1,\dots, N\}$, the first $i$ groups ($i$ groups with lowest median density) are assigned into the group ``spurious''. At each step, we track the separation and compactness in 1D density space of individual members across both groups. The more compact and well separated the members of both groupings are -- measured by the Cali{\'n}ski-Harabasz score~\citep{Calinski:1974} -- the more the classification at step $i$ agrees with our bi-modal assumption.

We use the classification at step $i$, which maximizes the Cali{\'n}ski-Harabasz score to characterize each cluster as ``real'' or ``spurious''. This classification is directly applied to each \texttt{SigMA} clustering solution, hence, before obtaining a consensus result across scaling factors and scale space (see last process in \texttt{SigMA} core in Fig.~\ref{fig:sigma_pipeline}).

\subsection{The \texttt{SigMA} pipeline}\label{sec:sigma_pipeline}
The proposed clustering method \texttt{SigMA} has many components that require sensible choices in order to work together properly. In previous sections~\ref{sec:sigma}\,--\,\ref{sec:spurious_removal} we have motivated and discussed our parameter choices which yield the final analysis pipeline. Figure~\ref{fig:sigma_pipeline} shows an overview of the full pipeline. It consists of two main parts, the \texttt{SigMA} core and two consensus clustering steps. In the following, we briefly summarize how these individual components come together.

The \texttt{SigMA} core outputs a clustering result for a given density level $i$ and velocity scale $c_j$. It iterates through the following steps: First, a $k$-NN density estimation is computed, see Sect.~\ref{sec:density_est}. Second, a gradient-based hill-climbing step is performed that produces individual modes and saddle point locations. The modal candidates (or clusters) at this point over-segment the data set, see Sect.~\ref{sec:model_candidates}. Third, at each saddle point a modality test is applied which determines whether two neighboring modal candidates should be merged or not, see Sect.~\ref{sect:modality_test}. Fourth, field star background is removed from each modal region, see Sect.~\ref{noise-removal}. Fifth, spurious clusters are removed from the extraction, see Sect.~\ref{sec:spurious_removal}. 

To guarantee stable results against velocity scaling factors (see Sect.~\ref{scaling-factors}) and density estimation (see Sect.~\ref{sec:scale-space}), we employ a consensus clustering approach, discussed in Sect.~\ref{scaling-factors}. From the cluster ensemble, we can extract a stability value for each source in our final catalog.

\section{Validation}
We follow a two-pronged approach to verify the proposed clustering method \texttt{SigMA}. First, in Sect.~\ref{sec:validation_qualitative} we validate our clustering technique qualitatively in a case study on the Sco-Cen OB association. We describe the results and comparisons to other studies in Sect.~\ref{sec:results}. Second, in Sect.~\ref{sec:validation_simulations} the algorithm is validated quantitatively on simulated data, where we compare \texttt{SigMA} to other established clustering methods used to identify co-moving groups.

\begin{table*}[!t] 
\begin{center}
\begin{small}
\caption{Overview of the clustering algorithms that we test with synthetic data (simulated cluster samples) and compare to \texttt{SigMA}. The results are listed in Table~\ref{tab:results_simulation_oc_and_dense}. \vspace{-2mm}}
\renewcommand{\arraystretch}{1.4}
\resizebox{0.85\textwidth}{!}{%
\begin{tabular}{llll}
\hline \hline
\multicolumn{1}{l}{Algorithm} &
\multicolumn{1}{l}{Main Reference} &
\multicolumn{1}{l}{Short description} &
\multicolumn{1}{l}{For example, used by\tablefootmark{b}} \\
\hline

DBSCAN & \citet{Ester:1996} & A density based  & \citet{Castro-Ginard2019, Castro-Ginard2020, Castro-Ginard2022}, \\ [-1mm]
 & & clustering algorithm & \citet{Zari2019}, \citet{Fuernkranz2019}, \\ [-1mm] 
 & & &  \citet{Hunt2021}  \\
 
\hline
HDBSCAN\tablefootmark{a} & \citet{Campello:2013} & A hierarchical density based  & \citet{Kounkel2019, Kounkel2020}, \\ [-1mm]
 & & clustering algorithm & \citet{Hunt2021}, \citet{Kerr2021} \\



\hline
\end{tabular}
} 
\renewcommand{\arraystretch}{1}
\label{tab:algorithms}
\tablefoot{
\tablefoottext{a}{See Appendix~\ref{app:methods} for further discussion on the HDBSCAN algorithm.}
\tablefoottext{b}{The reference lists are not exhaustive literature reviews, but are intended to highlight the relevance of the proposed comparison methods.}
}
\end{small}
\end{center}
\end{table*}


\subsection{Validation using astrophysical knowledge}\label{sec:validation_qualitative}
Two direct observables that can be identified in our application on Sco-Cen (Sect.\,\ref{sec:results}) serve as a validation test of the method. First, and apart from the youngest groups that are affected by dust extinction, the Gaia color-absolute-magnitude diagrams (equivalent to observational Hertzsprung-Russell Diagrams, HRDs) for the stars in each group show a narrow (coeval) distribution (see Fig.\,\ref{fig:hrd}, more detailed analysis of individual cluster age sequences will be discussed in future work). There is no procedural reason why this should be the case, the method does not know about the brightness and colors of the stars. Only a meaningful selection of co-moving stellar siblings can produce the observed narrow sequences in the HRDs. 

Another observable that serves as test is the prominence of massive stars associated in 2D projection with the \texttt{SigMA} identified groups, while they are often located at a central position within the concerned clusters (e.g., $\alpha$\,Sco, $\beta$\,Sco, $\delta$\,Sco, $\nu$\,Sco; see Sect.\,\ref{sec:results} and Tables~\ref{tab:averages-pos}--\ref{tab:averages-vel} \& \ref{tab:hip}). These massive stars are too bright to have reliable measurements in the Gaia archive and the brightest are not even in Gaia \citep[like Antares,][]{Ohnaka2013}, still, the method finds groups around them. Based on \textsc{Hipparcos} astrometry (Table\,\ref{tab:hip}) we find strong evidence that many of these bright stars share similar parallaxes and proper motions as the clusters they seem to belong to in projection. This is an astrophysically relevant result (massive stars do not form alone and are often found at central positions) and it serves as another direct validation of the method. 

\subsection{Validation using simulations}\label{sec:validation_simulations}
To objectively investigate the effectiveness of new clustering algorithms, synthetic data with known ground truth information facilitates the comparison to other clustering techniques. Since \texttt{SigMA} is tuned to astrophysical phase space data, we want to test its efficacy on simulations that approximate observational data as closely as possible. Therefore, simulated data should replicate the data model, content, volume, uncertainties, and selection effects of Gaia data as closely as possible. 

To our knowledge, the Gaia eDR3 mock catalog by \citet{Rybizki2020} best meets these criteria. In particular, the catalog contains a large, realistic open cluster sample with internal rotation and corresponding uncertainties. Although the cluster structure of the open clusters differ from Sco-Cen, access to ground truth data and realistic Gaia selection effects and content provides a firm validation basis. To test \texttt{SigMA}'s ability to separate groups in dense cluster environments, commonly found in young OB associations and star-forming regions such as Sco-Cen (Sect.\,\ref{sec:results}) or Orion \citep[e.g.,][]{Chen2020}, we aim to create a derived data set from the original mock catalog that mimics these conditions.

In the following sections, we describe the comparison of results in detail, particularly the data used, the algorithms against which we compare \texttt{SigMA}, and the validation results.

\subsubsection{Open cluster sample}\label{sec:open_cluster_sample}
The Gaia eDR3 mock catalog \citep{Rybizki2020} is extensive. As it aims to reproduce Gaia data realistically, it contains simulated measurements on over 1.5 billion sources and over 1,000 open clusters compiled from the catalogs of \citet{Kharchenko2013} and \citet{Cantat-Gaudin2018a}. To validate \texttt{SigMA} and compare different clustering algorithms, we need to reduce the data size to a manageable subset. Therefore, we limit the data to a range of $200\,$pc around the sun, which yields uncertainty characteristics similar to our Sco-Cen box sample. To increase comparability between our qualitative and quantitative tests, we apply the same error cuts to mock and real data\footnote{As no fidelity information is provided in the eDR3 mock catalog this quality flag was not reproduced on mock data.}, as described in Eq.(\ref{eq:gaia-quality}). Overall, these quality criteria result in the final test catalog size of 2,682,883 samples, of which 18,682 are part of 12 open clusters. The mock catalog does not contain the full 5D astrometric uncertainty covariance matrix which \texttt{SigMA} uses. We substitute missing values with real measurements from Gaia DR3. Each mock sample is randomly paired with a source within our Sco-Cen box (see Sect.~\ref{sec:data}) whose values it adopts.

This simulation now allows our proposed analysis method to be tested for accuracy. Importantly, we need to highlight its capacity to identify clusters in positional and kinematic data in contrast to established analysis methods. We limit our comparison to the two relevant clustering methods (DBSCAN, HDBSCAN) as listed in Table~\ref{tab:algorithms} which are considered as among the most promising candidates for stellar cluster analysis in a meta-study by~\citet{Hunt2021}.

Since \texttt{SigMA}'s parameters are tuned to deal specifically with Gaia data, we employ a grid search to find suitable parametrizations for each of the three clustering algorithms. This strategy provides a measure of the peak performance these clustering methods can achieve. A comparison against the best performance results allows for a discussion on methodological advantages and disadvantages rather than reflecting poor parameter selection. For a detailed discussion on the parameter search, see Appendix~\ref{app:grid_search}.

We will report the performance of the best model across our search to facilitate a fair comparison. The performance itself is measured using the following clustering validation metrics: Normalized Mutual Information score \citep[NMI,][]{Strehl:2002}, Adjusted Mutual Information \citep[AMI,][]{Vinh:2009}, and Adjusted Rand Index \citep[ARI,][]{Hubert:1985}. We also report on classification metrics that are easier to interpret such as: true positive rate or recall, precision, accuracy (ACC), balanced accuracy \citep[BACC,][]{Brodersen:2010}, and Matthews correlation coefficient \citep[MCC,][]{Matthews:1975}. In addition, we report the total number of identified clusters $N_{tot}$ as well as the number of non-noise clusters $N_{cluster}$ that are found to coincide with a toy cluster instead of field  members. Similar to precision and recall, we also report contamination and completeness. In contrast to precision and recall, we compute the average cluster contamination and completeness only for clusters that coincide with a true cluster, i.e., for the $N_{cluster}$ identified non-noise clusters. Thus, these measures are not influenced by large false positives. When $N_{cluster} = N_{tot}$ the completeness is exactly equal to recall and contamination becomes 1 - precision. The resulting numbers are summarized in Table~\ref{tab:results_simulation_oc_and_dense}. We automatically select the best-performing model from the grid search by maximizing the median across these seven metrics.

\begin{table*}[!ht] 
\begin{center}
\begin{small}
\caption{Test results on simulated cluster samples. 
\vspace{-2mm}}
\renewcommand{\arraystretch}{1.4}
\begin{tabular}{lcccccc}
\hline \hline
\multicolumn{1}{c}{} &
\multicolumn{3}{c}{open cluster sample\tablefootmark{a}} &
\multicolumn{3}{c}{compact cluster sample\tablefootmark{b}} \\
\cmidrule(lr){2-4}
\cmidrule(lr){5-7}
\multicolumn{1}{l}{} &
\multicolumn{1}{c}{\texttt{SigMA}} &
\multicolumn{1}{c}{DBSCAN} &
\multicolumn{1}{c}{HDBSCAN} &
\multicolumn{1}{c}{\texttt{SigMA}} &
\multicolumn{1}{c}{DBSCAN} &
\multicolumn{1}{c}{HDBSCAN} \\
\hline 

NMI & \textbf{0.95}  & 0.94  & 0.86 & \textbf{0.60} $\pm 0.02$ & $0.39 \pm 0.06$ & $0.33 \pm 0.16$ \\ 
AMI & \textbf{0.95} & 0.94 & 0.86 & \textbf{0.56} $\pm 0.01$ & $0.38 \pm 0.06$ & $0.32 \pm 0.16$ \\ 
ARI & \textbf{0.96} & 0.95 & 0.74 & \textbf{0.38} $\pm 0.03$ & $0.11 \pm 0.03$ & $0.08 \pm 0.06$ \\ 
Precision & 0.97 & \textbf{0.98} & 0.74 & \textbf{0.64} $\pm 0.05$ & $0.23 \pm 0.05$ & $0.22 \pm 0.15$ \\ 
Recall & \textbf{0.98} & 0.96 & 0.76 & \textbf{0.47} $\pm 0.04$ & $0.25 \pm 0.03$ & $0.20 \pm 0.06$ \\ 
Contamination & 0.03 & \textbf{0.02} & 0.08 & 0.24 $\pm 0.13$ & \textbf{0.13} $\pm 0.12$ & $0.25 \pm 0.22$ \\ 
Completeness & \textbf{0.98} & 0.96 & 0.92 & 0.76 $\pm 0.15$ & \textbf{0.79} $\pm$ 0.24 & $0.76 \pm 0.21$\\ 
ACC & \textbf{0.97} & 0.96 & 0.76 & \textbf{0.47} $\pm 0.04$ & $0.25 \pm 0.03$ & $0.20 \pm 0.06$ \\ 
BACC & \textbf{0.90} & 0.88 & 0.88 & \textbf{0.51} $\pm 0.04$ & $0.16 \pm 0.03$ & $0.18 \pm 0.07$ \\ 
MC & \textbf{0.96} & 0.95 & 0.76 & \textbf{0.55} $\pm 0.04$ & $0.23 \pm 0.04$ & $0.16 \pm 0.11$  \\ 
N$_\mathrm{tot}$ & \textbf{12} & \textbf{12} & 14 & \textbf{26.8} $\pm 2.0$ & $10.0 \pm 1.7$ & $24.2 \pm 4.8$ \\ 
N$_\mathrm{cluster}$ & \textbf{12} & \textbf{12} & \textbf{12} & \textbf{24.1} $\pm 1.7$ & $10.0 \pm 1.7$ & $12.0 \pm 5.5$\\ 

\hline
\end{tabular}
\renewcommand{\arraystretch}{1}
\label{tab:results_simulation_oc_and_dense}
\tablefoot{Bold faced numbers indicate the best performance given a specific evaluation metric. The three clustering methods, \texttt{SigMA}, DBSCAN, HDBSCAN, are applied to two data sets.  
\tablefoottext{a}{The open cluster sample, which is a subset of the Gaia mock EDR3 catalog~\citep{Rybizki2020}.}
\tablefoottext{b}{The compact cluster samples that mimic the cluster environment of Sco-Cen where groups are densely packed together. The ten compact cluster samples are generated in a random effects model and the resulting distribution varies substantially across realizations. To estimate the performance of clustering algorithms on the compact cluster sample, we average performances across ten individual samples. We report the mean and standard deviation of performance scores.}
}
\end{small}
\end{center}
\end{table*}


Only a fraction of sources, less than $1\%$, are located in clusters. Hence, many of the above-proposed validation metrics will report high values as long as most field stars are clustered in the same group. We remove correctly identified field stars before computing the validation metrics to prevent reporting on artificially inflated clustering scores. By removing this ``true negative'' component without removing field stars labeled as cluster members (false positives) and cluster members identified as field components (false negatives), the reported scores are a conservative estimate of the algorithms' actual performances. 

The results are summarized in Table~\ref{tab:results_simulation_oc_and_dense}. All algorithms are able to recover the 12 clusters within the data set. We find that the performance of \texttt{SigMA} and DBSCAN are essentially equal -- with relatively high evaluation scores --  while outperforming HDBSCAN. We find that HDBSCAN (within the parameters we searched) identifies fewer members while also identifying two false positives, i.e., two large extra clusters that entirely contain field stars.

The access to ground truth data also allows us to test internal measurements of contamination and completeness estimates. We find a true mean contamination rate of $2.6 \pm 0.7$\% across the twelve identified clusters. \texttt{SigMA}'s internal estimation is slightly lower than that at a mean contamination rate of $1.1 \pm 0.4$\%. We find an even better agreement between true and estimated completeness values. The true mean completeness rate is $98.3 \pm 0.7$\% which is almost identical to \texttt{SigMA}'s internal estimation of $98.4 \pm 0.2$\%.

Although the true contamination value is outside the $1 \sigma$ confidence interval, the estimated value is still very close to the true one in absolute terms. In the open cluster sample, the internal measurements provide a surprisingly good approximation given that we have not explicitly modeled signal and background in univariate density distribution but rather assumed a simple mixture of Gaussians. 

The high reported accuracy across all clustering methods highlights the nature of open clusters. They appear as salient over-densities in phase space, which makes their detection a fairly easy task. This situation contrasts with the complex structure that constitutes Sco-Cen. Distinguishing more densely packed groups from each other is a non-trivial task. \texttt{SigMA} was created with the intention of an interpretable cluster definition, which is put to the test especially in such environments. Therefore, our goal is to create a test data set that reproduces such densely packed piles to put our analysis tool through its paces.

\subsubsection{Tightly packed cluster environment}\label{sec:dense_clusters}

To our knowledge, there is no realistic (replicating data model, content, volume, uncertainties, and selection effects of Gaia data) simulation of Sco-Cen-like, densely packed associations that can be used to validate \texttt{SigMA} in densely packed cluster environments. Moreover, there are no similar (or any) star-forming regions where a consensus has been reached on the number of true groups along with their members. Hence, we will create a derivative toy data set from the eDR3 mock catalog, which simulates groupings in tightly packed arrangements. We will refer to this newly generated mock catalog as the ``compact cluster sample''. 

The biggest unknown in creating this sample set concerns the cluster details. In particular, their number, location, extent, and respective size. However, the application on Sco-Cen has already produced a cluster sample that can be considered when generating toy data, as it provides a candidate set of these quantities. 

Two conflicting objectives pose challenges for sampling with known cluster sizes. On the one hand, the cluster sample should avoid strong correlations to results on Sco-Cen (for a discussion of results, see Sect.~\ref{sec:results}). We want to avoid reproducing previous results as it would favor the \texttt{SigMA} clustering objective over other alternative formulations and inhibit an objective comparison across methods. On the other hand, the compact cluster sample should aim to represent reality faithfully. Since no ground truth exists either on stellar groups or in the form of dedicated simulations that reproduce such dense cluster structures, we anchor our simulations on reality by considering our Sco-Cen extraction and other literature results. To balance overconfidence in the given results with an accurate description of reality, we perturb the obtained cluster sizes to avoid an unduly high correlation with \texttt{SigMA} and literature results. 

In the following, we will describe how we use these general cluster details as a starting point to generate the compact cluster sample: (1) number of sources, (2a) mean position (in Heliocentric Galactic Cartesian coordinate frame, XYZ), (2b) mean space velocity (in Heliocentric Galactic Cartesian coordinate frame, UVW), mean statistical dispersion of objects in the cluster in (3a) positional and (3b) velocity space. 

To introduce small and medium-sized deviation from the selected sample, we treat the number of sources and statistical dispersion as a normal distribution centered on the measured value with a relative variance of $25\%$. We employ a different strategy to sample new cluster means in position and velocity. Typically, neighboring cluster centroids in the combined positional and velocity space are way within the relative variance of $25\%$. Thus, updated centroid positions would commonly lie outside the original cluster boundary, drastically interfering with the initial cluster distribution. Instead, we sample centroid positions from a $50\%$ subset of each cluster which introduces variations that guarantees to retain the overall structure. 

After sampling a set of cluster quantities, we pair each one with an open cluster from the mock eDR3 catalog. We use 15 open clusters within $250\,$pc from the sun. This selection provides access to a slightly more diverse cluster sample that shows similar measurement uncertainties to the initial Sco-Cen groups. As different cluster sizes show distinct morphological features --- a small cluster can typically not be reproduced by downsampling a large one to its size --- we aim to pair clusters based on member size. A given Sco-Cen cluster $c_S$ with $n_s$ sources has the following probability of being paired with one of the $N$ synthetic mock clusters $c_k$ with $n_k$ sources, where $k \in [1,\dots,N]$:
\begin{equation}
    p(c_k | c_s) = \frac{(n_s - n_k)^2}{\sum_{i} (n_s - n_i)^2}
\end{equation}
Thus, on average Sco-Cen clusters are paired with similar-sized clusters from the mock catalog while maintaining a non-zero probability to be paired with more unlike clusters. 

For each cluster, the mock counterpart is scaled to the randomly sampled dispersion in position and velocity space (separately), randomly downsampled to its corresponding (randomly determined) size and positioned at the corresponding mean in six-dimensional phase space. Subsequently, the synthetic clusters are embedded into the remaining field distribution. Finally, we project the space velocities to the tangential velocity plane, compute right ascension ($\alpha$, deg), declination ($\delta$, deg), and parallax ($\varpi$, mas), randomly remove about $62\%$ of radial velocity measurements and apply the coordinate and quality criteria from Eq.(\ref{eq:box}) \& (\ref{eq:gaia-quality}) to reflect the clustering conditions of Sco-Cen, as described in Sect.~\ref{sec:data}. 

To evaluate and compare \texttt{SigMA}'s clustering performance to alternative algorithms, we generate $10$ compact cluster samples and report mean performance scores across these realizations. The results are summarized in Table~\ref{tab:results_simulation_oc_and_dense}. Compared to results on the open cluster sample, \texttt{SigMA} shows a significantly higher score than competing algorithms, which on average achieve only half of \texttt{SigMA}'s performance. The performance of DBSCAN and HDBSCAN on the compact cluster sample is approximately similar. Compared to DBSCAN, we find that top-performing HDBSCAN runs again falsely identify groups of field stars as clusters in the data set. On average HDBSCAN finds as many false positives as true positives. \texttt{SigMA} on the other hand can on average identify about twice as many subgroups as other algorithms while keeping the relative number of false positives low. 

Although we can highlight \texttt{SigMA}'s performance in these compact cluster agglomerates compared to DBSCAN and HDBSCAN, the performance values are drastically worse than in the open cluster sample. We can partially attribute the poor performance to the tough clustering challenge created by the randomized process. By randomly perturbing mean cluster positions, neighboring clusters easily merge, decreasing the maximally achievable performance. We also find that some clusters (on average, 2-3 groups) can no longer be identified as their density is indistinguishable from the field. Further, the clusters' extent in 5D is scaled to approximate Sco-Cen deviations, which reliably reproduces the cluster core. However, in some cases, a considerable part of the cluster extends far (up to over 100pc) beyond traditional Sco-Cen boundaries. The density of these stars compared to the field and their distance to the cluster core makes them impossible to detect with the three tested algorithms. Although a very tough clustering challenge, we think it still provides a good test bed for algorithms applied to Sco-Cen-like cluster environments. 

Besides the clustering performance, we again compare true to inferred contamination and completeness estimates. We find a true mean contamination rate of $23.7 \pm 13.1$\% across the on average 24 identified clusters. \texttt{SigMA}'s internal estimation is significantly lower with a mean contamination rate of $6.8 \pm 3.4$\% (although $\sim 7$ times higher than in the open cluster sample). As discussed above, the merging of nearby clusters into a single indistinguishable cluster drastically increases the contamination; a factor that \texttt{SigMA} cannot account for. In contrast, \texttt{SigMA} can only control the contamination of low-density field stars and not cross-contamination (the major contributor) from other clusters. When we ignore cross-contamination from other clusters and focus purely on contamination from field stars, the true contamination fraction becomes $8.2 \pm 4.1$\%, which is close to the internal value. 

The true mean completeness rate is $76.2 \pm 15.2$\% while \texttt{SigMA}'s internal estimation is $89.1 \pm 2.0$\%. In this case \texttt{SigMA} underestimates the large source fraction far outside the central cluster region; a fraction that is possibly exaggerated considering the young nature of Sco-Cen sources. 

The strong agreement of \texttt{SigMA}'s internal metrics with true values in the open cluster sample is in stark contrast to the large discrepancy between estimated and true values in the compact cluster sample. While the contamination estimate yields satisfying results if only field star contamination is considered, the completeness estimate likely systematically underestimates the low-density cluster component of stellar clusters. Thus, internal contamination and completeness estimates should only be used as rough first approximations of the stellar content of detected clusters. To obtain a better understanding, especially of the completeness fraction, we call to consider additional membership analysis tools such as UPMASK~\citep{Krone-Martins2014}, BANYAN~\citep{Gagne2018c}, or \texttt{Uncover}~\citep{Ratzenboeck2020}.

\begin{figure*}[t!]
    \centering
    \includegraphics[width=1\textwidth]{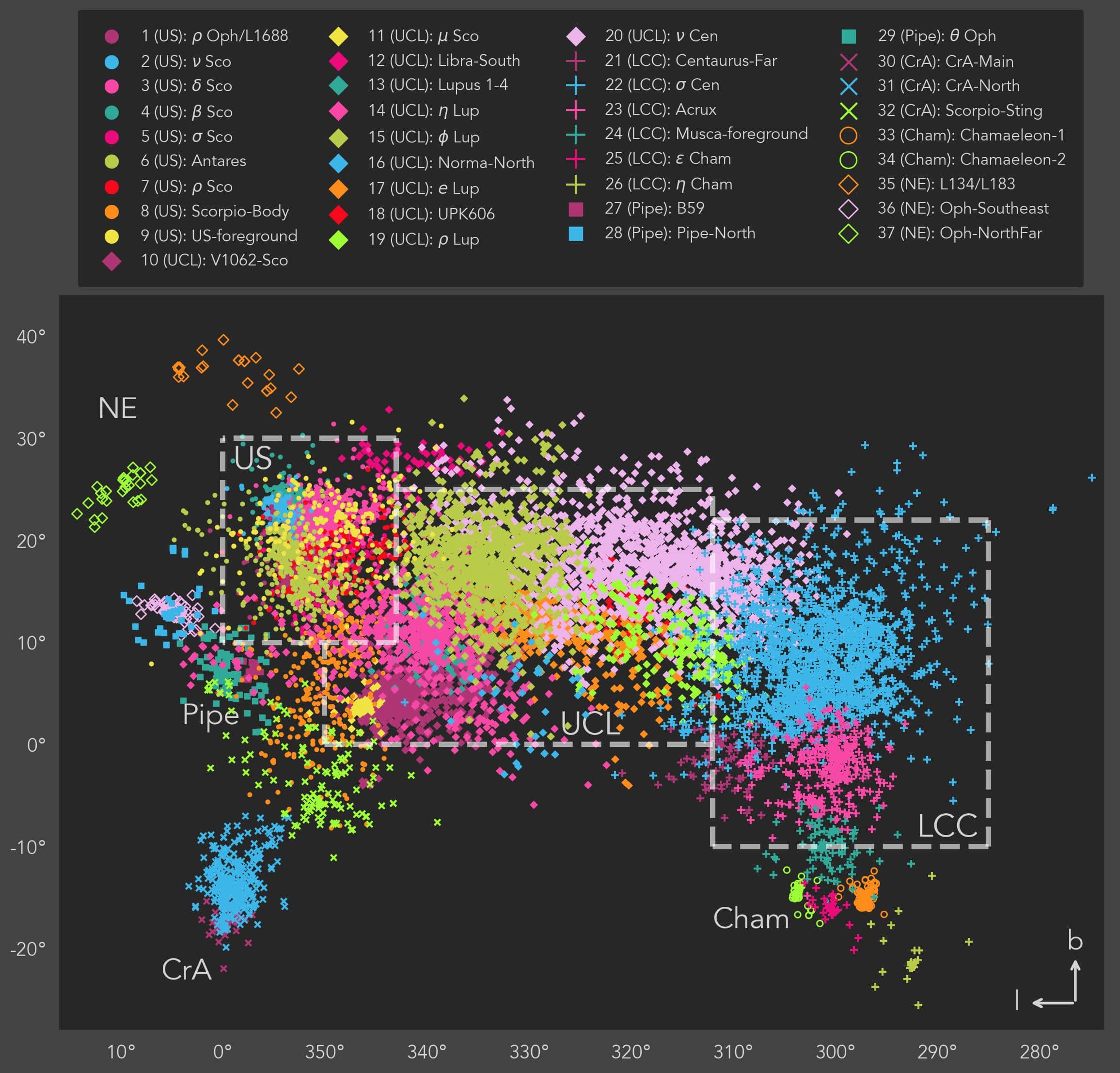}
    \caption{The distribution of the 37 \texttt{SigMA} clusters in Sco-Cen, projected in Galactic coordinates. Traditionally, the Sco-Cen OB association was separated into US, UCL, and LCC, marked with gray dashed lines. The clusters extracted with \texttt{SigMA} reveal a more complex substructure of Sco-Cen than initially proposed by \cite{Blaauw1946}, and they show a more extended spatial distribution that includes the CrA, Pipe, and Cham regions, and additional stellar groups toward the northeast (NE). The clusters are ordered in the legend by region, as given in Table\,\ref{tab:averages-pos}. 
    See here an \href{https://homepage.univie.ac.at/josefa.elisabeth.grossschedl/wp-content/uploads/2022/11/scocen_interactive_2D.html}{interactive 2D version} or Fig.\,\ref{fig:lb_small_muliples} for a separate view of each cluster. 
    For a better visualization of the clusters' distribution see the \href{https://homepage.univie.ac.at/josefa.elisabeth.grossschedl/wp-content/uploads/2022/11/scocen_interactive_3D-1.html}{interactive 3D version} (Fig.~\ref{fig:3D-map}). 
    } 
    \label{fig:lb-map}
\end{figure*}

\begin{figure*}[ht!]
    \centering
    \includegraphics[width=1\textwidth]{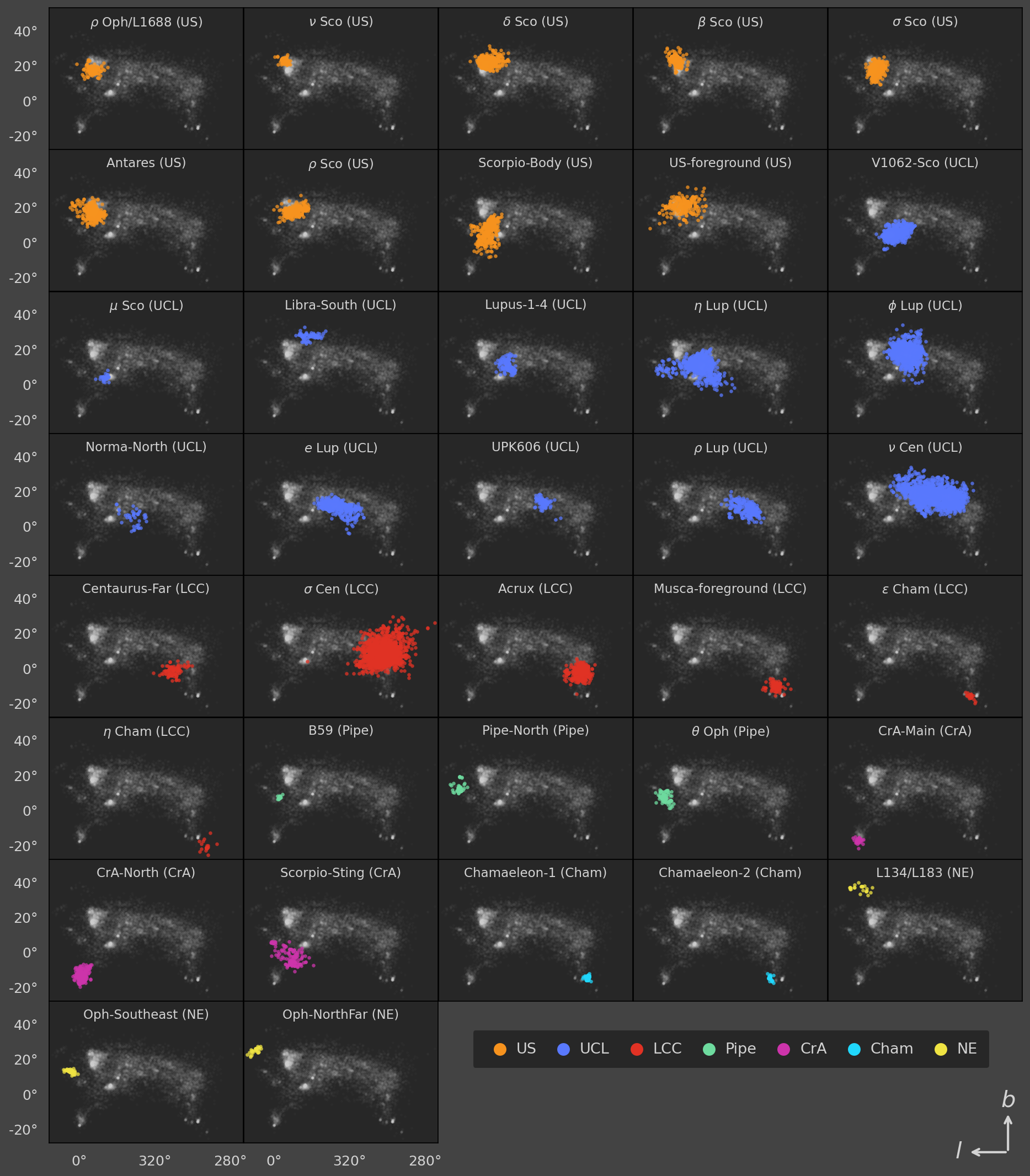}
    \caption{The distribution of \texttt{SigMA} clusters in Sco-Cen, projected in Galactic coordinates, stratified by group membership. Compared to Fig.~\ref{fig:lb-map}, the small multiples highlight the distribution of individual clusters in the Sco-Cen complex. The color coding represents the seven regions US (orange), UCL (blue), LCC (red), Pipe (green), CrA (magenta), Cham (cyan), and NE (yellow).} 
    \label{fig:lb_small_muliples}
\end{figure*}

\begin{figure*}
    \centering
    \includegraphics[width=1\textwidth]{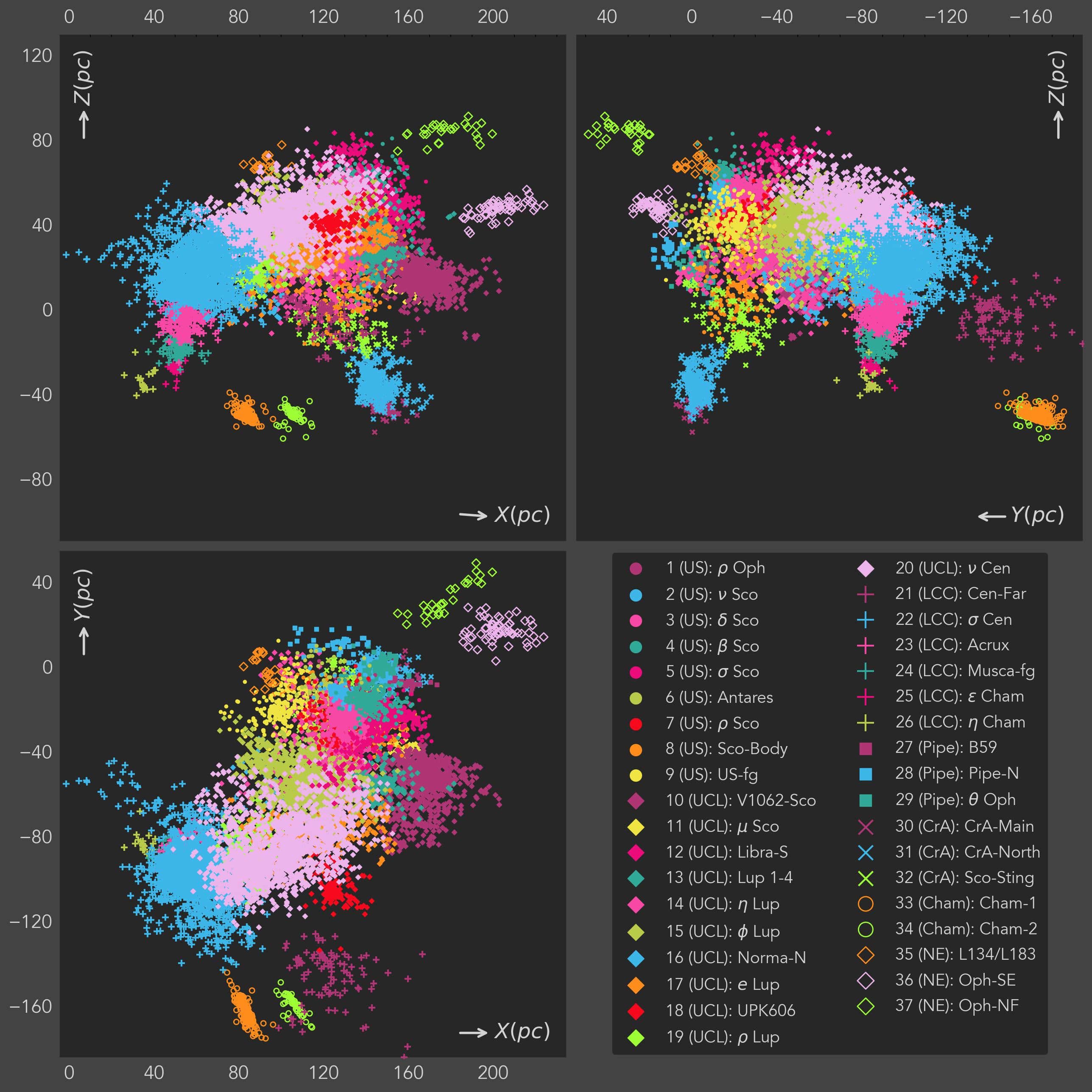}
    \caption{3D distribution of the 37 \texttt{SigMA} Sco-Cen clusters in Heliocentric Galactic Cartesian coordinates. The Sun is at (0,0,0). Colors and labels are as in Fig.~\ref{fig:lb-map}. See also the  \href{https://homepage.univie.ac.at/josefa.elisabeth.grossschedl/wp-content/uploads/2022/11/scocen_interactive_3D-1.html}{interactive 3D version} and Figs.\,\ref{fig:parameter-multis-1}--\ref{fig:parameter-multis-5}, which allow a better appreciation of individual cluster properties. By double-clicking on a cluster in the legend of the interactive version, the selected cluster can be isolated; by hovering over data points the cluster membership and observed $l$, $b$, $d$ position of a source gets visible. 
    }
    \label{fig:3D-map}
\end{figure*}

\begin{figure*}
    \centering
        \includegraphics[width=\textwidth]{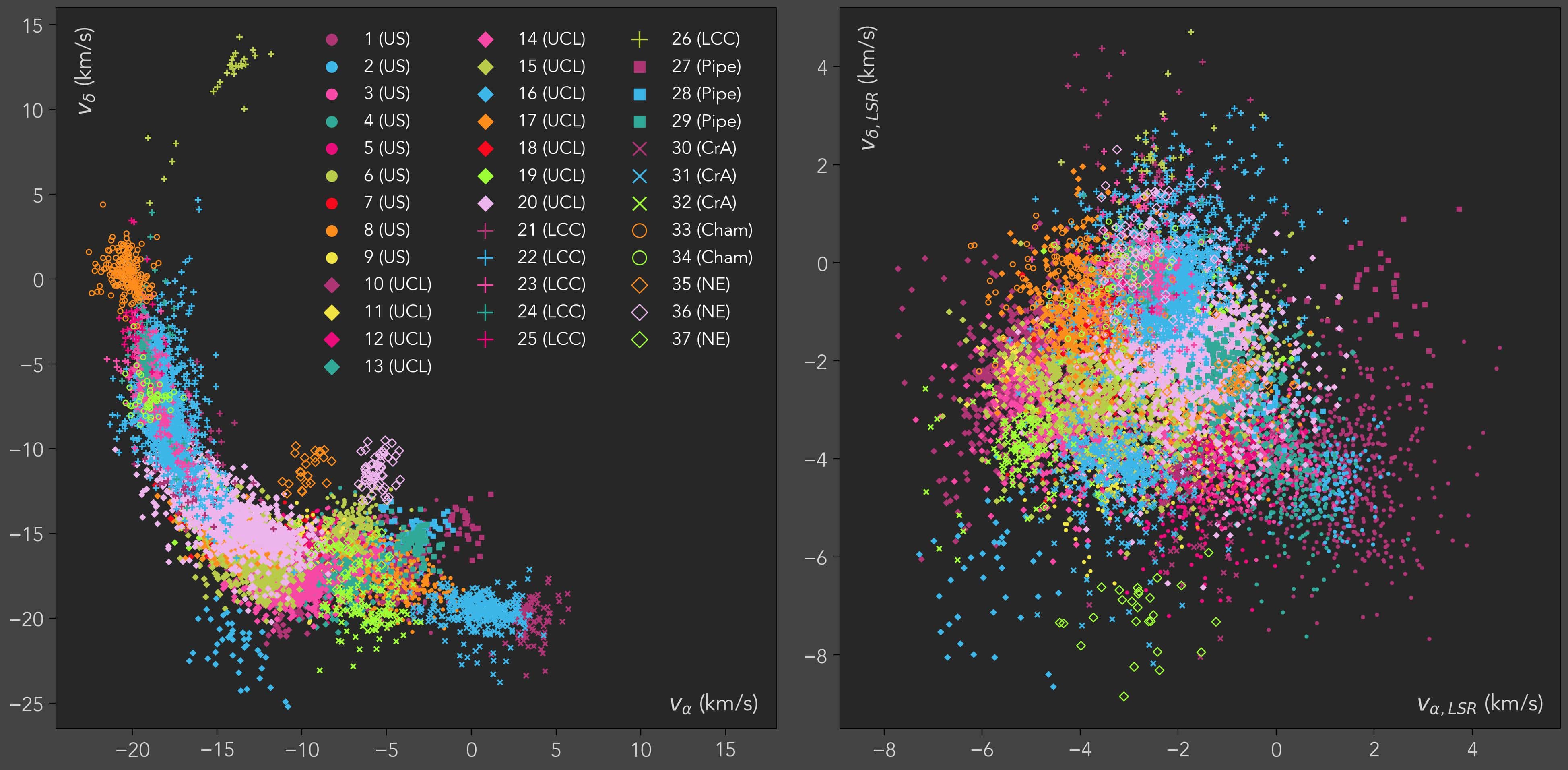}
    \caption{Tangential velocity distribution of the 37 \texttt{SigMA} clusters. Colors and labels are as in Fig.~\ref{fig:lb-map}.
    \textit{Left:} The observed tangential velocities along $\alpha$ and $\delta$ are strongly influenced by the Sun's reflex motion, while stellar clusters at similar distances and with similar space motions are arranged in a loop-like pattern.
    Sources at $l\sim\SI{0}{\degree}$ are located in the lower right part of the figure, and sources at $l\sim\SI{290}{\degree}$ in the upper left part of the figure (see Fig.\,\ref{fig:vtan_theoretical} in Appendix~\ref{app:vtan}). 
    \textit{Right:} The tangential velocities corrected for the Sun's motion, hence relative to the LSR. The correction reduces the projection effects of the observed tangential stellar motions. 
    See the \href{https://homepage.univie.ac.at/josefa.elisabeth.grossschedl/wp-content/uploads/2022/11/scocen_interactive_2D.html}{interactive 2D version} and Figs.\,\ref{fig:parameter-multis-1}--\ref{fig:parameter-multis-5} for a better appreciation of 2D kinematical properties of the clusters in Sco-Cen.
    }
    \label{fig:vtan}
\end{figure*}

\section{Application to Sco-Cen}\label{sec:results}

We apply \texttt{SigMA} to Gaia DR3 data inside a box of about $10^7$\,pc$^3$ containing the Sco-Cen OB association, as defined in Sect.~\ref{sec:data}. The box was chosen to include the classical Blaauw definition of Sco-Cen, including the classical sub-groups Upper-Scorpius (US), Upper-Centaurus-Lupus (UCL), and Lower-Centaurus-Crux (LCC), and to go beyond them and include the molecular cloud complexes of Pipe, Corona Australis (CrA), Chameleon (Cham), and three stellar groups to the Galactic northeast of Sco-Cen, which we put in the separate North East group (NE). Some of these regions were tentatively associated with Sco-Cen in the past  \citep[e.g.,][]{Lepine2003, Sartori2003, Preibisch2008, Bouy2015, Kerr2021}.   

In this paper, we discuss the \texttt{SigMA} extracted young stellar clusters in Sco-Cen, which are part of the $\lesssim 20$\,Myr Sco-Cen star formation event \citep{Pecaut2012}, and their connection to previous work. In a future work we plan to discuss in more detail the ages of the individual \texttt{SigMA} clusters and the star formation history of the Sco-Cen complex. 

In total \texttt{SigMA} extracts 60 clusters inside the defined search box. Of these, 23 clusters are older populations with ages\,$>$\,20\,Myr or which are kinematically unrelated. For example, the well studied IC\,2602 \citep[$\sim$30\,Myr, e.g.,][]{Randich1995, Stauffer1997, Dobbie2010, Damiani2019, Meingast2021}, or the Hyades, $\beta$\,Pictoris, Platais\,8, Platais\,9, Platais\,10, IC\,2391, Alessi\,9, Alessi\,13, Tucana-Horologium, Coma-Berenices, Volans-Carina, or NGC\,2451A \citep[e.g.,][]{Riedel2017, Gagne2018a, Gagne2018b, GagneFaherty2018, Sim2019, Fuernkranz2019, Cantat-Gaudin2020a, Meingast2021, Kerr2021, Galli2021, He2022b}. These clusters generally occupy distinct velocity spaces, different from the bulk motion of Sco-Cen. Moreover, the majority of these clusters are truncated by the borders of our defined box, hence they are incomplete, which is of no consequence to this study. In this work, we focus solely on the young Sco-Cen complex (1) to get a more complete picture of the substructure of this important nearby association, (2) to evaluate the differences to previous studies on Sco-Cen (Sect.\,\ref{sec:compare}), and (3) to highlight the capability of \texttt{SigMA} to untangle distinct clusters in a dense environment containing overlapping populations in space, which is especially true for young stellar associations like Sco-Cen. The 23 older or unrelated clusters are not discussed further here, although they might be related, or not, to Sco-Cen at larger scales \citep[e.g., ``blue streams'',][]{Bouy2015}. We will discuss these older clusters in future work. 

We find that 37 stellar clusters are associated spatially and kinematically with the Sco-Cen OB association, containing in total 13,103 stellar cluster members, which will be discussed in more detail in this section. Figures~\ref{fig:lb-map}~\&~\ref{fig:lb_small_muliples} show the distribution of the 37 Sco-Cen \texttt{SigMA} clusters projected in Galactic coordinates. Figure~\ref{fig:3D-map} shows the distribuiton of the clusters in 3D space in a Heliocentric Galactic Cartesian coordinate frame (see also the \href{https://homepage.univie.ac.at/josefa.elisabeth.grossschedl/wp-content/uploads/2022/11/scocen_interactive_3D-1.html}{interactive 3D version} and Figs.\,\ref{fig:parameter-multis-1}--\ref{fig:parameter-multis-5} for a better appreciation of individual clusters). The 37 clusters seem to form the continuous body of the Sco-Cen association, beyond Blaauw's original three subgroups boundaries. 

Figure~\ref{fig:vtan} shows the location of the \texttt{SigMA} clusters in the tangential velocity plane as observed from the Sun ($v_\alpha$/$v_\delta$) and also relative to the LSR ($v_{\alpha,\mathrm{LSR}}$/$v_{\delta,\mathrm{LSR}}$). Since the clusters partially occupy similar velocity spaces in the velocity planes, we also provide an \href{https://homepage.univie.ac.at/josefa.elisabeth.grossschedl/wp-content/uploads/2022/11/scocen_interactive_2D.html}{interactive 2D version} of this figure, allowing a better appreciation of 2D kinematical properties of the clusters in Sco-Cen (see also Figs.\,\ref{fig:parameter-multis-1}--\ref{fig:parameter-multis-5}). 
The 37 young clusters all fall on a connected loop-like pattern in tangential velocity space (Fig.\,\ref{fig:vtan}, left panel), a pattern largely created by the reflex motion of the Sun. This is highlighted in Fig.~\ref{fig:vtan_theoretical} in Appendix~\ref{app:vtan}, showing that these projected motions are expected for clusters at Sco-Cen positions and distances. 
To avoid this pattern caused by the Sun's motion, we additionally transform the tangential velocities to velocities relative to the LSR, using the standard solar motion from \citet{Schoenrich2010} (as mentioned in Sect.\,\ref{sec:data}). This is shown in Fig.\,\ref{fig:vtan} (right panel), where we can see that the clusters now occupy a more compact velocity space. In particular larger clusters, which are stretched over larger areas in the sky, show a smaller velocity dispersion after the LSR conversion (see also Table~\ref{tab:averages-vel}).
The 2D kinematical properties of individual clusters can be better appreciated when investigating the  \href{https://homepage.univie.ac.at/josefa.elisabeth.grossschedl/wp-content/uploads/2022/11/scocen_interactive_2D.html}{interactive 2D version} of the figure\footnote{Clusters can be viewed separately by double-clicking on the cluster name in the legend. By clicking once on another cluster it can be added to the visible clusters, and so on.}, where both velocity spaces can be compared directly. 

The Sco-Cen association, as extracted with \texttt{SigMA}, reaches well below the Galactic plane, as was indicated by previous works \citep[e.g.,][]{Kerr2021} and is now further confirmed here. 
This includes regions that are not traditionally associated with Sco-Cen, like Pipe, CrA, Cham, and clusters toward the Galactic northeast (NE), including a cluster connected to the L134/L183 clouds. Moreover, other well-known stellar clusters, traditionally not assigned to Sco-Cen but later suggested to be associated with it, were picked up by \texttt{SigMA}, like $\epsilon$\,Cha and $\eta$\,Cha \citep[e.g.,][]{Mamajek1999, Mamajek2000, Fernandez2008}, which are added here to the Sco-Cen complex.

The relatively young $\beta$~Pictoris stellar group \citep[$\beta$\,Pic, e.g.,][age $\sim$\,18--20\,Myr]{Fernandez2008, Crundall2019, Miret-Roig2020} was also picked up by \texttt{SigMA} in the selection box.
As mentioned above, we decided not to include this young local association as part of our final sample of 37 stellar clusters. The \texttt{SigMA} extraction of $\beta$\,Pic covers only one side of the known population as defined in \citet{Miret-Roig2020}. This is likely due to larger extent of $\beta$\,Pic in the sky (partially outside of our box boundaries) and due to the relatively close distance to the Sun (average distance of about 40\,pc), which makes it more difficult to extract members from the 5D phase space as used by \texttt{SigMA} in this work as the stars are distributed across the sky as seen from Earth (we are inside some of these nearby young associations). 

The majority of the 37 clusters can be related to previously identified clusters from the literature, which are often larger scale structures containing several of the \texttt{SigMA} clusters (see comparisons to the literature in Sect.\,\ref{sec:compare}). The rich sub-structure identified by \texttt{SigMA} also includes clusters with no clear counterpart in previous works. We decided to name such clusters after their location in the sky (based on constellation) or after the brightest star that is seen in projection to a cluster and where we feel confident that it is part of a cluster (see Sect.\,\ref{sec:overview-scocen-regions} and Tables\,\ref{tab:averages-pos} \& \ref{tab:hip}). We often find bright B-stars towards cluster centers at approximately the same distance and proper motion, in itself a validation of the ``\texttt{SigMA}'' algorithm, as many of these bright stars are not in Gaia (but only in \textsc{Hipparcos}). We use \textsc{Hipparcos} astrometry \citep{VanLeeuwen2007} to tentatively associate bright B-stars to the new clusters and list them and their astrometric properties in Table\,\ref{tab:hip}, showing the HIP ID and the \textsc{Hipparcos} astrometry. This table allows a direct comparison with the average properties of the \texttt{SigMA} clusters in Tables~\ref{tab:averages-pos}--\ref{tab:averages-vel}. For the cases where there is a reasonable match, we name the cluster with the name of the bright B-star\footnote{This approach seems valid in particular for US, since also other authors, like \citet{MiretRoig2022b} or \citet{BricenoMorals2022}, independently decided for similar cluster names.}. Additionally, we index the stellar clusters within this work from 1 to 37 as given in Col.~``\texttt{SigMA}'' in Table~\ref{tab:averages-pos}.

Figure~\ref{fig:hrd} shows the \texttt{SigMA} cluster members in a Gaia HRD, confirming the youth of most sources. In Appendix~\ref{ap:old-star-cont}, we give more details on the chosen photometric quality criteria and the selection conditions. We find an excess of older low-mass sources that visibly separate from the Sco-Cen population, potentially false positive Sco-Cen members. We use a 25\,Myr isochrone (to allow for random scatter) to separate ``younger'' Sco-Cen members from ``older'' populations or field stars as shown in Fig.~\ref{fig:hrd} (middle panel) as explained in Appendix~\ref{ap:old-star-cont}. This gives a rough estimate for a contamination fraction of about 4--10\%, depending on the photometric quality criteria. This contamination fraction is similar to the estimate in the methods section (Sect.~\ref{sect:contamination_completeness}). The influence of the stability that we assigned each cluster member can be seen in Fig.\,\ref{fig:stability}, which suggests that a cut at about 11\% would give a cleaner cluster membership selection and a lower contamination fraction (Appendix~\ref{ap:old-star-cont}), but also a less complete sample.

\begin{figure*}
    \centering
    \includegraphics[width=1\textwidth]{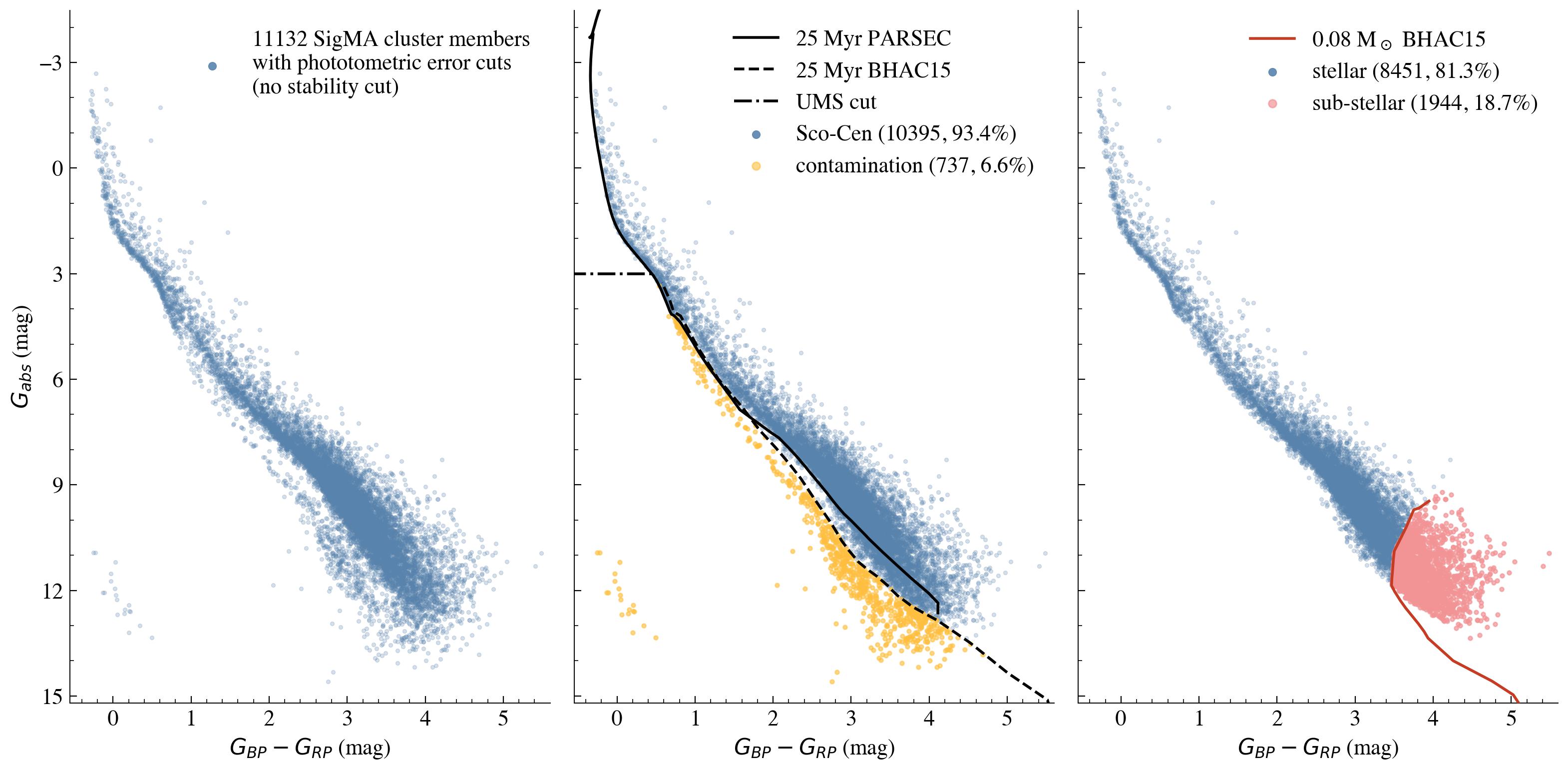}
    \caption{Gaia color-absolute-magnitude diagram $G_\mathit{abs}$ versus $G_\mathrm{BP}-G_\mathrm{RP}$ (HRD) of the SigMA Sco-Cen members.
    \textit{Left:} SigMA cluster members that pass the photometric quality criteria as given in Equ.\,(\ref{eq:phot-cut}).   
    \textit{Middle:} Potential contamination from older sources (orange), selected with a 25\,Myr isochrone from PARSEC (black line) and \citet{Baraffe2015} (black dashed line) and an additional cut at $G_\mathrm{abs}=3$\,mag (black dashed-dotted line), which excludes the upper--main-sequence. 
    The combined cuts indicate a contamination from older sources of about 6--7\% when using the given photometric quality criteria and no stability cut.
    \textit{Right:} Sub-stellar candidates (red dots) are selected with a $0.08\,M_\odot$ iso-mass line from \citet{Baraffe2015} (dark-red line), using only the younger source from the middle panel. This cut indicates that there are roughly 19\% of sub-stellar sources within the \texttt{SigMA} Sco-Cen members when applying the mentioned cuts and photometric quality criteria. 
    More details on the quality criteria, the selection borders, and the isochrone models are given in Appendix~\ref{app:hrd}. 
    }
    \label{fig:hrd}
\end{figure*}

When further investigating the young \texttt{SigMA} Sco-Cen members in the HRD in Fig.~\ref{fig:hrd} (right panel), we find that within the sample, there are about 19\% sub-stellar objects candidates (see also Appendix~\ref{ap:substellar}). In the future, more complete samples of the individual clusters can be obtained by using the known members as training sets \citep[e.g., as demonstrated with \texttt{Uncover} in][]{Ratzenboeck2020}. For instance, knowing the Brown Dwarf population will allow getting more complete initial mass functions beyond the hydrogen burning limit and a better characterization of the mass of the individual clusters \citep[e.g.,][]{MiretRoig2022a}. 

\begin{table*}[!ht] 
\begin{center}
\begin{small}
\caption{Overview of the 37 \texttt{SigMA} clusters in Sco-Cen, assigned to seven subregions (Col.~2). The median cluster positions are listed (see Table~\ref{tab:averages-vel} for the median velocities). \vspace{-1mm}}
\renewcommand{\arraystretch}{1.3}
\resizebox{0.99\textwidth}{!}{%
\begin{tabular}{clllrrrrrrrr}
\hline \hline
\multicolumn{1}{r}{\texttt{SigMA}} &
\multicolumn{1}{l}{Region} &
\multicolumn{1}{l}{Group Name} &
\multicolumn{1}{l}{Brightest star\tablefootmark{a}} &
\multicolumn{1}{r}{Nr.} &
\multicolumn{1}{c}{$l$} &
\multicolumn{1}{c}{$b$} &
\multicolumn{1}{c}{$\varpi$} &
\multicolumn{1}{c}{$d$} &
\multicolumn{1}{c}{$X$} &
\multicolumn{1}{c}{$Y$} &
\multicolumn{1}{c}{$Z$} \\
& & & & &
\multicolumn{1}{c}{(deg)} & 
\multicolumn{1}{c}{(deg)} &
\multicolumn{1}{c}{(mas)} & 
\multicolumn{1}{c}{(pc)} &
\multicolumn{1}{c}{(pc)} &
\multicolumn{1}{c}{(pc)} &
\multicolumn{1}{c}{(pc)} \\
\hline

1 & US & $\rho$ Oph/L1688 & * rho Oph & 535 & 353.20$^{+0.57}_{-0.81}$ & 17.09$^{+1.24}_{-0.81}$ & 7.20$^{+0.23}_{-0.28}$ & 139$^{+6}_{-4}$ & 132$^{+6}_{-4}$ & -16$^{+1}_{-2}$ & 41$^{+3}_{-2}$ \\
2 & US & $\nu$ Sco & * nu Sco & 150 & 354.46$^{+0.81}_{-1.03}$ & 22.86$^{+0.85}_{-0.92}$ & 7.18$^{+0.16}_{-0.20}$ & 139$^{+4}_{-3}$ & 128$^{+3}_{-3}$ & -12$^{+2}_{-3}$ & 54$^{+2}_{-2}$ \\
3 & US & $\delta$ Sco & * b Sco & 691 & 350.25$^{+2.11}_{-3.66}$ & 22.18$^{+1.50}_{-1.91}$ & 7.03$^{+0.20}_{-0.25}$ & 142$^{+5}_{-4}$ & 130$^{+5}_{-4}$ & -22$^{+5}_{-9}$ & 54$^{+4}_{-4}$ \\
4 & US & $\beta$ Sco & HD\,142883 & 285 & 353.21$^{+1.48}_{-1.26}$ & 23.53$^{+1.54}_{-3.08}$ & 6.49$^{+0.19}_{-0.25}$ & 154$^{+6}_{-4}$ & 141$^{+7}_{-5}$ & -17$^{+4}_{-4}$ & 61$^{+5}_{-6}$ \\
5 & US & $\sigma$ Sco & * c02 Sco & 544 & 351.13$^{+1.23}_{-2.34}$ & 17.87$^{+1.92}_{-2.31}$ & 6.29$^{+0.23}_{-0.26}$ & 159$^{+7}_{-6}$ & 149$^{+8}_{-7}$ & -24$^{+4}_{-6}$ & 49$^{+4}_{-6}$ \\
6 & US & Antares & HD\,146001 & 502 & 352.79$^{+2.18}_{-2.28}$ & 17.22$^{+2.83}_{-2.71}$ & 7.21$^{+0.45}_{-0.44}$ & 139$^{+9}_{-8}$ & 132$^{+9}_{-9}$ & -17$^{+5}_{-6}$ & 41$^{+5}_{-6}$ \\
7 & US & $\rho$ Sco & * rho Sco & 240 & 349.31$^{+3.85}_{-4.77}$ & 18.29$^{+1.80}_{-2.76}$ & 7.21$^{+0.47}_{-0.36}$ & 139$^{+7}_{-8}$ & 129$^{+8}_{-10}$ & -24$^{+9}_{-11}$ & 43$^{+5}_{-6}$ \\
8 & US & Scorpio-Body & HD\,150638 & 373 & 349.36$^{+3.32}_{-2.49}$ & 7.32$^{+4.09}_{-7.01}$ & 7.08$^{+0.91}_{-0.58}$ & 141$^{+13}_{-16}$ & 137$^{+11}_{-15}$ & -26$^{+9}_{-7}$ & 17$^{+13}_{-17}$ \\
9 & US & US-foreground & HD\,145964 & 276 & 348.81$^{+5.68}_{-3.80}$ & 21.04$^{+3.35}_{-3.46}$ & 9.05$^{+0.67}_{-0.74}$ & 110$^{+10}_{-8}$ & 102$^{+9}_{-7}$ & -20$^{+9}_{-6}$ & 39$^{+8}_{-7}$ \\
\hline 10 & UCL & V1062-Sco & * mu02 Sco & 1029 & 343.11$^{+1.35}_{-4.41}$ & 4.69$^{+1.53}_{-1.50}$ & 5.66$^{+0.23}_{-0.24}$ & 177$^{+8}_{-7}$ & 168$^{+7}_{-8}$ & -51$^{+5}_{-14}$ & 14$^{+5}_{-4}$ \\
11 & UCL & $\mu$ Sco & HD\,151726 & 54 & 346.19$^{+0.95}_{-0.65}$ & 3.90$^{+0.48}_{-0.75}$ & 6.07$^{+0.20}_{-0.09}$ & 165$^{+3}_{-5}$ & 160$^{+3}_{-5}$ & -40$^{+3}_{-1}$ & 11$^{+1}_{-2}$ \\
12 & UCL & Libra-South & HD\,138343 & 71 & 341.77$^{+3.12}_{-4.68}$ & 27.80$^{+1.07}_{-1.42}$ & 6.34$^{+0.30}_{-0.19}$ & 158$^{+5}_{-7}$ & 132$^{+6}_{-9}$ & -45$^{+9}_{-8}$ & 73$^{+4}_{-6}$ \\
13 & UCL & Lupus 1-4 & * LL Lup & 226 & 339.51$^{+1.15}_{-2.72}$ & 9.44$^{+3.59}_{-0.99}$ & 6.27$^{+0.16}_{-0.22}$ & 160$^{+6}_{-4}$ & 147$^{+6}_{-4}$ & -55$^{+3}_{-7}$ & 26$^{+11}_{-3}$ \\
14 & UCL & $\eta$ Lup & * eta Lup & 769 & 339.83$^{+6.08}_{-4.45}$ & 10.71$^{+3.13}_{-4.50}$ & 7.37$^{+1.04}_{-0.46}$ & 136$^{+9}_{-17}$ & 124$^{+8}_{-13}$ & -47$^{+15}_{-10}$ & 25$^{+8}_{-11}$ \\
15 & UCL & $\phi$ Lup & * phi02 Lup & 1114 & 334.17$^{+4.39}_{-4.22}$ & 17.69$^{+3.70}_{-3.78}$ & 7.65$^{+1.06}_{-0.78}$ & 131$^{+15}_{-16}$ & 112$^{+12}_{-16}$ & -54$^{+10}_{-9}$ & 40$^{+9}_{-10}$ \\
16 & UCL & Norma-North & HD\,143215 & 42 & 331.12$^{+4.01}_{-2.98}$ & 6.39$^{+2.19}_{-6.64}$ & 9.44$^{+0.80}_{-0.56}$ & 106$^{+7}_{-8}$ & 92$^{+6}_{-5}$ & -50$^{+9}_{-9}$ & 11$^{+4}_{-12}$ \\
17 & UCL & $e$ Lup & * e Lup & 516 & 327.40$^{+3.92}_{-6.80}$ & 11.48$^{+1.93}_{-2.35}$ & 6.89$^{+0.65}_{-0.49}$ & 145$^{+11}_{-13}$ & 120$^{+13}_{-20}$ & -76$^{+6}_{-9}$ & 29$^{+6}_{-7}$ \\
18 & UCL & UPK606 & HD\,125777 & 131 & 320.00$^{+2.17}_{-1.37}$ & 13.72$^{+1.36}_{-1.68}$ & 5.92$^{+0.27}_{-0.25}$ & 169$^{+7}_{-7}$ & 125$^{+7}_{-4}$ & -106$^{+9}_{-4}$ & 40$^{+3}_{-4}$ \\
19 & UCL & $\rho$ Lup & * rho Lup & 246 & 315.12$^{+6.44}_{-3.47}$ & 9.83$^{+3.78}_{-3.09}$ & 8.16$^{+0.55}_{-0.57}$ & 123$^{+9}_{-8}$ & 87$^{+7}_{-8}$ & -82$^{+9}_{-13}$ & 21$^{+7}_{-6}$ \\
20 & UCL & $\nu$ Cen & * nu Cen & 1737 & 318.50$^{+9.60}_{-8.02}$ & 17.45$^{+4.13}_{-4.08}$ & 7.21$^{+0.67}_{-0.69}$ & 139$^{+15}_{-12}$ & 99$^{+22}_{-19}$ & -87$^{+18}_{-12}$ & 41$^{+12}_{-9}$ \\
\hline 21 & LCC & Centaurus-Far & HD\,121808 & 99 & 310.69$^{+2.02}_{-3.30}$ & -1.18$^{+2.14}_{-2.74}$ & 5.25$^{+0.35}_{-0.31}$ & 190$^{+12}_{-12}$ & 122$^{+15}_{-12}$ & -142$^{+6}_{-15}$ & -4$^{+7}_{-9}$ \\
22 & LCC & $\sigma$ Cen & * sig Cen & 1805 & 301.56$^{+5.56}_{-5.32}$ & 8.35$^{+4.99}_{-4.74}$ & 8.71$^{+0.87}_{-1.05}$ & 115$^{+16}_{-10}$ & 60$^{+12}_{-11}$ & -96$^{+10}_{-13}$ & 17$^{+11}_{-10}$ \\
23 & LCC & Acrux & * zet Crux & 394 & 300.27$^{+3.20}_{-1.71}$ & -1.98$^{+2.05}_{-3.53}$ & 9.41$^{+0.47}_{-0.37}$ & 106$^{+4}_{-5}$ & 54$^{+5}_{-3}$ & -91$^{+7}_{-4}$ & -4$^{+4}_{-6}$ \\
24 & LCC & Musca-foreground & HD\,107947 & 95 & 300.64$^{+1.69}_{-2.18}$ & -10.27$^{+1.81}_{-1.92}$ & 9.79$^{+0.30}_{-0.40}$ & 102$^{+4}_{-3}$ & 52$^{+3}_{-4}$ & -86$^{+3}_{-4}$ & -18$^{+3}_{-4}$ \\
25 & LCC & $\epsilon$ Cham & * DX Cha & 39 & 300.34$^{+0.71}_{-0.27}$ & -15.97$^{+0.68}_{-0.76}$ & 9.81$^{+0.18}_{-0.26}$ & 102$^{+3}_{-2}$ & 50$^{+1}_{-1}$ & -85$^{+3}_{-3}$ & -28$^{+1}_{-2}$ \\
26 & LCC & $\eta$ Cham & * eta Cha & 30 & 292.49$^{+1.56}_{-0.48}$ & -21.60$^{+2.30}_{-0.32}$ & 10.14$^{+0.40}_{-0.13}$ & 99$^{+1}_{-4}$ & 35$^{+2}_{-1}$ & -85$^{+2}_{-1}$ & -36$^{+5}_{-1}$ \\
\hline 27 & Pipe & B59 & Em* AS 218 & 32 & 357.10$^{+0.38}_{-0.30}$ & 7.11$^{+0.59}_{-0.63}$ & 6.23$^{+0.12}_{-0.16}$ & 160$^{+4}_{-3}$ & 159$^{+4}_{-3}$ & -8$^{+1}_{-1}$ & 20$^{+2}_{-2}$ \\
28 & Pipe & Pipe-North & HD\,155427 & 42 & 4.92$^{+1.29}_{-1.15}$ & 12.85$^{+2.53}_{-1.98}$ & 7.69$^{+0.69}_{-0.30}$ & 130$^{+5}_{-11}$ & 126$^{+6}_{-12}$ & 11$^{+3}_{-3}$ & 29$^{+6}_{-5}$ \\
29 & Pipe & $\theta$ Oph & HD\,158704 & 98 & 359.71$^{+1.09}_{-2.75}$ & 7.05$^{+2.66}_{-1.95}$ & 6.79$^{+0.29}_{-0.20}$ & 147$^{+5}_{-6}$ & 146$^{+4}_{-6}$ & -1$^{+3}_{-7}$ & 19$^{+5}_{-6}$ \\
\hline 30 & CrA & CrA-Main & HD\,177076 & 96 & 359.87$^{+0.38}_{-0.66}$ & -17.65$^{+0.70}_{-0.32}$ & 6.46$^{+0.13}_{-0.13}$ & 155$^{+3}_{-3}$ & 147$^{+3}_{-3}$ & 0$^{+1}_{-2}$ & -47$^{+2}_{-1}$ \\
31 & CrA & CrA-North & HD\,172910 & 351 & 359.02$^{+1.17}_{-1.77}$ & -13.97$^{+2.74}_{-1.91}$ & 6.70$^{+0.25}_{-0.29}$ & 149$^{+7}_{-5}$ & 145$^{+7}_{-5}$ & -2$^{+3}_{-5}$ & -36$^{+8}_{-5}$ \\
32 & CrA & Scorpio-Sting & HD\,157864 & 132 & 350.59$^{+5.27}_{-3.11}$ & -3.04$^{+4.72}_{-3.77}$ & 7.49$^{+0.66}_{-0.52}$ & 134$^{+10}_{-11}$ & 131$^{+9}_{-11}$ & -22$^{+12}_{-6}$ & -7$^{+11}_{-9}$ \\
\hline 33 & Cham & Chamaeleon-1 & V* CR Cha & 192 & 297.22$^{+0.21}_{-0.52}$ & -15.52$^{+0.90}_{-0.37}$ & 5.25$^{+0.18}_{-0.10}$ & 191$^{+4}_{-6}$ & 84$^{+2}_{-3}$ & -164$^{+6}_{-4}$ & -50$^{+2}_{-2}$ \\
34 & Cham & Chamaeleon-2 & V* BF Cha & 54 & 303.69$^{+0.23}_{-0.37}$ & -14.72$^{+0.74}_{-0.38}$ & 5.08$^{+0.13}_{-0.16}$ & 197$^{+6}_{-5}$ & 105$^{+4}_{-3}$ & -159$^{+4}_{-5}$ & -50$^{+3}_{-4}$ \\
\hline 35 & NE & L134/L183 & HD\,141569 & 24 & 358.40$^{+5.78}_{-3.05}$ & 36.84$^{+0.79}_{-2.22}$ & 8.76$^{+0.24}_{-0.32}$ & 114$^{+4}_{-3}$ & 93$^{+4}_{-4}$ & -3$^{+9}_{-5}$ & 67$^{+4}_{-1}$ \\
36 & NE & Oph-Southeast & HD\,154922 & 61 & 4.54$^{+1.49}_{-1.01}$ & 13.06$^{+0.68}_{-0.63}$ & 4.80$^{+0.19}_{-0.24}$ & 208$^{+11}_{-8}$ & 202$^{+11}_{-8}$ & 16$^{+6}_{-4}$ & 48$^{+3}_{-4}$ \\
37 & NE & Oph-NorthFar & BD-06 4472 & 28 & 9.60$^{+2.56}_{-1.57}$ & 24.98$^{+1.32}_{-1.33}$ & 5.06$^{+0.24}_{-0.37}$ & 198$^{+15}_{-9}$ & 176$^{+16}_{-8}$ & 29$^{+11}_{-5}$ & 85$^{+2}_{-6}$ \\

\hline
\end{tabular}
} 
\renewcommand{\arraystretch}{1}
\label{tab:averages-pos}
\tablefoot{In Cols.~6--12 we list the medians of the positional parameters for each cluster including all cluster members (without considering any stability cut). The given lower and upper uncertainties represent the 1$\sigma$ scatter around the median. In this scatter the original measurement uncertainties of single stellar sources are not considered.
\tablefoottext{a}{Col.\,4 lists the brightest star that was selected as member by \texttt{SigMA}. The star annotation ``*'' is used as in the SIMBAD astronomical database \citep{Wenger2000} and it helps to distinguish stellar names from cluster names throughout the manuscript, since some clusters are named after bright stars. Further bright stellar member candidates, which were observed by \textsc{Hipparcos} (partially not in Gaia DR3), are listed in Table~\ref{tab:hip}.}
}
\end{small}
\end{center}
\end{table*} 
\begin{table*}[!ht] 
\begin{center}
\begin{small}
\caption{Median velocity parameters for the 37 \texttt{SigMA} clusters in Sco-Cen. \vspace{-1mm}}
\renewcommand{\arraystretch}{1.3}
\resizebox{0.85\textwidth}{!}{%
\begin{tabular}{llrrrrrrr}
\hline \hline
\multicolumn{1}{l}{\texttt{SigMA}} &
\multicolumn{1}{l}{Group Name} &
\multicolumn{1}{r}{Nr.} &
\multicolumn{1}{c}{$\mu_\alpha^*$} &
\multicolumn{1}{c}{$\mu_\delta$} &
\multicolumn{1}{c}{$v_\alpha$} &
\multicolumn{1}{c}{$v_\delta$} &
\multicolumn{1}{c}{$v_{\alpha,\mathrm{LSR}}$} &
\multicolumn{1}{c}{$v_{\delta,\mathrm{LSR}}$} \\
 & 
\multicolumn{1}{c}{(short version)} & 
 & 
\multicolumn{2}{c}{(mas/yr)} &
\multicolumn{2}{c}{(km/s)} &
\multicolumn{2}{c}{(km/s)} \\
\hline

1 (US) & $\rho$ Oph & 535 & -6.81$^{+1.36}_{-1.56}$ & -25.91$^{+1.78}_{-1.58}$ & -4.50$^{+0.89}_{-1.01}$ & -16.95$^{+0.91}_{-1.28}$ & 1.36$^{+0.84}_{-0.94}$ & -4.20$^{+1.01}_{-1.15}$ \\
2 & $\nu$ Sco & 150 & -8.47$^{+0.74}_{-0.81}$ & -24.50$^{+0.76}_{-0.98}$ & -5.62$^{+0.55}_{-0.56}$ & -16.22$^{+0.61}_{-0.73}$ & 1.21$^{+0.37}_{-0.43}$ & -4.52$^{+0.58}_{-0.60}$ \\
3 & $\delta$ Sco & 691 & -12.24$^{+2.04}_{-2.54}$ & -23.92$^{+1.27}_{-1.35}$ & -8.22$^{+1.40}_{-1.99}$ & -16.27$^{+0.88}_{-0.75}$ & -0.77$^{+0.88}_{-1.09}$ & -3.88$^{+0.57}_{-0.66}$ \\
4 & $\beta$ Sco & 285 & -9.39$^{+1.37}_{-1.07}$ & -21.69$^{+1.00}_{-1.73}$ & -6.87$^{+0.86}_{-0.77}$ & -15.86$^{+0.77}_{-1.50}$ & 0.26$^{+0.56}_{-0.59}$ & -4.13$^{+0.58}_{-0.96}$ \\
5 & $\sigma$ Sco & 544 & -10.68$^{+1.47}_{-1.37}$ & -21.79$^{+1.34}_{-1.87}$ & -8.12$^{+1.09}_{-0.87}$ & -16.49$^{+0.92}_{-1.23}$ & -1.68$^{+0.78}_{-0.61}$ & -3.48$^{+0.91}_{-1.31}$ \\
6 & Antares & 502 & -11.00$^{+1.59}_{-1.25}$ & -23.31$^{+2.15}_{-2.00}$ & -7.24$^{+0.87}_{-0.69}$ & -15.52$^{+1.29}_{-0.72}$ & -1.29$^{+0.84}_{-0.74}$ & -2.43$^{+0.86}_{-1.17}$ \\
7 & $\rho$ Sco & 240 & -15.98$^{+1.96}_{-2.08}$ & -24.14$^{+1.65}_{-1.72}$ & -10.44$^{+1.22}_{-1.18}$ & -15.82$^{+0.74}_{-0.85}$ & -3.64$^{+0.52}_{-0.59}$ & -2.78$^{+0.62}_{-0.78}$ \\
8 & Sco-Body & 373 & -8.04$^{+2.92}_{-1.91}$ & -26.82$^{+3.67}_{-3.55}$ & -5.30$^{+2.16}_{-1.70}$ & -17.64$^{+1.06}_{-1.08}$ & -1.38$^{+0.69}_{-0.39}$ & -2.83$^{+0.52}_{-0.67}$ \\
9 & US-fg & 276 & -19.97$^{+2.74}_{-2.86}$ & -31.25$^{+4.09}_{-3.02}$ & -10.63$^{+1.33}_{-0.96}$ & -16.28$^{+1.20}_{-1.07}$ & -3.33$^{+1.16}_{-0.73}$ & -3.86$^{+0.88}_{-0.77}$ \\
\hline 10 (UCL) & V1062-Sco & 1029 & -12.14$^{+0.96}_{-2.02}$ & -21.15$^{+0.93}_{-0.82}$ & -10.18$^{+0.75}_{-1.43}$ & -17.73$^{+0.75}_{-0.61}$ & -5.12$^{+0.41}_{-0.52}$ & -2.15$^{+0.66}_{-0.55}$ \\
11 & $\mu$ Sco & 54 & -11.75$^{+0.42}_{-0.83}$ & -22.61$^{+0.55}_{-0.37}$ & -9.22$^{+0.54}_{-0.53}$ & -17.61$^{+0.49}_{-0.44}$ & -5.28$^{+0.30}_{-0.25}$ & -2.11$^{+0.41}_{-0.31}$ \\
12 & Libra-S & 71 & -14.87$^{+1.55}_{-1.76}$ & -20.92$^{+1.22}_{-1.03}$ & -11.15$^{+1.11}_{-0.86}$ & -15.44$^{+0.45}_{-0.71}$ & -1.00$^{+0.53}_{-0.62}$ & -3.71$^{+0.43}_{-0.42}$ \\
13 & Lup 1-4 & 226 & -10.54$^{+1.51}_{-1.67}$ & -23.39$^{+1.01}_{-1.02}$ & -8.07$^{+1.18}_{-1.41}$ & -17.77$^{+0.95}_{-0.74}$ & -0.76$^{+0.79}_{-1.07}$ & -2.85$^{+0.75}_{-0.92}$ \\
14 & $\eta$ Lup & 769 & -17.63$^{+2.28}_{-2.01}$ & -27.84$^{+2.12}_{-3.93}$ & -11.21$^{+1.70}_{-1.34}$ & -17.83$^{+0.95}_{-0.83}$ & -4.15$^{+0.79}_{-0.85}$ & -3.18$^{+0.83}_{-0.65}$ \\
15 & $\phi$ Lup & 1114 & -20.99$^{+2.67}_{-4.71}$ & -25.60$^{+3.00}_{-4.16}$ & -13.31$^{+1.48}_{-1.24}$ & -15.82$^{+0.90}_{-1.38}$ & -3.37$^{+1.05}_{-1.00}$ & -2.45$^{+0.85}_{-1.03}$ \\
16 & Norma-N & 42 & -27.93$^{+2.68}_{-2.41}$ & -42.80$^{+2.50}_{-4.59}$ & -14.07$^{+1.61}_{-1.03}$ & -21.39$^{+1.22}_{-1.69}$ & -5.73$^{+0.82}_{-0.83}$ & -6.05$^{+0.73}_{-1.40}$ \\
17 & $e$ Lup & 516 & -20.80$^{+2.73}_{-4.62}$ & -21.67$^{+1.78}_{-1.60}$ & -14.44$^{+1.17}_{-1.46}$ & -15.07$^{+1.42}_{-0.67}$ & -3.81$^{+0.54}_{-0.50}$ & -1.05$^{+0.79}_{-0.60}$ \\
18 & UPK606 & 131 & -20.07$^{+1.27}_{-0.73}$ & -17.01$^{+0.71}_{-1.17}$ & -15.96$^{+0.87}_{-0.49}$ & -13.69$^{+0.44}_{-0.60}$ & -3.20$^{+0.44}_{-0.33}$ & -1.14$^{+0.47}_{-0.31}$ \\
19 & $\rho$ Lup & 246 & -26.22$^{+2.43}_{-2.95}$ & -23.13$^{+3.50}_{-2.37}$ & -15.51$^{+1.52}_{-1.06}$ & -13.14$^{+1.35}_{-1.36}$ & -1.95$^{+0.45}_{-0.73}$ & -1.18$^{+0.41}_{-0.40}$ \\
20 & $\nu$ Cen & 1737 & -23.33$^{+5.86}_{-4.69}$ & -20.27$^{+2.26}_{-2.61}$ & -15.28$^{+2.99}_{-1.84}$ & -13.53$^{+1.99}_{-1.64}$ & -1.78$^{+0.79}_{-0.73}$ & -1.71$^{+0.61}_{-0.82}$ \\
\hline 21 (LCC) & Cen-Far & 99 & -18.40$^{+1.87}_{-2.36}$ & -11.75$^{+2.75}_{-2.67}$ & -16.64$^{+1.24}_{-1.26}$ & -10.51$^{+2.22}_{-2.15}$ & -2.79$^{+1.28}_{-0.83}$ & 0.31$^{+1.72}_{-1.50}$ \\
22 & $\sigma$ Cen & 1805 & -33.23$^{+4.14}_{-3.71}$ & -13.67$^{+3.91}_{-5.10}$ & -18.27$^{+1.02}_{-0.69}$ & -7.70$^{+2.33}_{-2.30}$ & -2.20$^{+0.50}_{-0.67}$ & -0.41$^{+0.77}_{-0.63}$ \\
23 & Acrux & 394 & -37.73$^{+1.83}_{-1.67}$ & -11.36$^{+2.84}_{-4.43}$ & -19.10$^{+0.68}_{-0.40}$ & -5.69$^{+1.32}_{-2.00}$ & -2.75$^{+0.32}_{-0.37}$ & -0.17$^{+0.45}_{-0.35}$ \\
24 & Musca-fg & 95 & -39.37$^{+1.82}_{-1.57}$ & -9.33$^{+4.36}_{-3.95}$ & -19.20$^{+0.61}_{-0.33}$ & -4.49$^{+2.05}_{-2.02}$ & -2.85$^{+0.31}_{-0.22}$ & -0.23$^{+0.44}_{-0.40}$ \\
25 & $\epsilon$ Cham & 39 & -41.23$^{+2.27}_{-0.87}$ & -6.05$^{+2.04}_{-2.99}$ & -19.85$^{+0.73}_{-0.42}$ & -2.92$^{+1.01}_{-1.44}$ & -3.31$^{+0.50}_{-0.51}$ & -0.54$^{+0.49}_{-0.46}$ \\
26 & $\eta$ Cham & 30 & -30.16$^{+1.93}_{-4.40}$ & 26.86$^{+1.24}_{-5.90}$ & -14.06$^{+0.69}_{-2.04}$ & 12.55$^{+0.67}_{-3.18}$ & -2.58$^{+0.47}_{-0.25}$ & 2.09$^{+0.59}_{-0.51}$ \\
\hline 27 (Pipe) & B59 & 32 & -0.49$^{+0.70}_{-1.16}$ & -18.84$^{+0.41}_{-0.60}$ & -0.37$^{+0.53}_{-0.89}$ & -14.48$^{+0.71}_{-0.76}$ & 2.13$^{+0.70}_{-0.63}$ & -0.61$^{+0.68}_{-0.83}$ \\
28 & Pipe-N & 42 & -4.78$^{+1.51}_{-2.56}$ & -23.36$^{+0.62}_{-1.65}$ & -3.08$^{+1.04}_{-1.30}$ & -14.55$^{+0.79}_{-0.36}$ & -0.27$^{+0.44}_{-0.65}$ & -2.79$^{+0.38}_{-0.28}$ \\
29 & $\theta$ Oph & 98 & -4.71$^{+0.45}_{-0.94}$ & -21.85$^{+0.67}_{-2.08}$ & -3.29$^{+0.39}_{-0.73}$ & -15.41$^{+0.35}_{-0.93}$ & -1.16$^{+0.38}_{-0.40}$ & -2.03$^{+0.45}_{-0.70}$ \\
\hline 30 (CrA) & CrA-Main & 96 & 4.57$^{+1.19}_{-0.70}$ & -27.11$^{+0.88}_{-1.33}$ & 3.33$^{+0.98}_{-0.47}$ & -19.83$^{+0.63}_{-1.21}$ & -1.76$^{+0.83}_{-0.49}$ & -4.77$^{+0.67}_{-1.04}$ \\
31 & CrA-North & 351 & 0.92$^{+1.84}_{-2.34}$ & -27.60$^{+1.17}_{-1.00}$ & 0.65$^{+1.32}_{-1.59}$ & -19.51$^{+0.68}_{-0.63}$ & -3.16$^{+0.59}_{-0.80}$ & -4.25$^{+0.54}_{-0.62}$ \\
32 & Sco-Sting & 132 & -10.03$^{+2.98}_{-2.27}$ & -29.75$^{+1.78}_{-4.47}$ & -6.12$^{+1.33}_{-1.49}$ & -19.51$^{+1.49}_{-0.82}$ & -5.15$^{+0.52}_{-0.43}$ & -3.70$^{+0.65}_{-0.44}$ \\
\hline 33 (Cham) & Cham-1 & 192 & -22.55$^{+0.72}_{-0.91}$ & 0.38$^{+1.21}_{-1.11}$ & -20.31$^{+0.64}_{-0.71}$ & 0.35$^{+1.06}_{-1.01}$ & -3.96$^{+0.74}_{-0.70}$ & -0.79$^{+0.86}_{-0.79}$ \\
34 & Cham-2 & 54 & -20.16$^{+0.76}_{-0.94}$ & -7.55$^{+0.77}_{-0.64}$ & -18.95$^{+0.65}_{-0.73}$ & -7.07$^{+0.83}_{-0.53}$ & -3.24$^{+0.55}_{-0.57}$ & -0.22$^{+0.51}_{-0.64}$ \\
\hline 35 (NE) & L134/L183 & 24 & -17.63$^{+1.02}_{-0.97}$ & -20.28$^{+1.00}_{-1.52}$ & -9.75$^{+0.92}_{-0.62}$ & -11.13$^{+0.94}_{-0.92}$ & -0.83$^{+0.17}_{-0.36}$ & -2.44$^{+0.34}_{-0.17}$ \\
36 & Oph-SE & 61 & -5.65$^{+0.61}_{-0.54}$ & -11.44$^{+0.66}_{-0.75}$ & -5.63$^{+0.54}_{-0.60}$ & -11.58$^{+1.26}_{-0.63}$ & -2.77$^{+0.51}_{-0.46}$ & 0.30$^{+1.04}_{-0.69}$ \\
37 & Oph-NF & 28 & -8.57$^{+1.88}_{-1.46}$ & -16.75$^{+0.99}_{-1.59}$ & -7.78$^{+0.96}_{-1.11}$ & -15.89$^{+0.41}_{-0.96}$ & -2.83$^{+0.44}_{-0.85}$ & -7.11$^{+0.52}_{-0.79}$ \\

\hline
\end{tabular}
} 
\renewcommand{\arraystretch}{1}
\label{tab:averages-vel}
\tablefoot{In Cols.~4--9 we list the medians of the velocity parameters for each cluster including all cluster members (without considering any stability cut). The given lower and upper uncertainties represent the 1$\sigma$ scatter (velocity dispersion) around the median. In this scatter the original measurement uncertainties are not considered. See Table~\ref{tab:averages-pos} for the positional parameters.
}
\end{small}
\end{center}
\end{table*} 
\begin{table*}[!ht] 
\begin{center}
\begin{small}
\caption{\textsc{Hipparcos} astrometry from \citet{VanLeeuwen2007} of bright stellar members in Sco-Cen. The Gaia DR3 astrometry is given if available. \vspace{-3mm}}
\renewcommand{\arraystretch}{1.2}
\resizebox{1\textwidth}{!}{%
\begin{tabular}{llclrrrrrrrr}
\hline \hline
\multicolumn{1}{l}{HIP} &
\multicolumn{1}{l}{Name} &
\multicolumn{1}{l}{\texttt{SigMA}\tablefootmark{a}} &
\multicolumn{1}{l}{SpT} &
\multicolumn{5}{c}{\textsc{Hipparcos}} &
\multicolumn{3}{c}{Gaia DR3} \\
\cmidrule(lr){5-9}
\cmidrule(lr){10-12}
\multicolumn{1}{l}{} &
\multicolumn{1}{l}{} &
\multicolumn{1}{l}{} &
\multicolumn{1}{l}{} &
\multicolumn{1}{c}{$l$} &
\multicolumn{1}{c}{$b$} &
\multicolumn{1}{c}{$\varpi$} &
\multicolumn{1}{c}{$\mu_\alpha^*$} &
\multicolumn{1}{c}{$\mu_\delta$} &
\multicolumn{1}{c}{$\varpi$} &
\multicolumn{1}{c}{$\mu_\alpha^*$} &
\multicolumn{1}{c}{$\mu_\delta$} \\
& & & &
\multicolumn{2}{c}{(deg)} & 
\multicolumn{1}{c}{(mas)} &
\multicolumn{2}{c}{(mas yr$^{-1}$)} &
\multicolumn{1}{c}{(mas)} &
\multicolumn{2}{c}{(mas yr$^{-1}$)} \\
\hline

80473 & * rho Oph & 1 & B2V & 353.69 & 17.69 & 9.03$\pm$0.90 & -5.53$\pm$0.89 & -21.74$\pm$0.93 & 7.26$\pm$0.13 & -4.38$\pm$0.19 & -23.30$\pm$0.16 \\
79374 & * nu Sco (Jabbah) & 2 & B2IV & 354.61 & 22.70 & 6.88$\pm$0.76 & -7.65$\pm$0.71 & -23.71$\pm$0.47 & 7.06$\pm$0.22 & -7.44$\pm$0.25 & -28.20$\pm$0.16 \\
78401 & * del Sco (Dschubba) & 3 & B0.2IV & 350.10 & 22.49 & 6.64$\pm$0.89 & -10.21$\pm$1.01 & -35.41$\pm$0.71 &  &  &  \\
78820 & * bet Sco (Acrab) & 4 & B0.5V & 353.19 & 23.60 & 8.07$\pm$0.78 & -5.20$\pm$0.92 & -24.04$\pm$0.64 &  &  &  \\
80112 & * sig Sco (Alniyat) & 5 & B1III & 351.31 & 17.00 & 4.68$\pm$0.60 & -10.60$\pm$0.78 & -16.28$\pm$0.43 &  &  &  \\
79404 & * c02 Sco & 5 & B2V & 348.12 & 16.84 & 6.81$\pm$0.16 & -10.38$\pm$0.18 & -23.94$\pm$0.14 & 6.78$\pm$0.14 & -12.26$\pm$0.18 & -24.25$\pm$0.14 \\
81266 & * tau Sco & 6 & B0V & 351.53 & 12.81 & 6.88$\pm$0.53 & -9.89$\pm$0.61 & -22.83$\pm$0.55 &  &  &  \\
80763 & * alf Sco (Antares) & 6 & M1Ib+B2.5V & 351.95 & 15.06 & 5.89$\pm$1.00 & -12.11$\pm$1.22 & -23.30$\pm$0.76 &  &  &  \\
78104 & * rho Sco (Iklil) & 7 & B2IV/V & 344.63 & 18.27 & 6.91$\pm$0.19 & -15.68$\pm$0.21 & -24.88$\pm$0.19 & 7.32$\pm$0.29 & -17.03$\pm$0.33 & -23.66$\pm$0.27 \\
78265 & * pi Sco & 9 & B1V+B2V & 347.21 & 20.23 & 5.57$\pm$0.64 & -11.42$\pm$0.78 & -26.83$\pm$0.74 & 9.67$\pm$0.70 & -12.40$\pm$0.52 & -26.21$\pm$0.57 \\
81477 & V* V1062 Sco & 10 & Ap Si & 343.57 & 5.18 & 7.54$\pm$0.61 & -10.25$\pm$0.69 & -21.59$\pm$0.45 & 5.72$\pm$0.03 & -12.45$\pm$0.04 & -21.17$\pm$0.04 \\
82545 & * mu02 Sco (Pipirima) & 10 & B2IV & 346.20 & 3.86 & 6.88$\pm$0.12 & -11.09$\pm$0.13 & -23.32$\pm$0.11 & 5.66$\pm$0.28 & -12.11$\pm$0.30 & -22.57$\pm$0.27 \\
82514 & * mu01 Sco (Xamidimura) & 11 & B1.5IV+B & 346.12 & 3.91 & 6.51$\pm$0.91 & -10.58$\pm$0.87 & -22.06$\pm$0.74 &  &  &  \\
78533 & V* LL Lup & 13 & Ap Si & 339.64 & 11.41 & 8.61$\pm$0.69 & -12.65$\pm$0.75 & -24.14$\pm$0.68 & 6.18$\pm$0.03 & -12.18$\pm$0.03 & -24.02$\pm$0.02 \\
78384 & * eta Lup & 14 & B2.5IV & 338.77 & 11.01 & 7.38$\pm$0.18 & -16.96$\pm$0.19 & -27.83$\pm$0.19 & 7.59$\pm$0.43 & -17.59$\pm$0.49 & -27.26$\pm$0.37 \\
71865 & * b Cen & 15 & B2.5V & 325.90 & 20.10 & 9.62$\pm$0.18 & -29.92$\pm$0.14 & -30.68$\pm$0.13 & 10.03$\pm$0.31 & -29.83$\pm$0.37 & -31.91$\pm$0.52 \\
76945 & * psi02 Lup & 15 & B5V & 338.48 & 16.08 & 8.97$\pm$0.27 & -21.37$\pm$0.30 & -29.98$\pm$0.25 &  &  &  \\
75304 & * phi02 Lup & 15 & B4V & 333.84 & 16.75 & 6.28$\pm$0.20 & -18.24$\pm$0.22 & -20.72$\pm$0.16 & 6.49$\pm$0.22 & -18.47$\pm$0.30 & -20.41$\pm$0.29 \\
71860 & * alf Lup & 17 & B1.5III & 321.61 & 11.44 & 7.02$\pm$0.17 & -20.94$\pm$0.14 & -23.67$\pm$0.14 &  &  &  \\
74449 & * e Lup & 17 & B3IV & 327.83 & 11.43 & 6.47$\pm$0.21 & -22.01$\pm$0.18 & -21.75$\pm$0.19 & 7.24$\pm$0.20 & -21.67$\pm$0.20 & -21.90$\pm$0.18 \\
76371 & * d Lup & 17 & B3IVp & 331.02 & 8.76 & 7.62$\pm$0.43 & -20.53$\pm$0.33 & -21.23$\pm$0.31 & 6.85$\pm$0.13 & -20.79$\pm$0.15 & -20.79$\pm$0.11 \\
71536 & * rho Lup & 19 & B5V & 320.13 & 9.86 & 10.32$\pm$0.16 & -28.26$\pm$0.12 & -28.82$\pm$0.13 & 9.35$\pm$0.20 & -28.92$\pm$0.22 & -29.21$\pm$0.32 \\
72800 & V* V1019 Cen & 20 & B7II/III & 327.93 & 19.11 & 6.63$\pm$0.22 & -20.48$\pm$0.23 & -19.20$\pm$0.22 & 6.96$\pm$0.17 & -20.60$\pm$0.16 & -20.35$\pm$0.20 \\
67464 & * nu Cen & 20 & B2IV & 314.41 & 19.89 & 7.47$\pm$0.17 & -26.77$\pm$0.12 & -20.18$\pm$0.08 & 8.05$\pm$0.35 & -26.14$\pm$0.42 & -21.23$\pm$0.94 \\
63945 & * f Cen & 22 & B5V & 305.47 & 14.34 & 8.36$\pm$0.25 & -29.85$\pm$0.18 & -15.17$\pm$0.15 & 8.16$\pm$0.16 & -29.56$\pm$0.12 & -15.69$\pm$0.10 \\
60823 & * sig Cen & 22 & B3V & 299.10 & 12.47 & 7.92$\pm$0.18 & -32.36$\pm$0.15 & -12.51$\pm$0.13 & 7.92$\pm$0.23 & -32.17$\pm$0.18 & -13.17$\pm$0.16 \\
60009 & * zet Cru & 23 & B2.5V & 299.32 & -1.36 & 9.12$\pm$0.45 & -33.80$\pm$0.48 & -10.15$\pm$0.43 &  &  &  \\
60718 & * alf Cru (Acrux) & 23 & B0.5IV & 300.13 & -0.36 & 10.13$\pm$0.50 & -35.83$\pm$0.47 & -14.86$\pm$0.43 &  &  &  \\
58484 & * eps Cha & 25 & B9Vn & 300.21 & -15.62 & 9.02$\pm$0.36 & -40.34$\pm$0.38 & -8.30$\pm$0.40 &  &  &  \\
42637 & * eta Cha & 26 & B9IV & 292.40 & -21.65 & 10.53$\pm$0.16 & -28.89$\pm$0.16 & 27.21$\pm$0.16 & 10.17$\pm$0.07 & -29.43$\pm$0.18 & 26.83$\pm$0.12 \\
84970 & * tet Oph & 29 & B2IV & 0.46 & 6.55 & 7.48$\pm$0.17 & -7.37$\pm$0.18 & -23.94$\pm$0.10 &  &  &  \\
86670 & * kap Sco & 32 & B1.5III & 351.04 & -4.72 & 6.75$\pm$0.17 & -6.05$\pm$0.21 & -25.54$\pm$0.13 &  &  &  \\

\hline
\end{tabular}
} 
\renewcommand{\arraystretch}{1}
\label{tab:hip}
\tablefoot{Shown are mostly B-type stars that are either part of the \texttt{SigMA} selected clusters, or which are the name-givers of some clusters. This is not a complete list of B-stars in Sco-Cen.
The \textsc{Hipparcos} astrometry can be compared to average parameters of the \texttt{SigMA} clusters as derived from Gaia astrometry in Tables~\ref{tab:averages-pos}--\ref{tab:averages-vel}, to evaluate possible correlations. The \textsc{Hipparcos} parallaxes should be treated with caution, since significant deviations to Gaia parallaxes are possible (on the order of about 60\,pc when converted to distances), while proper motions show deviations on the order of about $\pm2$\,mas\,yr$^{-1}$, when comparing sources which are both in \textsc{Hipparcos} and Gaia DR3 within about 500\,pc from the Sun.
\tablefoottext{a}{Col.~3 gives the index of the \texttt{SigMA} cluster that is hosting the given star.}
}
\end{small}
\end{center}
\end{table*}

\subsection{Overview of the seven subregions in Sco-Cen} \label{sec:overview-scocen-regions}

In the following, and to help compare \texttt{SigMA} results with the literature, we give a brief overview for each sub-region within Sco-Cen (US, UCL, LCC, Pipe, CrA, Cham, and NE). We then give a more detailed comparison to recent works in Sect.\,\ref{sec:compare}. The listed seven subregions include four regions that are not a traditional part of the Sco-Cen OB association, namely CrA, Pipe, Cham, and NE clusters, while we find them to be be part of Sco-Cen because they are co-moving with the complex itself, and have ages below of about the 20\,Myr age cut we used for the association. Even if we assign each stellar group to one of the seven subregions, we stress that this classification should not be seen as physically distinct regions inside Sco-Cen, but simply to help compare our results with the literature.

\subsubsection{Upper Scorpius (US)} \label{sec:us}

Toward US we identify nine clusters containing in total 3596 stellar sources, which are partially extending beyond the traditional borders (Fig.~\ref{fig:lb-map}). Of these nine clusters, seven appear higher surface density and tend to be associated with prominent B-stars, as already pointed out above, namely $\rho$\,Oph/L1688, $\beta$\,Sco, $\delta$\,Sco, $\nu$\,Sco, $\sigma$\,Sco, Antares, and $\rho$\,Sco (see Tables\,\ref{tab:averages-pos}--\ref{tab:hip}). The remaining two clusters appear more extended, which we name US-foreground and Scorpio-Body. 

The clusters $\rho$\,Oph/L1688, Antares, and $\rho$\,Sco show significant overlap in the same volume in space, while separating in velocity space. In a recent paper \citep{Grasser2021} we studied the $\rho$\,Oph/L1688 cluster with Gaia EDR3 data and identified two kinematically distinct populations within the same volume (Pop\,1 and Pop\,2). These two populations coincide with the $\rho$\,Oph/L1688 and Antares clusters, respectively. In detail, the cross-matched Pop\,1 sample contains $\sim$93\% of the $\rho$\,Oph/L1688 group and few matches with other clusters (Antares, $\sigma$\,Sco, $\beta$\,Sco, $\delta$\,Sco). The cross-matched Pop\,2 sample contains $\sim$75\% of the Antares group and few matches with other clusters ($\rho$\,Sco, $\sigma$\,Sco, US-foreground). 
\citet{Luhman2022a} point out that ``new'' $\rho$\,Oph/L1688 members in \citet{Grasser2021} have already been identified previously by other literature as being part of US. We clarify here that the ``new'' sources in  \citet{Grasser2021} refer to sources that have not been assigned previously as members of the young $\rho$\,Oph/L1688 star-forming event. The two intertwining distinct populations within the same volume have been mentioned for the first time in \citet{Grasser2021}. In this work, we add another stellar population to the region, $\rho$\,Sco, which also seems to occupy a similar volume in space, partially overlapping with the two populations while having distinct velocities from these.  

The group US-foreground is located in front of the more compact clusters, visible in 3D space (Fig.\,\ref{fig:3D-map}), hence the chosen name. Finally, the Scorpio-Body group extends from US toward the Galactic South, beyond the traditional borders of US, with a significant fraction located in UCL and in the direction of CrA (Sect.\,\ref{sec:cra}). It spans the Scorpius constellation's central body, hence the name. The nine clusters toward US reveal a complex star formation history, which will be further discussed in a follow-up paper.

\subsubsection{Upper Centaurus Lupus (UCL)} \label{sec:ucl}

We identify rich substructure within UCL separated into 11 \texttt{SigMA} clusters (5935 stellar sources), as listed in Table\,\ref{tab:averages-pos}. The most prominent cluster in the region is V1062\,Sco \citep[][]{Roeser2018}, lying towards the far side of Sco-Cen. This cluster was picked up easily by visual selection methods (e.g., by \citealp{Damiani2019} or \citealp{Luhman2022a}; see Sects.\,\ref{sec:damiani}, \ref{sec:luhman}). We identify a second cluster close to V1062\,Sco, which we call $\mu$\,Sco, since its members are scattered around the bright B-star *\,mu01\,Sco. We find that the positions and velocities of the two \texttt{SigMA} clusters are very similar, and members of both clusters are part of V1062-Sco-selections in previous work (Sect.\,\ref{sec:compare}), also named UPK\,640 in \citet{Cantat-Gaudin2020a}. 
The star *\,mu01\,Sco, which is the name giver of $\mu$\,Sco lies in the center of the cluster, while the star *\,mu02\,Sco is part of the \texttt{SigMA} selected members for V1062\,Sco, located at the periphery of this cluster. This suggests a possible connection between the two clusters, but this statement is tentative.  

Lupus\,1--4 appears correlated with regions of high dust column-density, matching with previous selections of Lupus--3 and 4 stellar members \citep[e.g.,][]{Damiani2019, Kerr2021}, which are merged in the \texttt{SigMA} selection. 
The average distance to Lupus\,1--4 matches well with cloud distance estimates from \citet[][derived from \citealp{Leike2020}]{Zucker2021}, who report a distance between 155--198\,pc, or an average of about 165\,pc for the Lupus 1--4 clouds \citep[see also][]{Teixeira2020}. 

At the heart of UCL lie the clusters $\eta$\,Lup, $\phi$\,Lup, and $e$\,Lup, which likely belong to the oldest parts of Sco-Cen, probably the clusters where the first supernovae in Sco-Cen originated from \citep{Zucker2022}. To the North of the traditional UCL borders we find a clustering, which has not been isolated in previous works, named Libra-South, based on its location within that constellation. 

There is one cluster slightly in front and to the south of the main UCL body, called Norma-North, named after its location in that constellation. This is a new clustering, which does not have a clear counterpart in the literature.
Another \texttt{SigMA} cluster lies to the far side of UCL and the Galactic west of the Lupus constellation. This cluster correlates with UPK\,606 when compared to \citet{Cantat-Gaudin2020a} (see also \citealp{Kerr2021} and Table~\ref{tab:kerr}). 
Finally, to the Galactic west, UCL connects with LCC via the clusters $\nu$\,Cen and $\rho$\,Lup.

\subsubsection{Lower Centaurus Crux (LCC)} \label{sec:lcc}

We find six \texttt{SigMA} clusters (2462 stellar sources) toward the LCC region (see Table\,\ref{tab:averages-pos}), which is now reaching farther below the Galactic plane compared to most of the work in the literature. For the \texttt{SigMA} extraction, the young local associations $\epsilon$\,Cha and $\eta$\,Cha are part of LCC, located at the Southern most tip, confirming the results of \citet{Mamajek1999, Mamajek2000} or \citet{Fernandez2008}.
Toward LCC, \texttt{SigMA} extracts a cluster that seems unrelated to the main body of LCC, which we name Centaurus-Far, since it lies about 60\,pc to the back of it, at a distance similar to that of the Chamaeleon clouds (Sect.~\ref{sec:cham}). This cluster was already identified in \citet{Kerr2021}, as part of the TLC21 group (Cham-group) as EOM3, and named Cen-South (see Sect.\,\ref{sec:kerr} and Table\,\ref{tab:kerr}). Consequently, the main body of LCC is composed of five subgroups, which seem to constitute an age gradient \citep[e.g.,][]{Kerr2021} from north to south ($\sigma$\,Cen, Acrux, Musca-foreground, $\epsilon$\,Cha, and $\eta$\,Cha), which will be analyzed in future work.

\subsubsection{Pipe Nebula} \label{sec:pipe}

Although not traditionally considered part of the Sco-Cen association, we find three \texttt{SigMA} clusters toward the Pipe nebula (172 stellar sources), including B59, Pipe-North, and $\theta$\,Oph. The group B59 seems to be closely related to the star forming B59 cloud \citep[e.g.,][]{Lombardi2006, Brooke2007, Roman-Zuniga2007, Roman-Zuniga2010}. This is supported not only by projection in the sky towards cluster and cloud but also by the cloud distance between 147--163\,pc \citep{Zucker2021}, compatible with the cluster distance of about 160\,pc. The $\theta$\,Oph  cluster, surrounding the B2 star *\,tet\,Oph, is located at about the same distance to B59 and is close to the stem of the Pipe Nebula cloud, giving ground to studies of a possible interaction between the B2 star and the cloud \citep{Gritschneder2012}.
Pipe-North lies slightly in front of the other two clusters (at about 130\,pc) and to the Galactic north of the Pipe Nebula, as the name suggests. 

\subsubsection{Corona Australis (CrA)} \label{sec:cra}

The possible physical connection between CrA and the Sco-Cen association was already pointed out in previous studies \citep[e.g.,][]{Mamajek2001, Preibisch2008, Kerr2021} and confirmed by our work.
We count three \texttt{SigMA} clusters to the CrA region, containing in total 579 sources.
We identify a distinct cluster projected on top of the CrA molecular cloud and the embedded Coronet clusters, which we call the CrA-Main group. The cluster distance fits to the cloud distance of about 136--179\,pc \citep{Zucker2021}.
To the Galactic north we identify a second more extended group, called CrA-North, which was already discussed in \citet{Galli2020} or \citet{Esplin2022}. Additionally, we identify a third group to the Galactic northwest of the two other clusters, apparently building a bridge to the main body of Sco-Cen. This group we name Scorpio-Sting since its projected location matches the sting of the Scorpio constellation. Sco-Sting has only one clear counterpart in the literature, namely the TLC22/EOM7 group in \citet{Kerr2021} (see Sect.\,\ref{sec:kerr} and Table\,\ref{tab:kerr}), while they identify a smaller sub-sample of this group (12 members in \citealp{Kerr2021} versus 132 members in this work).

\subsubsection{Chamaeleon (Cham)} \label{sec:cham}

The well-known star-forming molecular clouds of Chamaeleon are seen through the same line-of-sight as the southern tip of LCC, but lie clearly towards the back of LCC when seen in 3D (Fig.~\ref{fig:3D-map}). We identify two clusters with a total of 246 stellar sources in Chamaeleon 1 \& 2, which are likely directly related to the two molecular clouds of the same name and are already characterized with Gaia \citep[e.g.][see also Sect.\,\ref{sec:kerr}]{Roccatagliata2018, Galli2020, Kerr2021}. Due to their youth, position, and tangential velocities we assume that the Cham clusters and clouds are part of the Sco-Cen star formation event, but this must be confirmed by tracebacks of the young population \citep[see, e.g.,][]{Grossschedl2021}. Similar suggestions appear in \citet{Lepine2003} or \citet{Sartori2003}. 

\subsubsection{Northeast clusters (NE)} \label{sec:L134}

We identify three extra clusters to the Galactic north and east of Sco-Cen, which we like to discuss separately in this section, which are L134/L183, Oph-southeast, and Oph-north-far. We assign these clusters to a separate region, which we call the Northeast clusters (NE), based on their location relative to Sco-Cen in Galactic coordinates, since they do not fit to any other of the Sco-Cen subregions. 

The cluster L134/L183 is a small, newly identified stellar group to the Galactic north of US (with 24 stellar members).  This stellar group is likely associated with the small molecular clouds L134 and L183 \citep*[or MBM\,36 and 37,][]{Magnani1985}, that are currently non star-forming \citep{Pagani2003, Pagani2004,  Pagani2005}. The distances to the clouds in \citet{Zucker2019} are about 105--120\,pc, which match the cluster distance of about 114\,pc. The presence of the young stellar group close by the clouds suggests that (1) the clouds are remnants of a larger cloud that formed the newly identified \texttt{SigMA} cluster and (2) that the newly identified sources might be playing a role in the observed ``cloudshine'' phenomenon towards this cloud \citep{Steinacker2010, Steinacker2015}. 

The cluster Ophiuchus Southeast (Oph-SE, 61 members) lies at a similar projected position as Pipe-North, while being at a farther distance, about 50\,pc in the back (hence, we do not count it to the Pipe region). This stellar group was already selected by \citet{Kerr2021} as TLC\,4 with 31 members (Sect.\,\ref{sec:kerr}). Finally, the group Ophiuchus Northfar (Oph-NF, 28 members) appears to be a newly identified stellar group, located at a similar distance as Oph-SE. This new group needs more investigations in the future, since the stability of the selected members, as determined by the \texttt{SigMA} algorithm, is generally very low (stability < 11\%).


\begin{table*}[!ht] 
\begin{center}
\begin{small}
\caption{Overview of the recent Literature to which we compare the \texttt{SigMA} Sco-Cen clustering results in more detail. \vspace{-2mm}}
\renewcommand{\arraystretch}{1.2}
\resizebox{1\textwidth}{!}{%
\begin{tabular}{lllllcccc}
\hline \hline
\multicolumn{1}{l}{Reference} &
\multicolumn{1}{l}{Sect.} &
\multicolumn{1}{l}{Data} &
\multicolumn{1}{c}{Studied Area} &
\multicolumn{5}{c}{Number statistics} \\
\cmidrule(lr){5-9}
%
\multicolumn{1}{l}{} &
\multicolumn{1}{l}{} &
\multicolumn{1}{l}{} &
\multicolumn{1}{c}{} &
\multicolumn{1}{l}{Nr.\,in\,Ref.\tablefootmark{a}} & 
\multicolumn{1}{c}{\texttt{SigMA}\tablefootmark{b}} &
\multicolumn{1}{c}{Matches\tablefootmark{c}} &
\multicolumn{1}{c}{Recovered\tablefootmark{d}} &
\multicolumn{1}{c}{Rejected\tablefootmark{d}} \\
\hline

\citet{Damiani2019}\tablefootmark{e} & \ref{sec:damiani} & DR2 &  
 $(l = 360^\circ$ to $280^\circ$, $b = 0^\circ$ to $30^\circ)$ OR
 & 10,421  & 11,796 & 9328 & 89.5\% & 10.5\% \\
 & &  & $(l = 315^\circ$ to $280^\circ$, $b = -10^\circ$ to $0^\circ)$ & 1732 clustered (17\%)  &  & 1719 & 99.2\% & 0.8\% \\
 & & &  FOV = 2750\,deg$^2$, $d<\SI{200}{pc}$ & 8689 diffuse (83\%) &  & 7609 & 87.6\%  & 12.4\% \\
\hline

\citet{Kerr2021}\tablefootmark{f} & \ref{sec:kerr} & DR2  & The whole TLC22 stellar group & 7394 & 12,796 & 6270 & 84.8\% & 15.2\%  \\ 
 &  &   &  TLC22 without EOM\,1--5 & 7138 & 12,796 & 6270 & 87.8\% & 12.2\%  \\
 &  &   &  22 EOMs of TLC22 (without EOM\,1--5) & 3453 & 12,796 & 3447 & 99.8\% & 0.2\%  \\
 \hline
 
\citet{Schmitt2022}\tablefootmark{g} & \ref{sec:schmitt} & EDR3 \& & \citet{deZeeuw1999} borders: & 6150 & 11,348 & 3385 & 55.0\% & 45.0\% \\
 & & eROSITA & US $(l = 343^\circ$ to $360^\circ$, $b = 10^\circ$ to $30^\circ)$ OR  & $\sim$69\% vel-clustered &  & 3385 & $\sim$80\% & $\sim$20\%  \\
 &  &   & UCL $(l = 312^\circ$ to $350^\circ$, $b = 0^\circ$ to $25^\circ)$ OR  & $\sim$26\% vel-diffuse &  & 0 & &  \\
 &  &  &  LCC $(l = 285^\circ$ to $312^\circ$, $b = -10^\circ$ to $22^\circ)$, & $\sim$5\% IC\,2602 &  & 0 & &   \\
 & &  &  FOV = 2050\,deg$^2$, $d \sim 60$--$\SI{200}{pc}$  &  &  &  & &   \\
\hline

\citet{Luhman2022a} & \ref{sec:luhman} & EDR3 & $l = 2^\circ$ to $285^\circ$, $b = -12^\circ$ to $35^\circ$, & 10,509 & 12,215 & 9838 & 93.6\% & 6.4\%   \\
 & &  & FOV = 3252\,deg$^2$, $d \sim 90$--$\SI{250}{pc}$ &  &  & & &   \\
\hline

\citet{Zerjal2021arxiv} & \ref{sec:zerjal} & DR2 & $l = 40^\circ$ to $240^\circ$, $b = -60^\circ$ to $70^\circ$, & 8185 & 12,943  & 7671  & 93.7\% & 6.3\%  \\
 &  &  & FOV = 36,400\,deg$^2$, $d \sim 83$--$\SI{200}{pc}$ &  &  &  & &  \\
\hline

\citet{Squicciarini2021}\tablefootmark{h} & \ref{sec:squicc} & EDR3 & $\alpha = 236^\circ$ to $251^\circ$, $\delta = -29^\circ$ to $-16^\circ$ & 2745  & 2717 & 2575  & 93.8\% & 6.2\% \\
(only US,  & &  & FOV = 195\,deg$^2$, $d \sim 125$--$\SI{175}{pc}$ & 1442 clustered (53\%) &  & 1435  & 99.5\% & 0.5\%  \\
subsample with RVs)  & &  &  & 1303 diffuse (47\%) &  & 1140  & 87.5\% & 12.5\% \\
\hline

\citet{MiretRoig2022b}\tablefootmark{i} & \ref{sec:miret} & DR3 & $\alpha = 235^\circ$ to $252^\circ$, $\delta = -30^\circ$ to $-17^\circ$  & 2810 & 3089  & 2683  & 95.5\% & 4.5\% \\
(only US,   & & & FOV = 221\,deg$^2$, $d \sim 80$--$\SI{200}{pc}$ & 2190 5D (78\%)  &  & 2145  & 97.9\% & 2.1\%  \\
subsample with RVs) & & & & 670 6D (24\%) & & 667 & 99.6\% & 0.4\%   \\
 & & & & 620 Rest (22\%) & & 538  & 86.8\% & 13.2\%  \\




\hline
\end{tabular}
} 
\renewcommand{\arraystretch}{1}
\label{tab:literature}
\tablefoot{
\tablefoottext{a}{Number of stellar members from the given reference. If there was a distinction in the literature between members in a more clustered or diffuse mode (which are generally differently defined in each reference), then the numbers are given below.}
\tablefoottext{b}{Number of stellar cluster members from  \texttt{SigMA} in the given studied area (volume), out of the total 13,103 \texttt{SigMA} stellar cluster members.}
\tablefoottext{c}{Number of matches between the given reference and the \texttt{SigMA} clusters. If a distinct comparison with clustered or diffuse sources was given in the literature, then the numbers and matches with these are given below.}
\tablefoottext{d}{The fraction of recovered and rejected (or not-recovered) sources when comparing the sample sizes from the literature (Col.\,$a$) with the number of \texttt{SigMA} matches (Col.\,$c$).}
\tablefoottext{e}{For \citetalias{Damiani2019} we only give the number of sources within their clustered or diffuse populations within $1000/\varpi_\mathrm{EDR3}<200$\,pc after a cross-match with Gaia EDR3, and without IC\,2602.}
\tablefoottext{f}{For \citetalias{Kerr2021} we do not give the surveyed area, since they extracted the clusters from all-sky data within 333\,pc from the Sun. We show a comparison to their whole TLC22 group (their main Sco-Cen group), then the TLC22 without the older EOM groups, and finally the TLC22 members without older groups and which are in on of the younger EOM clusters.}
\tablefoottext{g}{The X-ray selected sources from \citetalias{Schmitt2022} included velocity-clustered and velocity-diffuse sources. The separation of these was applied by us by hand, guided by Fig.~7 in \citetalias{Schmitt2022}. Hence, the fractions are only given roughly. The fraction of potential IC\,2602 members is also given. The SigMA clusters have only matches with their velocity-clustered population.}
\tablefoottext{h}{\citetalias{Squicciarini2021} only studied the US region, finding sources in a more clustered mode and sources in a more diffuse mode, while the latter are simply the residuals of their clustering procedure. They study a subsample of sources with $v_r$ information in the 6D phase space ($\sim$28\%), which is not further discussed in this work.}
\tablefoottext{i}{Similarly as \textit{h}, \citetalias{MiretRoig2022b} only studied the US region. They study a subsample of sources with $v_r$ information in the 6D phase space ($\sim$30\%). The numbers of sources in the \citetalias{MiretRoig2022b} bona-fide-5D and bona-fide-6D samples are listed separately, with ``Rest'' being sources in neither of the two. Note that the 6D sample is contained withing the 5D sample.}
}
\end{small}
\end{center}
\end{table*}

\subsection{Comparison with previous work} \label{sec:compare}

In the following we compare the \texttt{SigMA} selected stellar clusters with recent results from the literature (Table~\ref{tab:literature}), including eight publications. The studies by \citet{Damiani2019}, \citet{Schmitt2022}, \citet{Luhman2022a}, and \citet{Zerjal2021arxiv} discuss the whole Sco-Cen region, slightly extending beyond the traditional Sco-Cen borders, while excluding the regions to the Galactic South (CrA and Cham). The first three of these studies select members within broad selection borders decided by hand, which we call in this paper \textit{visual selection methods}. \citet{Squicciarini2021}, \citet{MiretRoig2022b}, and \citet{BricenoMorals2022} focus only on the US region and extract clusters using a combination of Gaia astrometry and radial velocities. 
\citet{Kerr2021} present an all-sky study of young stars within 333\,pc, hence covering the new extended view of the Sco-Cen association, using an unsupervised machine learning approach, which is more similar to our work then the aforementioned studies. 

The literature samples are cross-matched with the \texttt{SigMA} clusters using the Gaia DR3 \texttt{source\_id}, as specified in Appendix~\ref{app:box}. We provide an overview of the discussed literature samples in Table~\ref{tab:literature}, giving the total number of sources of each literature sample, the total number of sources of \texttt{SigMA} Sco-Cen cluster members within the respective studied areas, and the number of total matches. Finally, we list the fraction of sources that we recover or reject (or do not recover) when compared to the individual literature samples.

\subsubsection{Comparison with Damiani et al. (2019)} \label{sec:damiani}

\citet[][hereafter DPP19]{Damiani2019} analyze Sco-Cen using Gaia DR2 data and a traditional approach, selecting by hand over-densities in velocity and position space, followed by selecting pre--main-sequence (PMS) stars from an HRD. Their field of view (FOV) goes slightly beyond the traditional borders of the association (see Table~\ref{tab:literature}). They discuss eight compact clusters, which are prominently peaked in projection and in velocity space (hence, easier to identify with \textit{visual selection methods}); these are UCL-1, UCL-2, UCL-3, Lupus\,3, LCC-1, US-far, US-near, and the well studied IC\,2602. Although \texttt{SigMA} easily detects IC\,2602, we do not discuss this cluster since its age ($\sim$ 30 Myr) excludes it as a part of the recent Sco-Cen star formation event, as mentioned above.  \citetalias{Damiani2019} also discuss four diffuse populations (D1, D2a, D2b, US-D2), which are generally distributed across large parts of the traditional Blaauw Sco-Cen OB association. Moreover, their catalog includes sources, which have not been assigned to any group (labeled with ``N'' in Table~\ref{tab:damiani}).

The \citetalias{Damiani2019} catalog contains in total \num[group-separator={,}]{14437} sources, of which 1734 are in their seven clustered Sco-Cen populations (350 in IC\,2602), 8727 are in their four diffuse populations, and the rest 3626 have not been assigned to any population (labeled with ``N''). When cross-matching the \citetalias{Damiani2019} Gaia DR2 sample with DR3 astrometry, we find that 201 stars (1.4\%) are rejected when applying the distance criteria from \citetalias{Damiani2019} ($d<200$\,pc), due to updated parallaxes in DR3. The majority of these sources have not been assigned to any group or belong to one of the diffuse populations. When now considering only the sources in the clustered and diffuse populations within 200\,pc (and without IC2602), then there are 10,425 potential Sco-Cen members in \citetalias{Damiani2019}, or 10,421 when additionally applying the box and quality criteria form Sect.\,\ref{sec:data}. 

There are in total 9635 cross-matches between the \texttt{SigMA} clusters and \citetalias{Damiani2019}, while 9328 of these (89.5\% out of 10,421) belong to either the clustered or diffuse populations (307 are not assigned, ``N''). Of the 9328 cross-matches, 7609 belong to one of the four diffuse populations. Comparing this number to their total diffuse population (8689 within 200\,pc), we find that about 88\% are a match with the  \texttt{SigMA} clusters, calling into question the existence of any physically meaningful diffuse stellar population. In most cases, more than one \citetalias{Damiani2019} group (both clustered or diffuse) fits to one of our clusters, and vice versa (see Table\,\ref{tab:damiani}). In particular, their diffuse groups each contain sub-parts of about 10--20 of the \texttt{SigMA} clusters. 

Focusing on the 1732 \citetalias{Damiani2019} sources in compact clusters, there are 1719 matches with \texttt{SigMA} clusters (99\%) within 200\,pc. The better consensus considering their compact samples highlights the higher robustness of these samples (see also Table~\ref{tab:literature}). Focusing on individual samples, we find that their US-near and US-far can not be assigned clearly to only one of the \texttt{SigMA} clusters (see Table\,\ref{tab:damiani}). US-near correlates best with $\rho$\,Oph/L1688 (containing fractions of $\delta$\,Sco, $\nu$\,Sco, Antares, and $\beta$\,Sco), and US-far with $\sigma$\,Sco (containing fractions of Antares, $\beta$\,Sco, $\delta$\,Sco, and $\rho$\,Sco). In particular, Antares is distributed almost equally among these two clusters. The Antares group is partially occupying the same volume as $\rho$\,Sco and in particular $\rho$\,Oph/L1688 (see Sect.\,\ref{sec:us} and \citealp{Grasser2021}). This highlights the capability of \texttt{SigMA} to untangle young populations that share the same volume but have different space motions.
The rest of the \citetalias{Damiani2019} compact clusters correlate best with \texttt{SigMA} clusters as follows: UCL-1 with V1062\,Sco and $\mu$\,Sco, UCL-2 with UPK\,606, UCL-3 with $\phi$\,Lup, LCC-1 with Acrux, and Lup\,III with Lupus\,1--4. 
Finally, about 9\% of the \texttt{SigMA} cluster members correlate with unassigned sources in \citetalias{Damiani2019} (N), within their FOV and our box criteria.

Concerning the different approaches, comparing the \citetalias{Damiani2019} visual section method and the \texttt{SigMA} unsupervised clustering method, we first note that the method used by \citetalias{Damiani2019} starts with a selection of stars by hand in velocity space, followed by a selection by hand of PMS stars on the HRD. Such an approach will generally find more candidates than an unsupervised method and it will deliver the most prominent clusters. However, somewhat less dense clusters can not be identified easily, when compared to unsupervised machine learning tools, like \texttt{SigMA}, and their method is less sensitive to possible spatial and kinematical structure in the Sco-Cen population. For example, a look at Figs.~2, 3, and 4 in \citetalias{Damiani2019} will make clear that the total number of member candidates using this approach is a strong function of the size of the selection shapes used in tangential velocity space and the HRD. These selection borders will necessarily select a larger number of true positives than an unsupervised method, while also the total number of false positives is likely higher.  

When focusing on a comparison of the total number of Sco-Cen members in \citetalias{Damiani2019} stellar clusters (compact and diffuse within 200\,pc, 10,421 sources) to the number of matched \texttt{SigMA} cluster members (9328 within 200\,pc) (see Table~\ref{tab:literature}), we find there are 1093 sources only in \citetalias{Damiani2019}, implying that we could be missing about 10\% of possible members if all 10,421 sources were good members. We did not perform a detailed comparison but find that the 1093 sources also contain sources that seem to be older than the \texttt{SigMA} clusters when investigated in an HRD (similar as in Fig.~\ref{fig:hrd}), hence the difference based on this comparison is likely lower than 10\% \citep[see also comparison with][]{Luhman2022a}. As mentioned above, we expect \texttt{SigMA} to be missing possible candidates when compared with a method that selects broad regions in various 2D planes of the phase space, but also expect the \texttt{SigMA} sample to be less contaminated. Nevertheless, the \texttt{SigMA} sample contains in total 11,796 sources inside the \citetalias{Damiani2019} FOV, implying that, in the end, we find about 10\% more Sco-Cen members. This could partially be caused by the different data sets, DR2 versus DR3, while the different methodologies likely cause more severe disagreements. A deeper analysis is needed, although not warranted in this paper.

\subsubsection{Comparison with Kerr et al. (2021)} \label{sec:kerr}

Recently, \citet[][hereafter KRK21]{Kerr2021} presented a study of nearby young stellar populations within 333\,pc from the Sun. They use the HDBSCAN clustering algorithm (see Sect.\,\ref{sec:hdbscan}) on Gaia DR2 parallaxes and proper motions on a pre-selected sample of PMS stars with ages $\lesssim$ 50\,Myr. They identify 27 \textit{top-level clusters} (TLC), including Chameleon as TLC\,21 and the Sco-Cen association as TLC\,22. The latter was further broken down into another 27 sub-groups based on the \textit{excess of mass} (EOM) method, selecting the most persistent clusters in the clustering tree. Three of these EOM sub-groups (EOM\,12 Lupus; EOM\,17 US; and EOM\,27 LCC) were further broken down into \textit{leafs}, which are nodes of the clustering tree. 

The TLC\,22 covers the main Sco-Cen association and TLC\,21 the Chamaeleon region. Additionally, there were cross-matches with members of the group TLC\,4, which is called Ophiuchus Southeast in \citetalias{Kerr2021}. These three TLC groups combined show a similar extent to our Sco-Cen extraction. \texttt{SigMA} finds in total a slightly lower number of groups toward Sco-Cen (37 in this work versus 45 in \citetalias{Kerr2021}), while the TLC\,22 sub-groups in \citetalias{Kerr2021} also include older or unrelated populations (e.g., $\beta$\,Pic, IC\,2602, Platais\,8, and EOM-2 \& 5), which are not included in our final Sco-Cen sample, as outlined above. Consequently, only 39 of the \citetalias{Kerr2021} groups toward Sco-Cen fall within the 37 selected \texttt{SigMA} clusters from this work. 

In Table\,\ref{tab:kerr}, we show an overview of the matches of \texttt{SigMA} groups with corresponding \citetalias{Kerr2021} groups. Overall, the \texttt{SigMA} Sco-Cen groups are more richly populated compared to the \citetalias{Kerr2021} groups. In most cases, there is at least some overlap between our groups and the TLC\,22 main Sco-Cen group (and with TLC\,21, Cham; or TLC\,4, Oph-SE), while some of our groups also distinctly correspond to EOM subgroups (or leafs). For about 40\% of the \texttt{SigMA} groups, a clear accordance with a single EOM group (or leaf group) is not possible due to overlaps with more than one \texttt{SigMA} group or due to no or only insignificant overlap (see also Table~\ref{tab:kerr}). 

Some differences between the \texttt{SigMA} and \citetalias{Kerr2021} clustering results might arise from the different data input since we use Gaia DR3 and \citetalias{Kerr2021} use DR2, while this would only create minor deviations. Although both HDBSCAN and \texttt{SigMA} approximate the hierarchical cluster tree, we expect discrepancies in clustering results. The primary reason for this difference is the cluster tree pruning strategy discussed in Appendix~\ref{sec:hdbscan}. The EOM heuristic prioritizes large clusters over their children when they maintain a long lifetime in the density hierarchy. The resulting children fail to exceed the parent's EOM. Conversely, our pruning strategy does not depend on cluster lifetimes but only cares about substantial density valleys between neighboring density peaks.

The additional leaf separations in \citetalias{Kerr2021} were applied to the Lupus, US, and LCC regions (in TLC\,22, EOM-12,17,27) since they found that there are substructures that have not been identified by the EOM method. Some leaf clusters match quite well with \texttt{SigMA} clusters; in contrast, the \texttt{SigMA} clusters are significantly richer and mostly more extended, and in some cases, they are differently separated (see details in Table~\ref{tab:kerr}). Compared to the EOM heuristic or \texttt{SigMA}'s multi-modality considerations, leaf clusters do not come with statistical guarantees. The clustering result is highly susceptible to random density fluctuations since leaf nodes are extracted only considering the minimum cluster size criterion \citep{Stuetzle:2010}; see Sect.~\ref{sec:hdbc} for more details. Without any additional pruning strategy, which deals with spurious clusters, leaf clustering results need to be taken with a grain of salt. Nevertheless, some of the leafs in US  ($\rho$\,Oph/L1688, $\nu$\,Sco, $\delta$\,Sco, $\beta$\,Sco) show good agreement with the \texttt{SigMA} US cluster separations, indicating the robustness of these clusters (see also Sect.~\ref{sec:miret}).

When comparing all members in the TLC22 group (7394), which encompasses the main Sco-Cen association, with our Sco-Cen \texttt{SigMA} extraction (12,796 members without Cham or Oph-SE) we find 6270 cross-matches in total (Table~\ref{tab:literature}). Hence, 1124 ($\sim$15\% of TLC22) sources are only in TLC22, and 6526 are only in \texttt{SigMA} ($\sim$50\% of \texttt{SigMA}). 
We find that the \citetalias{Kerr2021} TLC22 sample contains at least 256 sources from older stellar groups, which gets apparent from their Table~6 (EOM 1--5, including $\beta$\,Pic, IC\,2602, and Platais\,8), and 456 sources of TLC22 match with sources that are in older \texttt{SigMA} clusters. Combined, this leaves 6895 potential younger TLC22 Sco-Cen members, hence 625 possible extra sources ($\sim$8\% of TLC22). For these extra sources, a clear separation of the younger Sco-Cen stellar groups as discussed in this work, and the somewhat older groups is not straightforward, since about 50\% of the sources in the TLC22 group have not been assigned to a separate sub-cluster (EOM or leaf). The somewhat older sources can also be estimated when investigating an HRD or the velocity space. In the HRD no clear separation of older or younger sources can be identified.
In tangential velocity space, there are sources that have slightly deviating motions from expected Sco-Cen motions or which coincide with velocity spaces of the \citetalias{Kerr2021} older EOM groups or older \texttt{SigMA} groups. Taking all this into account, the fraction of TLC\,22-only sources is likely below 8\%.

The reason for these extra potential Sco-Cen members in the \citetalias{Kerr2021} TLC22 group is similar to the mentioned reasons above (e.g., in Sect.~\ref{sec:damiani}). The TLC22 group represents a cluster root, enveloping the whole Sco-Cen region and somewhat beyond, and no additional substructure was extracted (yet). In the following step \citetalias{Kerr2021} use the EOM and leaf methods to identify individual clusters, while in this step they lose almost 50\% of the original TLC22 group, as mentioned above. 
Focusing only on the TLC22 members that are in one of the 22 younger EOM sub-groups, we find that we recover 99.8\% of these sources as Sco-Cen members. Finally, the TLC22 group seems to be overall more incomplete compared to the \texttt{SigMA} Sco-Cen extraction, since we find in total more members ($\sim$40\%, 12,796 versus 7394), and also somewhat different substructure. At the same time, the sub-clusters themselves are significantly richer compared to \citetalias{Kerr2021}. 

In conclusion, the comparison with \citetalias{Kerr2021} highlights the differences that can arise with different unsupervised machine learning tools. 
Compared to applications of HDBSCAN, we find that \texttt{SigMA} is able to extract similar sub-structure, however, with only one clustering step (no sub-steps like EOM or leafs are needed), while at the same time extracting significantly higher numbers of members per cluster.

\subsubsection{Comparison with Schmitt et al. (2022)} \label{sec:schmitt}

Recently, \citet[][hereafter SCF22]{Schmitt2022} used eROSITA\footnote{Extended ROentgen Survey with an Imaging Telescope Array. A wide-field X-ray telescope on-board the Russian-German ``Spectrum-Roentgen-Gamma'' (SRG) observatory.} \citep{Merloni2020} to search for low-mass Sco-Cen members by cross-correlating the eRASS1 source catalog with the Gaia EDR3 catalog. They discuss 6190 X-ray observed sources within the traditional borders \citep{Blaauw1964a, deZeeuw1999} which could be Sco-Cen members. They include sources within a distance range of 60\, to 200\,pc (Table~\ref{tab:literature}), restricted to low-mass stars ($G_\mathrm{BP}-G_\mathrm{RP}>1$, following \citealp{Pecaut2013}). The 6190 sources include 40 double Gaia sources that match the same eROSITA source. We will only discuss the 6150 single Gaia sources. 

Since X-ray emitting sources are expected to be young \citep[e.g.,][]{Schmitt1997, Neuhauser1997-du, Feigelson1999, Bouvier2014}, the sources detected by eROSITA in the direction of Sco-Cen, as discussed in \citetalias{Schmitt2022}, are potential members of Sco-Cen. They found X-ray sources down to about 0.1 M$_\odot$, and, unexpectedly, they also found the existence of a population of young X-ray emitting stars that appear to be more diffuse in velocity space\footnote{We applied the separation of kinematically clustered and diffuse populations by hand in $v_\alpha$/$v_\delta$ space, as indicated in Fig.~7 of \citetalias{Schmitt2022}.}, calling into question search approaches relying on kinematic selections. 

We cross-matched the 6150 \citetalias{Schmitt2022} X-ray selected sources with the \texttt{SigMA} selection in the same FOV (containing 11,348 \texttt{SigMA} sources, see Table~\ref{tab:literature}). We find in total 3385 cross-matches, while none of these belong to their velocity-diffuse population. The latter is expected since \texttt{SigMA} only selects clusters confined in position-velocity space, which naturally excludes any such velocity-diffuse sources. \citetalias{Schmitt2022} claim that the diffuse population is largely composed of young stars, only somewhat older compared to the kinematically confined Sco-Cen members. We confirm the general youth of the sources by inspecting the two populations in an HRD. However, we see a relatively clear age separation between the velocity-clustered and velocity-diffuse populations. X-ray sources that occupy similar velocity spaces as the Sco-Cen members have ages between 0.1--20\,Myr, while X-ray sources that are velocity-diffuse have ages between 10--1000\,Myr, with the majority at about 30--100\,Myr. While these are technically young stars, they seem too old to be related to the Sco-Cen association. 

The origin of this co-spatial but velocity diffuse population remains mysterious. Since these sources are older than Sco-Cen, they are unlikely to result from stellar interactions in Sco-Cen (an a priori unlikely process given the low stellar density of Sco-Cen). The diffuse population, or the co-eval part, could be related to a relatively older star-formation episode, sharing today the volume space of Sco-Cen, a plausible scenario in the Milky Way \citep{Fuernkranz2019}. We posit here that the \citetalias{Schmitt2022} velocity-diffuse young sources are unlikely to be part of Sco-Cen, but represent a mystery that needs to be solved. As \citetalias{Schmitt2022} points out, the sensitivity of eROSITA will allow in the near future to detect virtually all young Sco-Cen low-mass members. A combination of eROSITA future releases and Gaia data in Sco-Cen will be crucial to increase statistics and better understand the relation between observed X-ray luminosity with distance, age, stellar masses, and the origin of the velocity-diffuse population. 

When concentrating on the velocity-coherent sample in \citetalias{Schmitt2022} (without IC\,2602), we find that there are about 20\% in the whole \citetalias{Schmitt2022} sample that could be additional Sco-Cen candidate members. These are only in \citetalias{Schmitt2022} and have similar velocities as \texttt{SigMA} Sco-Cen members. When investigating possible older-star contaminants in an HRD (similar to Fig.~\ref{fig:hrd}), this fraction would reduce to about 10\%. These extra potential members might result from the broad selection conditions in \citetalias{Schmitt2022}, based on all X-ray detected sources within the Blaauw borders in a distance range of 60\, to 200\,pc (Table~\ref{tab:literature}). These broad conditions, which do not attempt to identify any underlying clustered structure, will naturally pick up more candidate members, although more false positives too, as discussed in Sect.~\ref{sec:damiani} and \ref{sec:remarks}.

\subsubsection{Comparison with Luhman (2022)} \label{sec:luhman}

\citet[][hereafter L22A]{Luhman2022a} recently investigated the Sco-Cen region using Gaia EDR3 data to identify 10,509 kinematic candidate members of Sco-Cen (see Table~\ref{tab:literature}). \citetalias{Luhman2022a} includes selections for US, UCL/LCC, V1062\,Sco, Ophiuchus, and Lupus (the Southern parts of Sco-Cen are not discussed in \citetalias{Luhman2022a}), and concentrates on established stellar groups in Sco-Cen to guide the selection. The visual selection approach of \citetalias{Luhman2022a} is not suitable to separate the underlying kinematical substructure of the Sco-Cen population. For example, it is clear from Fig.~4 in \citetalias{Luhman2022a} (bottom panel) that the UCL/LCC group contains several over-densities in $l/b$ space, but these are not extracted or identified. The \citetalias{Luhman2022a} selection is based on global kinematic criteria, extracting candidates exhibiting proper motions similar to expected proper motions of known members.

Cross-matching the 10,509 \citetalias{Luhman2022a} Sco-Cen candidate members with the \texttt{SigMA} clusters gives a total of 9838 matches (93.6\%), 671 \citetalias{Luhman2022a} only sources (6.4\%), and 2377  \texttt{SigMA} only sources within the \citetalias{Luhman2022a} studied area (Table~\ref{tab:literature}), where the \texttt{SigMA} sample contains 12,215 sources in total. A more detailed comparison of the \texttt{SigMA} clusters with the \citetalias{Luhman2022a} subgroups\footnote{Subgroups are given in Table 1 of \citetalias{Luhman2022a} in columns ``kin'' and ``pos''.}, which are generally larger scale groups, shows no clear correlation between single groups. Virtually each \texttt{SigMA} cluster has several matches with various \citetalias{Luhman2022a} subgroups (and vice versa).
When investigating the 671 \citetalias{Luhman2022a} only sources, we find that about 50\% of these sources do not show significant signs of being older than 20\,Myr, and the majority of the sources do not show significant deviating motions from \texttt{SigMA} Sco-Cen cluster velocities. These extra \citetalias{Luhman2022a} sources, or part of them, could be Sco-Cen members, meaning we might be missing up to about 6\% of the candidates in \citetalias{Luhman2022a}. This is not surprising because methods based on visual selection, using broad selection borders, will naturally find more candidates, but also more false positives, as also discussed in Sect.~\ref{sec:damiani}. Nevertheless, the \texttt{SigMA} samples contain in total more Sco-Cen member candidates within the same field of view.

\subsubsection{Comparison with \v{Z}erjal et al. (2021)} \label{sec:zerjal}

\citet[][hereafter ZIC21]{Zerjal2021arxiv} present another clustering result for the Sco-Cen association using \textsc{Chronostar}, a clustering tool developed by \citet{Crundall2019}. This is a Bayesian tool to kinematically decompose stellar groups using the full 6D kinematic data, also performing a kinematic age determination. They identify eight distinct kinematic components containing in total 9556 sources\footnote{18 sources are outside of our box and quality criteria, leaving 9538 \citetalias{Zerjal2021arxiv} sources. Moreover, there are 25 sources that are outside of the \citetalias{Zerjal2021arxiv} parallax range due to updates from DR2 to DR3.}.
The 9556 stellar members are both within dense and also diffuse stellar groups. They also include two known clusters that we excluded from the final Sco-Cen sample, which are IC\,2602 and Platais\,8 (H and I in \citetalias{Zerjal2021arxiv}). Without these clusters their sample contains 8185 stars, of which 7671 (94\%) match with the \texttt{SigMA} groups.

The groups are C--US, E--US-multi\footnote{\citetalias{Zerjal2021arxiv} define group E as a complex, multi-population component.}, \mbox{D--UCL-V1062-Sco}, \mbox{F--UCL-V1062-Sco}, \mbox{G--UCL-East}, \mbox{T--UCL-West}, A--LCC-North, and \mbox{U--LCC-South}.
Hence, with their method they are splitting US, UCL, LCC, and also V1062\,Sco, each into two parts.
We list all matches of \texttt{SigMA} clusters with \citetalias{Zerjal2021arxiv} in Table~\ref{tab:damiani}.
Generally, it can be seen that there is significant mixing of various groups in both the \citetalias{Zerjal2021arxiv} and the \texttt{SigMA} groups. In particular, the \citetalias{Zerjal2021arxiv} groups encompass larger areas, often containing several or up to 20 of the \texttt{SigMA} groups. Concerning the groups D and F, we find that both match with V1062-Sco and $\mu$\,Sco, while D has a slightly higher correlation with V1062-Sco and F with $\mu$\,Sco. 

Within their FOV also other groups exist (see Fig.~8 in \citetalias{Zerjal2021arxiv}), like CrA or the Cham cloud regions, while they were not selected as kinematic members of Sco-Cen.
We speculate that their initial sample by~\citet{Gagne2018c}, which includes sources from US, UCL, and LCC, limits their ability to select these additional groups. Their low signal-to-noise ratio, positional distance, and slightly deviating motions from bulk Sco-Cen sources\footnote{Possibly caused by internal feedback mechanisms in the history of Sco-Cen \citep[e.g.,][]{Zucker2022}.} likely prevented their classification as Sco-Cen groups.

Their method fits a mixture of Gaussians to data. Instead of allowing arbitrary covariance matrices, \textsc{Chronostar} constrains 6D Gaussian distributions in XYZUVW to the following form. Present-day observations are assumed to follow a ballistically evolved Gaussian in Galactic potential. The free-fitting parameters are a cluster's mean birth position in phase space, its birth positional and kinematic variance (the covariance matrix is assumed uncorrelated), and its age. The fitting is done via a modified EM algorithm where the number of components is determined via the BIC. 
\citetalias{Zerjal2021arxiv} state that they see evidence for substructure in several groups. Thus, it has to be investigated if the groups found in \citetalias{Zerjal2021arxiv} will eventually break in subgroups. When compared to \texttt{SigMA} we can already see that the individual \citetalias{Zerjal2021arxiv} groups generally contain more than one \texttt{SigMA} cluster (see Table\,\ref{tab:damiani}).

\subsubsection{Comparison with Squicciarini et al. (2021)} \label{sec:squicc}

\citet[][hereafter SGB21]{Squicciarini2021} studied 2745 potential US members (see Table~\ref{tab:literature}) by selecting subgroups solely based on kinematics. They divided the region into eight groups which they call the clustered population (1442 stars), and into an older diffuse population (1303), which is, however, differently defined than the velocity-diffuse population in \citetalias{Damiani2019} or \citetalias{Schmitt2022}. 

When comparing the \citetalias{Squicciarini2021} selection to the \texttt{SigMA} clusters, we find that there are 2575 cross-matches ($\sim$94\%) in total out of the 2745 sources in \citetalias{Squicciarini2021}. Hence we miss 170 \citetalias{Squicciarini2021} US candidate members, while 13 sources are lost due to our box and quality criteria (Sect.\,\ref{sec:data}).  
Focusing on the \citetalias{Squicciarini2021} candidate members in the clustered populations (1442), there are only seven sources that are only in \citetalias{Squicciarini2021} and not in \texttt{SigMA}, while we miss 163 of the  \citetalias{Squicciarini2021} diffuse members, which are overall more uncertain members. In total we find almost a factor two more clustered sources with \texttt{SigMA} in the same FOV as \citetalias{Squicciarini2021}, 2717 versus 1442 (see Table~\ref{tab:literature}). 
The 2575 sources match with nine of the \texttt{SigMA} clusters, while only seven \texttt{SigMA} clusters have a significant number of matches.

We list the cross-matches of \texttt{SigMA} with \citetalias{Squicciarini2021} in Table\,\ref{tab:squicc}. We highlight more significant matches here.
Groups\,1, 2, 3, and 4 match best with $\rho$\,Oph/L1688, $\nu$\,Sco, $\delta$\,Sco, and $\beta$\,Sco, respectively, while Group\,6 also has significant matches with $\beta$\,Sco. 
Group\,5 matches best with $\sigma$\,Sco, while the majority of $\sigma$\,Sco is in the \citetalias{Squicciarini2021} diffuse population.
Group\,7 and Group\,8 match best with Antares, while the majority of Antares is also in their diffuse population. Generally, the Antares group seems to split up into more than one cluster, also in other previous work.
The two groups US-foreground and $\phi$\,Lup have only few matches with the diffuse population. 
The \citetalias{Squicciarini2021} diffuse population is largely contained within the \texttt{SigMA} groups $\sigma$\,Sco, Antares, $\rho$\,Sco, and $\delta$\,Sco, with some diffuse members distributed among each mentioned group (see Table\,\ref{tab:squicc}). This indicates that the  diffuse population is not a separate older group but it contains stars that were not clustered by the methodology in \citetalias{Squicciarini2021}, while they are clustered in \texttt{SigMA}. 

The differences in the final cluster definition in US likely arise from the different clustering methodologies. To better understand the \citetalias{Squicciarini2021} approach we outline the basics here. \citetalias{Squicciarini2021} use a semi-automated approach based on iterative k-means clustering on a 4D sample, using 2D sky positions and 2D tangential velocities.
The authors propagate the sky positions $15$\,Myr into the past and future, producing a new 4D data set at each step; tangential velocities are constant throughout individual data sets. By studying the sky distribution of each slice, \citetalias{Squicciarini2021} visually identify over-densities. These over-densities are extracted via k-means clustering in 4D space at a given time step. Subsequently, the clustered data points are removed from the data set, and the process of looking for over-densities starts anew. The clustering process terminates when the authors cannot find any apparent density peaks in the sky distribution.

Besides the feature space difference, \texttt{SigMA} has significant differences compared to \citetalias{Squicciarini2021}'s iterative clustering approach. First, the k-means algorithm cannot deal with the observed non-convex cluster shapes in projected coordinates. The extracted clusters are 4D Voronoi cells\footnote{As far as we know, scaling between sky coordinates and tangential velocities was not considered.} which can have very elongated shapes. Second, \citetalias{Squicciarini2021} analyze 2D projections of the high-dimensional data to identify clusters visually. Thus, cluster selection is influenced by projection effects and human judgment. Conversely, \texttt{SigMA} employs a modality test directly in the high-dimensional phase space, taking into account multi-dimensional relationships between data axes.
These rather different approaches to extract clusters in US make it clear that the results can not be compared at face value, while fractions of the most robust clusters ($\rho$\,Oph/L1688, $\nu$\,Sco, $\delta$\,Sco, $\beta$\,Sco, $\sigma$\,Sco, and Antares) have been identified by either method.

\subsubsection{Comparison with Miret-Roig et al. (2022)} \label{sec:miret}

\citet[][hereafter MR22]{MiretRoig2022b} recently applied a Gaussian mixture model to Gaia DR3 data of the US region. They include radial velocities, when available, to select the stellar groups. They identify seven stellar groups within a FOV of about 220\,deg$^2$ that is centered on US, containing 2810 sources (see also Table\,\ref{tab:literature}). A cross-match with the \texttt{SigMA} members gives 2683 matches, hence we miss 127 sources, of which 23 are outside of our box and quality criteria (Sect.~\ref{sec:data}). This leaves 104 potential members that the \texttt{SigMA} algorithm did not select. Within the same FOV \texttt{SigMA} contains 3089 Sco-Cen members.

When comparing the \texttt{SigMA} groups to the individual seven \citetalias{MiretRoig2022b} groups in more detail, we find largely good agreement, especially for $\rho$\,Oph, $\nu$\,Sco, $\delta$\,Sco, and $\beta$\,Sco (see also Table~A.1 in \citetalias{MiretRoig2022b} and Table~\ref{tab:squicc}). 
The \texttt{SigMA} Antares and $\sigma$\,Sco groups seem to be mixed in \citetalias{MiretRoig2022b}, with significant fractions in both \citetalias{MiretRoig2022b}'s $\alpha$\,Sco and $\sigma$\,Sco. 
Finally, the \citetalias{MiretRoig2022b} $\pi$\,Sco group, which lies largely in the foreground of the other US groups, coincides largely with two of the \texttt{SigMA} groups: US-foreground (which we identified as a foreground population to the traditional US) and $\rho$\,Sco. The latter is overlapping in space with Antares and $\rho$\,Oph. The larger volume investigated with \texttt{SigMA} allows selecting a more complete sample of US-foreground, since the \citetalias{MiretRoig2022b} FOV cuts off the outer edges of what they call $\pi$\,Sco. 

Since \citetalias{MiretRoig2022b} also use radial velocity information for a subsample ($\sim$30\%), the classification of these 6D selected sources might be more robust compared to a 5D selection. Investigating the bona-fide 6D selected members in \citetalias{MiretRoig2022b}, we still find some mixing of \texttt{SigMA} clusters within \citetalias{MiretRoig2022b} clusters (and vice versa), especially concerning Antares and $\sigma$\,Sco. These differences need more investigations in the future, and dedicated cluster studies are called for \citep[e.g., with \texttt{Uncover},][]{Ratzenboeck2020}.

\subsubsection{Comparison with Brice\~{n}o-Morales \& Chanam{\'e} (2022)} \label{sec:briceno}

\citet[][hereafter BMC22]{BricenoMorals2022} present another recent study on US, where they seem to find a similar substructure. We do not include a detailed comparison with \citetalias{BricenoMorals2022}, since we do not have access to their final catalog\footnote{The paper only appeared on the arXiv after the first submission of our paper draft, and we did not (yet) get access to their data tables.}. Therefore, we only give a qualitative comparison here.

\citetalias{BricenoMorals2022} use Gaia EDR3 data to obtain a clustering solution for US by first combining the convergent point method \citep{Perryman1998} with a Gaussian Mixture fit \citep{Pedregosa2011} to identify kinematic groups with a Bayesian approach, and second, they use OPTICS \citep{Ankerst1999} to identify spatial substructure in XYZ. 
This is complemented with age estimates based on Gaia photometry. They select in total 4028 US members, while their astrometrically clean sample contains 3114 sources, but most of them seem to be contained within a diffuse population ($\sim$66\%). In the same FOV\footnote{\citetalias{BricenoMorals2022}  FOV: $343^\circ < l < 360^\circ$, $10^\circ < b < 30^\circ$, $\varpi > 5$\,mas.} we find 3418 clustered Sco-Cen members.

They find that US contains three main kinematic groups, and about 34\% are contained in seven spatial substructures. They find a clear correlation between cluster density and cluster age, suggesting that these substructures expand at a measurable rate. Finally, they argue that four potential supernovae happened in the US region based on literature information \citep[e.g.,][]{Breitschwerdt2016, Neuhaeuser2020, Forbes2021}, cluster mass estimates, and on certain ``voids'' that are identified in the 3D distribution of their cluster extraction. Notably, we do not see such voids in the distribution of the \texttt{SigMA} clusters.

Based on Table~2 in \citetalias{BricenoMorals2022} we identify the following qualitative matches of \texttt{SigMA} clusters with \citetalias{BricenoMorals2022} clusters, when comparing average positions and velocities. 
The clusters $\rho$\,Oph/L1688, $\beta$\,Sco, $\delta$\,Sco and $\nu$\,Sco seem to match well with the same named clusters in \citetalias{BricenoMorals2022}.
Their $\alpha$\,Sco seems to match our $\sigma$\,Sco cluster (not with Antares), while they mention that both stars, *\,alf\,Sco and *\,sig\,Sco, are likely members of this cluster. This connection is of interest, since also other works seem to mix the \texttt{SigMA} Antares and $\sigma$\,Sco clusters, which needs further investigations in the future. 
Their $\pi$\,Sco seems to match best with US-foreground.
Their $\omega$\,Sco seems to match approximately with Antares.
A detailed cross-match with their data is needed to make a more quantitative statement, which is not possible at this stage. 


\subsubsection{Concluding remarks on the comparison with the literature} \label{sec:remarks}

Although \texttt{SigMA} finds the larger number of Sco-Cen member candidates of all the methods presented in the literature, visual selection methods \citep[e.g.,][]{Damiani2019, Luhman2022a} contain about 6--10\% candidate members not detected by \texttt{SigMA} (Table~\ref{tab:literature}). 
This is mainly due to these methods ``select-by-eye" approach, in projected sub-spaces of the multi-dimensional phase space, to identify Sco-Cen candidates. Unsupervised machine learning methods, on the other hand, find more spatial and kinematical substructures in the Sco-Cen population and produce samples with lower contamination levels compared to visual selection methods. More importantly, the \texttt{SigMA} method reveals a more complex velocity structure across the entire Sco-Cen, critical for a physical description the formation process of OB associations such as Sco-Cen. When comparing \texttt{SigMA} to other unsupervised methods, which studied the whole Sco-Cen area \citep[in particular][]{Kerr2021}, the \texttt{SigMA} clusters are generally richer in members. 

Focusing on the US region, we find generally good agreement between cluster selections from \citetalias{Squicciarini2021},  \citetalias{Kerr2021}, \citetalias{MiretRoig2022b}, \citetalias{BricenoMorals2022}, and \texttt{SigMA}. Not surprisingly, the denser clusters in US ($\rho$\,Oph, $\nu$\,Sco, $\delta$\,Sco, $\beta$\,Sco) have been recovered well by the different approaches. The Antares, $\sigma$\,Sco, $\rho$\,Sco, and US-foreground clusters are slightly more dispersed and they have less clear matches across the methods (e.g., merged or distributed differently in different samples). The newly identified velocity substructure in the US region, as revealed by Gaia data, is relevant to understand the star formation processes at play in Sco-Cen and will continue to be an obvious target for future studies and surveys. Adding radial velocity information will be critical to further characterize these clusters, as already presented, for example, by \citet{MiretRoig2022b}.

The somewhat different clustering results for the US region, considering the five mentioned studies, call for a re-analysis of this important young region, to better understand its star formation history.
In conclusion, \texttt{SigMA} finds significant numbers of sources not present in other samples (up to 15\% more), while missing candidates when compared with visual selection methods (of the order of 10\%). More work is needed to understand the sources \texttt{SigMA} misses. A possible way forward toward a most complete sample of Sco-Cen members is to use \texttt{SigMA} cluster members and 3D velocities (by including radial velocities) as training sets to the \texttt{Uncover} method, a validated bagging classifier of one-class support vector machines (see the application in \citealp{Ratzenboeck2020} to the Meingast-1 stellar stream, \citealp{Meingast2019}).

In the near future, improved membership lists will allow a more precise analysis of the star formation history of Sco-Cen, the initial mass function of each cluster, and the dynamical state of the Sco-Cen complex.

\section{Summary}\label{sec:conclusion}

In this paper, we present \texttt{SigMA}, a method that explores the topological properties of a density field to define significant structure. To test and validate \texttt{SigMA}, we apply it to Gaia DR3 data of the nearest OB association to Earth, Sco-Cen. The main results of this work can be summarized as follows:

   \begin{enumerate}
   
      \item \texttt{SigMA} is a novel clustering method that interprets density peaks, which are separated by dips, as significant clusters. Using a graph-based approach, the technique detects peaks and dips directly in the multi-dimensional phase space.
      
      \item \texttt{SigMA} is fine-tuned to large-scale surveys in astrophysics. In this context, this new method is able to identify co-spatial and co-moving groups with non-convex shapes and variable densities, with a measure of significance. \texttt{SigMA} is able to properly incorporate 5D astrometric uncertainties, does not require any photometric pre-filtering of stellar populations, and scales to millions of points.
   
      \item  \texttt{SigMA} is capable of finding clusters in Gaia DR3 data, reaching stellar volume densities as low as 0.01\,sources/pc$^3$ and tangential velocity differences of about 0.5\,km\,s$^{-1}$ between clusters. 
      
      \item We validate \texttt{SigMA} on two simulated data sets and highlight its merits in relation to established clustering techniques. Our comparison shows that \texttt{SigMA} can significantly outperform competing methods, especially in environments where clusters are densely packed, such as the Sco-Cen OB association.
      
      \item
      \texttt{SigMA} identifies more than \SI{13000}{} Sco-Cen members located in 37 individual clusters of co-spatial and co-moving young stars. The HRD for each cluster shows a  well-defined sequence. Because \texttt{SigMA} is not aware of a star's brightness nor color, the well-defined sequences constitute a validation test to the ability of \texttt{SigMA} to extract coeval populations.
      A large fraction of clusters is seen toward well-known Sco-Cen massive stars, too bright to be in Gaia DR3, and we (tentatively) associated with them. Because \texttt{SigMA} is not aware of these massive stars, their  association with \texttt{SigMA} clusters also constitutes a validation test to \texttt{SigMA}.
      
      \item
      When comparing the 37 \texttt{SigMA} stellar populations in Sco-Cen to previous results from the literature we find mostly agreement, however, several discrepancies exist. When compared to visual selection methods used recently on Gaia data of Sco-Cen we find that we might be missing roughly 10\% of candidates, while at the same time finding a higher total number of stellar members. Unsupervised methods like \texttt{SigMA} find more spatial and kinematical substructure for the same data set and produce samples with lower contamination levels.

  \end{enumerate}

In the future, in particular, when combined with auxiliary radial velocity surveys, a detailed comparative study of the different clustering methods is fully warranted. The application of \texttt{SigMA} to upcoming Gaia data releases promises the unveiling of detailed cluster distributions like the one presented here but for all the near star-forming regions. Reconstructing an accurate and high-spatial-resolution Star Formation History of the last 50\,Myr in the Local Milky Way with Gaia data is within reach.

\begin{acknowledgements}
    We thank the anonymous referee for their detailed and very helpful comments.
    We thank N\'uria Miret-Roig, Maru{\v{s}}a {\v{Z}}erjal, Vito Squicciarini, and J.\,H.\,M.\,M.\,Schmitt for providing their data ahead of final publication.
    S.\,Ratzenb{\"o}ck acknowledges funding by the Austrian Research Promotion Agency (FFG, \url{https://www.ffg.at/}) under project number FO999892674. Additionally, S.\,Ratzenb{\"o}ck extends his gratitude to the research network Data Science @ Uni Vienna and Petra Sch{\"o}nfelder for excellent administrative support. 
    J.\,Gro{\ss}schedl acknowledges funding by the Austrian Research Promotion Agency (FFG) under project number 873708.
    This work has used data from the European Space Agency (ESA) mission {\it Gaia} (\url{https://www.cosmos.esa.int/gaia}), processed by the {\it Gaia} Data Processing and Analysis Consortium (DPAC, \url{https://www.cosmos.esa.int/web/gaia/dpac/consortium}). Funding for the DPAC has been provided by national institutions, in particular the institutions participating in the {\it Gaia} Multilateral Agreement. 
    This research has made use of Python, \url{https://www.python.org}, of \textit{Astropy}, a community-developed core Python package for Astronomy    \citep{Astropy2013, Astropy2018}, NumPy \citep{Walt2011}, 
    Matplotlib \citep{Hunter2007}, 
    Galpy \citep{Bovy2015}, 
    and Plotly \citep{plotly}. 
    This research has used the SIMBAD database operated at CDS, Strasbourg, France \citep{Wenger2000}, of the VizieR catalog access tool, CDS, Strasbourg, France \citep{Ochsenbein2000}, and of ``Aladin sky atlas'' developed at CDS, Strasbourg Observatory, France \citep{Bonnarel2000, Boch2014}. This research has made use of TOPCAT, an interactive graphical viewer and editor for tabular data \citep{Taylor2005}.
\end{acknowledgements}

\bibliographystyle{aa.bst}
\bibliography{sebastian}

\begin{appendix}

\section{Gaia DR3 data retrieval and details on quality criteria}\label{app:box}

The Gaia DR3 data was downloaded from the Gaia Archive\footnote{\url{https://gea.esac.esa.int/archive/}} using the following ADQL query:

\begin{tiny}
\begin{verbatim}
SELECT * FROM gaiadr3.gaia_source 
WHERE (1000./parallax*COS(l*PI()/180)*COS(b*PI()/180))>-50.
AND (1000./parallax*COS(l*PI()/180)*COS(b*PI()/180))<250.
AND (1000./parallax*SIN(l*PI()/180)*COS(b*PI()/180))>-200.
AND (1000./parallax*SIN(l*PI()/180)*COS(b*PI()/180))<50.
AND (1000./parallax*SIN(b*PI()/180))>-95.
AND (1000./parallax*SIN(b*PI()/180))<100.
AND parallax_over_error>4.5
\end{verbatim}
\end{tiny}

The first six expressions give the XYZ box conditions, while X is positive toward the Galactic center, Y is positive in direction of Galactic rotation, and Z points toward the Galactic north-pole. 
XYZ can also be calculated using \texttt{astropy.coordinates.SkyCoord} from Astropy \texttt{v4.0}.
The parameter fidelity\_v2 from \citet{Rybizki2022} was retrieved with the following ADQL query, using the Topcat TAP Query and the GAVO service\footnote{\url{https://dc.zah.uni-heidelberg.de/}, German Astrophysical Virtual Observatory}:

\begin{tiny}
\begin{verbatim}
SELECT gaia.*
FROM gedr3spur.main AS gaia
JOIN tap_upload.t1 AS mine
USING (source_id)
\end{verbatim}
\end{tiny}

We support our quality criteria choices from Sect.~\ref{sec:data} as follows. Using fidelity\_v2\,$>$\,0.5 is suggested as separator into ``good'' and ``bad'' sources by \citet{Rybizki2022} \citep[see also][]{Zari2021}. Using a threshold of 0.9 would give slightly cleaner data, however only few sources lie in the range between 0.5 and 0.9 ($\sim$2\% in the box), and we decide for the less conservative value. 
The additional cut using \texttt{parallax\_over\_error}, which is similar to the signal-to-noise ratio ($\mathrm{S/N}_\varpi$), is used to reduce further parallax uncertainties. The choice of the threshold $\mathrm{S/N}_\varpi > 4.5$ is further supported by \citet{Rybizki2022}, where they apply different classifiers (high- and low-S/N classifiers) for sources above and below this threshold. To avoid inhomogeneous data we decide to include this S/N threshold. 
Moreover, we want to avoid too high parallax errors. As mentioned, we use the inverse of the parallax to estimate the distance to a source, which gets unreliable if the uncertainties are too large. For more distant sources, or intrinsically faint sources with high parallax errors, the distance estimate becomes a non-trivial inference problem \citep[e.g.,][]{Luri2018, Bailer-Jones2021}.
In Table~\ref{tab:typ-unc} we list the typical uncertainties of the various parameters for sources inside the box after the applied quality criteria, and also separately for sources that are selected as members of the 37 Sco-Cen clusters (given in brackets in Table~\ref{tab:typ-unc}).  
It can be seen that the majority of the sources in the box (2$\sigma$) have parallax uncertainties below 0.6\,mas. 

For \texttt{SigMA} we use the 5D phase space, as mentioned in Sect.~\ref{sec:data}, for which we use tangential velocities in km\,s$^{-1}$. The proper motions ($\mu_\alpha^* = \mu_\alpha \cos(\delta)$, $\mu_\delta$) are transformed from mas\,yr$^{-1}$ to tangential velocities in km\,s$^{-1}$ as follows, where the conversion constant 4.74047 is in units of km\,yr\,s$^{-1}$.
\begin{equation}
    \label{eq:tan-vel}
    \begin{aligned}
    v_{\alpha} = 4.74047 \cdot \mu_\alpha^*/\varpi \\
    v_{\delta} = 4.74047 \cdot \mu_\delta/\varpi
    \end{aligned}
\end{equation}

Moreover, we correct the tangential motions for the Sun's reflex motion, resulting in tangential motions relative to the LSR, using the values by \citet{Schoenrich2010}. This conversion is accomplished with the help of Astropy, by defining the below sky coordinates.
For a conversion of heliocentric proper motions to proper motions relative to the LSR, the radial velocity can be set to an arbitrary value in the sky-coordinate definition of Astropy (here set to $0$), since different RV values do not change the outcome of this conversion.

\begin{tiny}
\begin{verbatim}
from astropy.coordinates import ICRS, LSR
from astropy import units as u

skyicrs = ICRS(ra = ra * u.deg,
               dec = dec * u.degree,
               distance = 1000./parallax * u.pc,
               pm_ra_cosdec = pmra * u.mas/u.yr,
               pm_dec = pmdec * u.mas/u.yr,
               radial_velocity = 0. * u.km/u.s)

pma_lsr = skyicrs.transform_to(LSR()).pm_ra_cosdec.value
pmd_lsr = skyicrs.transform_to(LSR()).pm_dec.value

v_a_lsr = 4.74047 * pma_lsr / parallax
v_d_lsr = 4.74047 * pmd_lsr / parallax
\end{verbatim}
\end{tiny}

\begin{table}[!t] 
\begin{center}
\begin{small}
\caption{Typical parameter uncertainties. The listed uncertainties give the threshold below which the given percentage of sources can be found in the whole box (or in the \texttt{SigMA} Sco-Cen selection, given in brackets). \vspace{-3mm}}
\renewcommand{\arraystretch}{1.6}
\resizebox{0.88\columnwidth}{!}{%
\begin{tabular}{lrrr}
\hline \hline
\multicolumn{1}{l}{} &
\multicolumn{3}{c}{1, 2, 3 $\sigma$ percentiles} \\ 
\cmidrule(lr){2-4}
\multicolumn{1}{l}{Uncertainties\tablefootmark{a}} &
\multicolumn{1}{c}{68.3\%} &
\multicolumn{1}{c}{95.5\%} &
\multicolumn{1}{c}{99.7\%}  \\ 
\hline





$\sigma_\varpi$ (mas) $<$ & 0.13 (0.08) & 0.56 (0.34) & 1.17 (0.94)  \\
$\mathrm{S/N}_\varpi$ $>$ & 47.6 (83.2) & 9.3 (20.8) & 4.8 (7.5)  \\
$\sigma_d$ (pc) $<$  & $^{+3.8\,(+1.8)}_{-3.6\,(-1.7)}$ & $^{+26.0\,(+7.4)}_{-21.0\,(-6.7)}$ & $^{+65.2\,(+23.8)}_{-43.7\,(-18.0)}$ \\
$\sigma_{\mu_{\alpha}}$ (mas yr$^{-1}$) $<$ & 0.13 (0.09) & 0.60 (0.39)  & 1.47 (1.06)    \\ 
$\sigma_{\mu_{\delta}}$ (mas yr$^{-1}$) $<$ & 0.12 (0.08) & 0.54 (0.32)  & 1.34 (0.89)    \\ 
$\sigma_{v_{\alpha}}$ (km s$^{-1}$) $<$ & $^{+0.4\,(+0.2)}_{-0.4\,(-0.2)}$ & $^{+3.2\,(+0.6)}_{-2.7\,(-0.6)}$ & $^{+12.1\,(+2.1)}_{-8.9\,(-1.7)}$  \\
$\sigma_{v_{\delta}}$ (km s$^{-1}$) $<$ & $^{+0.4\,(+0.2)}_{-0.4\,(-0.2)}$ & $^{+3.3\,(+0.8)}_{-2.8\,(-0.7)}$ & $^{+12.1\,(+2.4)}_{-8.9\,(-1.9)}$  \\

$\sigma_{v_{r}}$ (km s$^{-1}$) $<$ & 3.3 (5.8) & 8.3 (13.0) & 26.2 (37.3) \\

$\sigma_X$ (pc) $<$  & $^{+2.3\,(+1.4)}_{-2.2\,(-1.4)}$ & $^{+20.1\,(+6.3)}_{-16.5\,(-5.7)}$ & $^{+54.6\,(+20.3)}_{-36.8\,(-16.0)}$ \\
$\sigma_Y$ (pc) $<$  & $^{+1.6\,(+0.6)}_{-1.5\,(-0.6)}$ & $^{+13.4\,(+2.9)}_{-11.2\,(-2.7)}$ & $^{+41.5\,(+12.1)}_{-28.1\,(-9.8)}$ \\
$\sigma_Z$ (pc) $<$  & $^{+0.9\,(+0.4)}_{-0.9\,(-0.4)}$ & $^{+6.5\,(+1.8)}_{-5.4\,(-1.7)}$ & $^{+19.6\,(+6.4)}_{-13.4\,(-5.1)}$ \\



\hline
\end{tabular}
} 
\renewcommand{\arraystretch}{1}
\label{tab:typ-unc}
\tablefoot{
\tablefoottext{a}{Given are the uncertainties for the following parameters: parallax, distance, proper motions in direction of $\alpha$ and $\delta$, tangential velocities in direction of $\alpha$ and $\delta$, radial velocities from Gaia DR3, and XYZ. The RVs are available for a subsample of 367,127 sources (37\%) inside the box (or 4967, 40\%, in the \texttt{SigMA} Sco-Cen selection). For the derived parameters we give lower and upper uncertainty thresholds, since the errors, gained via a Monte Carlo approach, get asymmetric for large parallax errors.}
}
\end{small}
\end{center}
\end{table}


In Sect.~\ref{sec:compare} we compare the \texttt{SigMA} clusters with recent literature samples. To this end we cross-match the samples using the Gaia DR3/EDR3 \texttt{source\_id}. This cross-match is straight forward for the samples of \citet{Schmitt2022}, \citet{Squicciarini2021}, \citet{Luhman2022a}, and \citet{MiretRoig2022b} who used Gaia DR3/EDR3 data, which allows a  direct match with the \texttt{source\_id}. In the case of \citet{Damiani2019}, \citet{Kerr2021}, and \citet{Zerjal2021arxiv}, who used Gaia DR2 data, we first retrieve the Gaia DR3 \texttt{source\_id} using the \texttt{gaiadr3.dr2\_neighbourhood} catalog from the Gaia Archive, since the DR2 and DR3 \texttt{source\_id}s are not generally the same. Such a cross-match delivers few sources that have several possible matches of DR3 with DR2 sources \citep[see][]{Torra2021, Brown2021}. In such cases we choose the closer match, using the provided \texttt{angular\_distance} parameter.

\section{Related work: cluster analysis}\label{app:methods}
In the following we present a cross section of related work upon which \texttt{SigMA} rests. From the vast corpus of data mining and statistics literature we focus specifically on identifying stable groups using density-based clustering methods.    

\subsection*{Hierarchical, density-based clustering}\label{sec:hdbc}
The strength of level-set formulation, see Eq.~(\ref{eq:level_set}) and further discussions in Sect.~\ref{sect:nonparam_clustering}, lies in the natural emergence of a cluster tree, a clustering hierarchy which arises from sweeping the density threshold $\lambda$ from $\infty \to -\infty$. Under a continuous change of $\lambda$, the number of connected components changes when the threshold passes through a critical point in $f$, thus $\nabla f = \Vec{0}$. A new cluster is born when $\lambda$ reaches the height of a mode in $f$. On the other hand, a cluster dies when $\lambda$ traverses a saddle point or a local minimum, in which case the two connected components merge into a single one. The cluster creation and merging process is schematically shown in Fig.~\ref{fig:level_set}. 

However, estimating the connected components of level-sets, while easy in one dimension, gets nontrivial in higher dimensions. Consequently, algorithmic realizations of the \citet{Hartigan:1975} level-set idea rely on graph heuristics and graph theory in which connected components arise naturally. Early implementations by \citet{Azzalini:2007} and \citet{Stuetzle:2010} and subsequent theoretical analyses~\citep{Chaudhuri:2010, Kpotufe:2011,Chaudhuri:2014} adopt a graph $G(\lambda)$ over the data samples where vertices and/or edges are filtered according to $\lambda$, thus $\{\Vec{x} \in X~ : ~ \hat{f}(\Vec{x}) \ge \lambda \}$\footnote{Edges are commonly assigned the minimum density sampled along the path connecting two vertices.}. 

However, the use of graphs to represent the connectivity comes with its own limitations. This scheme guarantees that two samples from one connected component of $G(\lambda)$ are to be found in a connected component in $L(\lambda)$. However, as \citet{Stuetzle:2010} point out, the reverse implication is not necessarily given. This means, samples from the same connected component in $L(\lambda)$ may end up in different connected components of $G(\lambda)$. Since density estimates are inherently noisy, usually too many clusters arise from this iterative filtration procedure. To counteract this over-clustering, the resulting graph cluster tree is usually pruned in a post-processing step during which spurious clusters are identified and merged back into the ``mother cluster''~\citep{Stuetzle:2010, Kpotufe:2011,Chaudhuri:2014}.

\subsection*{The HDBSCAN algorithm} \label{sec:hdbscan}
A well-known algorithm belonging to the family of hierarchical level-set methods is the HDBSCAN algorithm \citep{Campello:2013} (Hierarchical Density-Based Spatial Clustering of Applications with Noise), which recently has been gaining attention in the astronomical community \citep[e.g.,][]{Kounkel2019, Kounkel2020, Hunt2021, Kerr2021}. In order to prevent over-clustering, the authors introduce the \emph{minimum cluster size} parameter which provides an interpretable pruning strategy. 

At each cluster split decision, the smaller cluster created is merged back into the ``mother cluster'' if it has less than \emph{minimum cluster size} points, otherwise, a new cluster is created. To obtain a flat clustering result from the cluster tree, HDBSCAN estimates the stability of a cluster in the hierarchy via the concept of \emph{relative excess of mass} (EOM). Similar to the concept of excess mass~\citep{Muller:1991}, it measures the lifetime and size of a cluster. The heuristic favors more prominent and stable clusters that live longer in the cluster tree. For example, a group that persists for a long time as a single connected component should be preferred over the two small clusters it breaks into and which quickly vanish.

However, the EOM criterion tends to produce too large clusters in practice. If a large group persists in the hierarchy for a long enough time, its children are unlikely to exceed the parent's EOM. Alternatively, the HDBSCAN implementation by \citet{McInnes:2017} offers the opportunity to extract the leaf nodes from the cluster tree. Since the leaf nodes are extracted only considering the \emph{minimum cluster size} criterion, the resulting clusters lack any stability guarantee; thus, the clustering result is highly susceptible to random density fluctuations. In general, these methods suffer from complex and hard-to-interpret pruning procedures and parameters, which affect the confidence and interpretability of the clustering result.

\subsection*{Topological methods} \label{sec:topological-methods}
Extracting a flat clustering from the cluster tree requires a notion of cluster stability. As discussed, the concept of relative excess of mass, which inherently depends on the pruning process, can lead to too coarse clusters. 
A related pruning heuristic comes from considering the topological persistence of each mode in $\hat{f}$, introduced by~\citet{Chazal:2013}. Persistence is defined as the lifespan of each connected component. The notion of persistence is shown to be stable under small perturbations to the initial density $f$~\citep{Edelsbrunner:2000, Zomorodian:2005, Ghrist:2008}. 


A variation on the persistence formulation is proposed by \citet{Ding:2016}, who instead of thresholding the cluster lifetime, use cluster \emph{saliency} $\nu$, defined by the ratio of birth and death density, as a cluster stability criterion. By varying $\nu$ between 0 and 1 the cluster tree is revealed and the most stable and long-lived configuration is chosen as an appropriate clustering result.

While easy to interpret, these stability parameters can get quite tedious to select in practice. In the large data and cluster regime, the separation between stable and unstable clusters becomes less apparent. In these limiting cases, selecting the input parameters again warrants a proper parameter search.

\subsection*{Extracting stable and significant clusters}
Compared to the notion of persistence, there is also growing research to apply statistical methods that test the modality structure of the data. These methods offer the advantage of an interpretable and meaningful parameter $\alpha$, defining the significance level of a corresponding hypothesis test. The null hypothesis H$_0$ commonly assumes that the data, or subsets of it, are sampled from an uni-modal density, whereas the alternative hypothesis H$_1$ suggests multi-modality. The null hypothesis is rejected at a significance level $\alpha$ if the $p$-value from the corresponding test procedure exceeds this significance level.

We identify first applications of hypothesis test procedures in the clustering literature in the context wrapper methods around the \emph{k}-means and EM frameworks. G-means~\citep{Hamerly:2004} employs the Anderson-Darling statistic to test the hypothesis that each cluster is generated from a Gaussian distribution. 
Instead of testing on a per-cluster basis, Pg-means~\citep{Feng:2007} tests the whole Gaussian mixture model (GMM) at once. 
Dip-means~\citep{Kalogeratos:2012} proposes an incremental clustering scheme for selecting $k$ in $k$-means which employs Hartigan’s dip statistic~\citep{Hartigan:1985}. 
In case the distance distribution of one or more points to their co-cluster members exhibit a significant multimodal structure, the cluster is split. 

Skinny-dip~\citep{Maurus:2016} also implements Hartigan's dip test and applies it to one-dimensional linear projections of the data set. 
Distinct density peaks are to be identified based on the gradient of the projected cdf. By projecting the data iteratively into multiple axes, the samples are partitioned into clusters. Skinny-dip is specifically able to handle background noise very well, however, it considers noise samples to be uniformly distributed and clusters to be axis-parallel.

These algorithms, however, are intrinsically tied to convex or Gaussian cluster assumptions. The recently proposed M-dip~\citep{Chronis:2019} is able to deal with arbitrary oriented and shaped clusters, which applies a simulation strategy to approximate values for smallest density dips of uni-modal data sets of the same size and density. However, we do not want to depend on simulations but instead directly obtain a measure of significance from given data.  

\section{Testing for uni-modality}\label{app:modality_test}
Here we highlight the work of~\citet{Burman:2008} more closely, who's modality test procedure we adopt in this work. The modality procedure is tied to the notion of a density dip along a path between two points in the data set. In the following, we aim to define the concept of such a path formally.

\subsection*{A formal description of the test procedure}
We consider directed, continuous paths from $\Vec{x}_1$ to $\Vec{x}_2$ through input space $\mathcal{X}$. By assuming there exists a parametrization $\Vec{r}(t)$, with $t \in [0,1]$, the path becomes the image of $\Vec{r}(t)$. With this map, we can uniquely express every point on the path via the parameter $t$. For example, its start and endpoints are given by $\Vec{x}_1 = \Vec{r}(0)$ and $\Vec{x}_2 = \Vec{r}(1)$, respectively.

Let $f$ be the underlying density function and $\Vec{x}_1$ and $\Vec{x}_2$ two candidate modes of $f$. We assume, without loss of generality, that $f(\Vec{x}_1) < f(\Vec{x}_2)$. If all possible paths undergo a density dip when moving from $\Vec{x}_1$ to $\Vec{x}_2$, both points are found in two distinct modal regions:
\begin{equation}
    \exists \,t \in (0,1) : f(\Vec{r}(t)) < f(\Vec{x}_1)  
\end{equation}

Conversely, if we can find a path between $\Vec{x}_1$ and $\Vec{x}_2$ where all points have a higher density than $\Vec{x}_1$, both points are part of the same modal region:
\begin{equation}\label{eq:1_app1}
    f(\Vec{r}(t)) \ge f(\Vec{x}_1) \quad  \forall \,t \in (0,1]
\end{equation}
Eq. (\ref{eq:1_app1}) describes the case of single-modality, which constitutes the null hypothesis we aim to reject. For general pairs of modal candidates it becomes:
\begin{equation}\label{eq:2_app1}
    f(\Vec{r}(t)) \ge \text{min}( f(\Vec{x}_1), f(\Vec{x}_2) ) \quad  \forall \,t \in (0,1]
\end{equation}
An equivalent and useful formulation is obtained by taking the logarithm on both sides; after that the left side is subtracted from the inequality.
\begin{equation}
    \label{eq:2_app_loghypothesis}
      \text{SB}(t) := -\text{log} f(\Vec{r}(t)) + \text{min}( \text{log} f(\Vec{x}_1),~\text{log} f(\Vec{x}_2) ) \\
\end{equation}
Using the variable $\text{SB}(t)$, we can formulate the null hypothesis as follows:
\begin{equation}\label{eq:3_app_h0}
    \text{H}_0 : \text{SB}(t) \le 0 \qquad \forall~t \in (0,1)
\end{equation}
Rather than testing H$_0$ across the full path a pointwise test H$_{0,t} : \text{SB}(t) \le 0$ for some values of $t$ is employed.

Since we do not have access to the underlying density $f$, we cannot test the hypothesis in Eq. (\ref{eq:3_app_h0}) directly. Instead, we have a data set of $d$-dimensional random variables drawn from $f$. Given proper normalization of the coordinate axes (see Sect.~\ref{scaling-factors}), \citet{Burman:2008} show that the following expression is asymptotically standard normal distributed and converges -- up to a constant factor -- to SB$(t)$ as the number data samples approaches infinity:
\begin{equation}
    \label{eq:4_app}
    \widehat{\text{SB}}(t) = d \sqrt{k/2} \left[ \text{log}~d_k(\Vec{r}(t)) - \text{max} ( \text{log}~d_k(\Vec{x}_1),~\text{log}~d_k(\Vec{x}_2) \right]
\end{equation}
Here $d_k(\Vec{x})$ denotes the distance to the $k$'th nearest neighbor of the point $\Vec{x}$. The distance is an approximation to the density $f$. Due to their inverse proportionality the sign is flipped between Eq. (\ref{eq:2_app_loghypothesis}) and Eq. (\ref{eq:4_app}); and the minimum is replaced with the maximum function.

Since the corresponding test statistic $\widehat{\text{SB}}(t)$ is approximately standard normally distributed, the null hypothesis is rejected at significance level $\alpha$ if 
\begin{equation}
    \label{eq:5_reject_cond}
    \widehat{\text{SB}}(t) \ge \Phi^{-1}(1-\alpha)
\end{equation}
where $\Phi$ is the standard normal cdf. Therefore, if any $t \in (0,1)$ fulfills condition~(\ref{eq:5_reject_cond}), H$_0$ is rejected. 

Due to the employment of $k$ nearest neighbor technique, this test procedure applies naturally to multivariate data without the need of projecting the data onto a one-dimensional line, as is the case for most modality tests. Furthermore, nearest neighbor queries have access to very efficient algorithms such as the kd-tree~\citep{Bentley:1975} which reduces neighbor searches to only $\mathcal{O}(\text{log}N)$ distance computation. Thus, these considerations allow us to study the modality structure of the data set at Gaia data scales without careful projection loss considerations. 

\begin{figure}[ht!]
    \centering
    \includegraphics[width=\columnwidth]{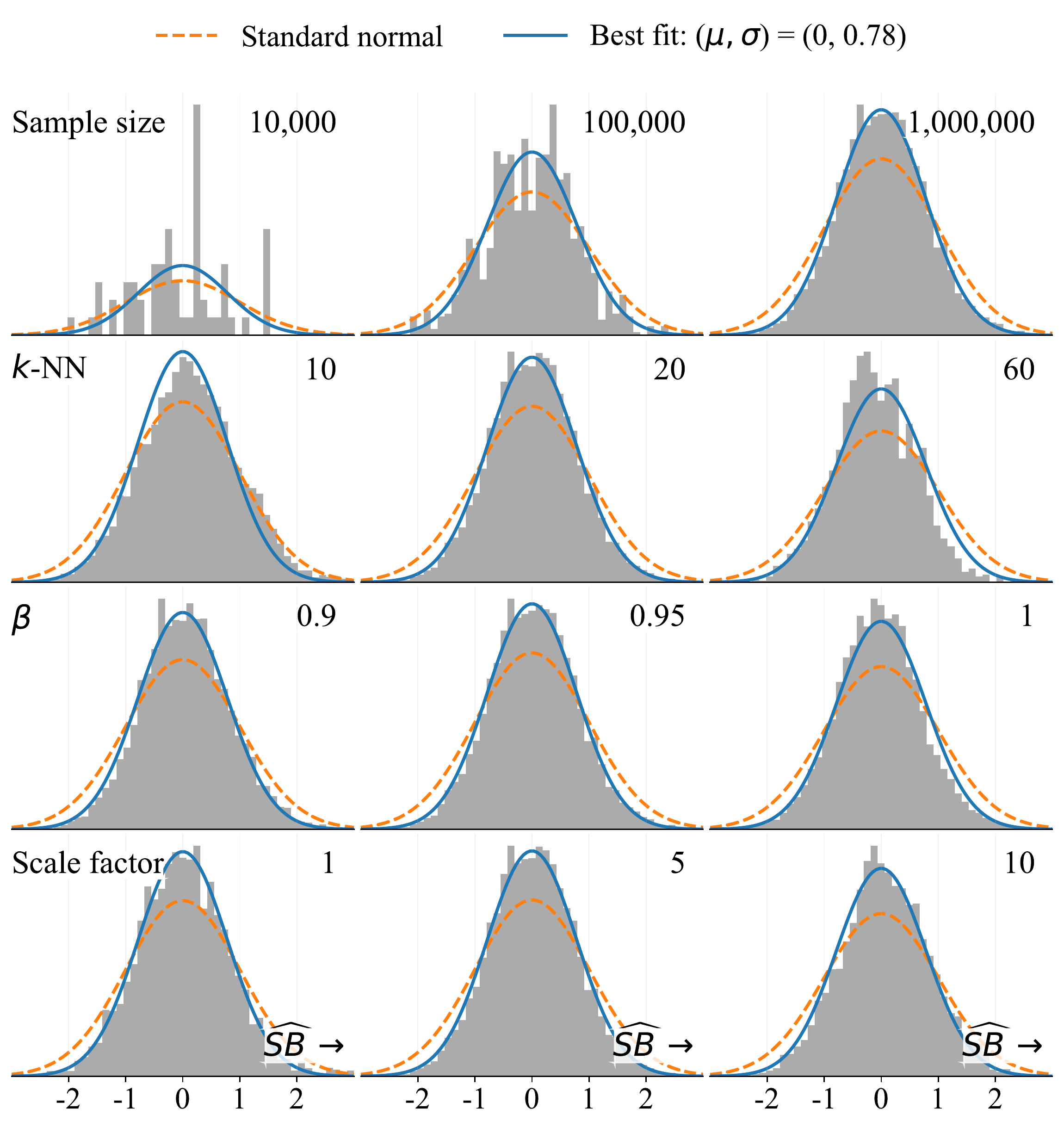}
    \caption{Test statistic distribution $\widehat{\text{SB}}$ of our reduced test procedure on a uni-modal data set. The distribution is stable under variation of different parameters and respective values (as shown in corresponding rows and columns).}
    \label{fig:test_statistic}
\end{figure}

\subsection*{Empirical results}
\citet{Burman:2008} describe the iterative application of the test procedure to modal candidates to cluster the data into significant modal regions. However, the test is employed along the straight line path connecting two modes, which limits the procedure to convex cluster shapes only. Moreover, to detect significant dips reliably, enough samples need to be tested along the path. 

We provide a natural extension to the presented procedure, which applies to arbitrary cluster shapes while reducing the number of test evaluations to a single one. In Sect.~\ref{sec:sigma} we describe the modification in more detailed and argue that the reduction of the original method to a single pointwise evaluation at the saddle point leaves the test unchanged. 

To substantiate this statement, we empirically validate our method on simulated data. In particular, we aim to show that these changes do not affect the distribution of the test statistic under the null hypothesis. According to \citet{Burman:2008} the test statistic $\widehat{\text{SB}}$ is standard normally distributed under $H_0$, which assumes uni-modality between two points in the input space. 

Figure~\ref{fig:test_statistic} shows the test statistic distribution $\widehat{\text{SB}}$ of our reduced test procedure on a uni-modal data set. To faithfully test the distribution under H$_{0}$ we require a uni-modal data set with Gaia and Sco-Cen-like positions and kinematic properties, realistic errors, and suitable size. Since the Gaia eDR3 mock catalog~\citep{Rybizki2020} reproduces kinematic features found in the Milky Way, its uni-modality in guaranteed. 

Instead, we directly use data within the box defined around Sco-Cen in Eq.(\ref{eq:box}). To remove any local over-densities we shuffle the observations randomly in each respective feature individually. This procedure leaves marginal distributions unchanged while fully decorrelating the data. Since all marginal distributions are uni-modal, the joint distribution (which is implicitly constructed as a factorization of marginals) is equally uni-modal. 

To gauge parameter effects on the test statistic distribution we vary the sample size and three \texttt{SigMA} parameters across different values\footnote{While varying one parameter, the remaining ones are set to their default values as described in Sect.~\ref{sect:parameter_selection}}. The parameters are, the overall sample size, the $k$ parameter of $k$-nearest neighbor density estimation method (see Sect.~\ref{sec:density_est}), the $\beta$ parameter of the underlying $\beta$-skeleton graph (see Sect.~\ref{sec:graph}), and the velocity scaling factor (see Sect.~\ref{scaling-factors}). 

As shown in Fig.~\ref{fig:test_statistic}, varying the given parameters within a sensible range does not modify the test statistic distribution. Due to its stability, a single test statistic distribution under H$_0$ can be universally assumed over different parametrizations of \texttt{SigMA}.  

The distribution of $\widehat{\text{SB}}$ on uni-modal data follows closely a zero-mean Gaussian. However, in our tests, the standard deviation differs slightly from unity as stated by~\citet{Burman:2008}. We update its value in our tests to the average standard deviation across our test of $\sigma = 0.78$.

\section{Scaling factor distribution}\label{app:scaling_factors}
Here we discuss the derivation of the scaling factor distribution, which we use to weigh the velocity sub-space in the clustering process. For a more detailed motivation see Sect.~\ref{scaling-factors}.

We replace the scaling factor variable $c_v$ with $y$ to simplify and shorten the reading flow. Additionally, compared to the main text, we denote the distance to a cluster with $r$ instead of $d$. This notation makes the integration alongside the differential $dr$ easier to read (otherwise the differential would be $dd$). 

Our goal is to obtain the distribution $f(y \mid r_0 \le r \le r_1)$, which describes the behavior of the scaling factor $y$ for a given range of distances to groups of interest. A simple way to find this distribution is to interpret the empirical linear model $g(r)$ and associated Gaussian uncertainties as an improper probability function $f(r,y)$\footnote{Since the marginal distribution $f(r) \propto 1$ is a uniform distribution over $\mathbb{R}^{+}$, the joint distribution $f(r,y)$ is improper as it does not integrate to unity.}.

As we are dealing with an improper pdf, we consider the following proportionality condition and handle the normalization of the left hand side later.
\begin{align}
    \begin{aligned}
    f(y \mid r_0 \le r \le r_1) &\propto \int_{r_0}^{r_1} f(r,y) \,dr \\
                                &\propto \int_{r_0}^{r_1} f(y \mid r) f(r) \,dr
    \end{aligned}
\end{align}

Since $f(r) \propto 1$ is independent on the distance $r$ we can add it to the yet unknown constant normalization factor and move it out of the integral. Hence we can write the target distribution as:
\begin{equation}\label{eq:app_cond_dist_1}
    f(y \mid r_0 \le r \le r_1) \propto \int_{r_0}^{r_1} f(y \mid r) \,dr
\end{equation}

\begin{figure}[t]
    \centering
    \includegraphics[width=0.98\columnwidth]{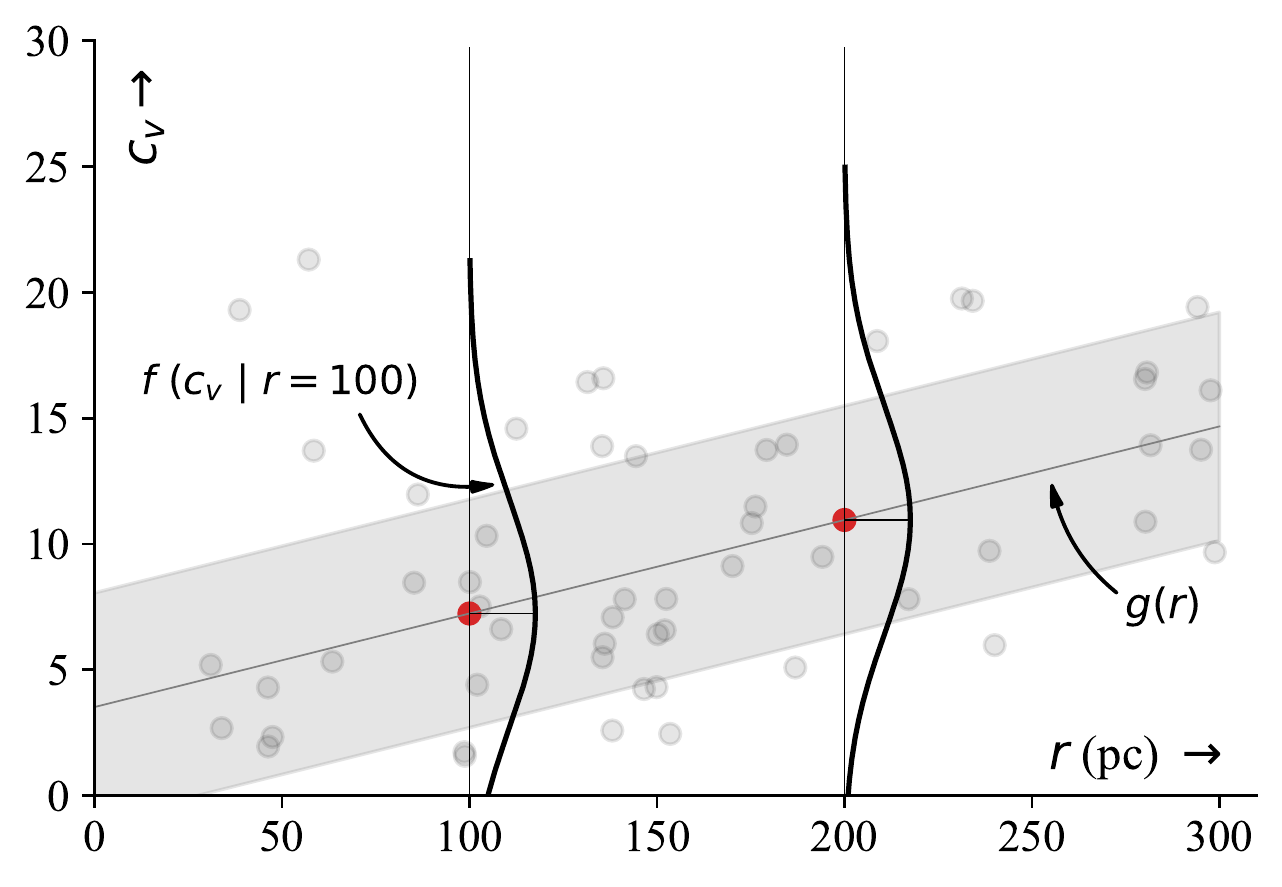}
    \caption{Scaling factor determination via the empirical distance-scaling relationship. The scaling factor distribution for groups at a distance between $100 - 200$ pc depends on the conditional distribution of scaling factors at a given distance $f(c_x/c_v \mid r)$.}
    \label{fig:probablity_integration}
\end{figure}

Thus, to obtain an analytic solution to Eq. (\ref{eq:app_cond_dist_1}) we need an expression for the conditional pdf $f(y \mid r)$. Assuming that the data are Gaussian distributed around the linear model with a constant standard deviation $\sigma$\footnote{We observe the standard deviation to be approximately constant for the range of interest; $r \in [100, 200]$}, we can write the following expression:
\begin{equation}\label{eq:app_cond_dist_2}
    f(y \mid r) \propto \text{exp} \left( - \frac{(y - g(r))^2}{2\sigma^2} \right) 
\end{equation}
Figure~\ref{fig:probablity_integration} schematically shows the integrating process where the conditional pdfs $f(y \mid r)$ are shown for $r=100$ and $r=200$.

By substituting Eq.~(\ref{eq:app_cond_dist_2}) into Eq.~(\ref{eq:app_cond_dist_1}) and solving the integral we obtain:
\begin{equation}\label{eq:app_cond_dist_3}
    f(y \mid r_0 \le r \le r_1) \propto \text{erf} \left( \frac{y - g(r_0)}{\sqrt{2}\sigma} \right) - \text{erf} \left( \frac{y - g(r_1)}{\sqrt{2}\sigma} \right)
\end{equation}
where erf($x$) is the error function. To normalize the probability density in Eq. (\ref{eq:app_cond_dist_3}), we compute its integral. Since both summands are of the same type, we only need to solve the following integral:
\begin{equation}\label{eq:app_integral_erf_1}
    I(c) = \int_{-\infty}^{+\infty} \text{erf} \left( \frac{y - c}{\sqrt{2}\sigma} \right) \,dy
\end{equation}
The variable $c$ represents the constants $g(r_0)$ and $g(r_1)$. The integral in Eq. (\ref{eq:app_integral_erf_1}) evaluates to:
\begin{equation}\label{eq:app_integral_erf_2}
    I(c) = \text{exp} \left( \frac{-(y - c)^2}{2 \sigma^2} \right) \sqrt{\frac{2 \sigma^2}{\pi}} 
    + (y - c)~\text{erf} \left( \frac{y - c}{\sqrt{2} \sigma} \right) ~\Biggr|_{-\infty}^{+\infty}
\end{equation}
Thus, the integral of Eq. (\ref{eq:app_cond_dist_3}) can be expressed in the following form:
\begin{equation}\label{eq:app_integral_erf_3}
        \int_{-\infty}^{+\infty} f(y \mid r_0 \le r \le r_1) = n\times[I(g(r_0)) - I(g(r_1))] \overset{!}{=}1
\end{equation}
The factor $n$ represents the normalization factor. Rearranging the resulting terms by function type, we get the following:
\begin{align}
    \begin{aligned}\label{eq:scaling_derivation_integral_1}
    I(g(r_0)) - I(g(r_1)) = h(y) + l(y) ~\Biggr|_{-\infty}^{+\infty}
    \end{aligned}
\end{align}
The functions $h(y)$ and $l(y)$ describe a sum of exponential and error functions, respectively. The functions are defined in the following:
\begin{align}
    \begin{aligned}\label{eq:h_l_definition}
        h(y) &=  \sqrt{\frac{2 \sigma^2}{\pi}} \left[ \text{exp} \left( \frac{-(y - g(r_0))^2}{2 \sigma^2} \right) - \text{exp} \left( \frac{-(y - g(r_1))^2}{2 \sigma^2} \right)  \right] \\
        l(y) &= (y - g(r_0))~\text{erf} \left( \frac{y - g(r_0)}{\sqrt{2} \sigma} \right) - (y - g(r_1))~\text{erf} \left( \frac{y - g(r_1)}{\sqrt{2} \sigma} \right)
    \end{aligned}
\end{align}
We evaluate the summands of this primitive integral at the border individually. First, sum Gaussians in $h(y)$ goes to zero as $y$ approaches negative and positive infinity:
\begin{equation}
    \lim_{y \to \pm \infty} h(y) = 0
\end{equation}
The sum of error functions can be rearranged into the following form:
\begin{align}
    \begin{aligned}
        l(y) &= y \left[ \text{erf}\left(\frac{y-g(r_0)}{\sqrt{2}\sigma}\right) - \text{erf}\left(\frac{y-g(r_1)}{\sqrt{2}\sigma}\right) \right]\\ 
        &+ g(r_1)~\text{erf}\left(\frac{y-g(r_1)}{\sqrt{2}\sigma}\right)
        - g(r_0)~\text{erf}\left(\frac{y-g(r_0)}{\sqrt{2}\sigma}\right)
    \end{aligned}
\end{align}
Evaluating $l(y)$ at the borders results in the following:
\begin{equation}
    \lim_{y \to \pm \infty} y \left[ \text{erf}\left(\frac{y-g(r_0)}{\sqrt{2}\sigma}\right) - \text{erf}\left(\frac{y-g(r_1)}{\sqrt{2}\sigma}\right) \right] = 0
\end{equation}
and
\begin{align}
    \begin{aligned}
        \lim_{y \to \pm \infty} &\left[ g(r_1)~\text{erf}\left(\frac{y-g(r_1)}{\sqrt{2}\sigma}\right)
            - g(r_0)~\text{erf}\left(\frac{y-g(r_0)}{\sqrt{2}\sigma}\right) \right] = \\
            &= \pm (g(r_1) - g(r_0))
    \end{aligned}
\end{align}
This last term is the only non-zero contribution to the integral. Its evaluation at the lower edge results in the same but negative value to the upper edge. Thus, the area under the curve is twice that value. The normalization factor $n$ then becomes the following:
\begin{align}\label{eq:app_integral_erf_5}
    \begin{aligned}
        I(g(r_0)) - I(g(r_1)) &= 2 [g(r_1) - g(r_0)]\\
        n &= \frac{1}{2 [g(r_1) - g(r_0)]} =: \frac{1}{2 \Delta g}
    \end{aligned}
\end{align}
The function value difference, $\Delta g$, is always positive since $g(r)$ is a strictly monotonically increasing function and $r_1 > r_0$, see Fig.~\ref{fig:probablity_integration}. Thus, $n$ is a proper normalization factor that is non-zero and positive for all pairs $r_0$ and $r_1$.
Using this normalization constant, the conditional pdf can be written as:
\begin{align}\label{eq:app_cond_dist_final}
    f(y \mid r_0 \le r \le r_1) = \frac{1}{2 \Delta g} \left[ \text{erf} \left( \frac{y - g(r_0)}{\sqrt{2}\sigma} \right) - \text{erf} \left( \frac{y - g(r_1)}{\sqrt{2}\sigma} \right)
    \right]
\end{align}

\begin{figure}[!t]
    \centering
    \includegraphics[width=0.98\columnwidth]{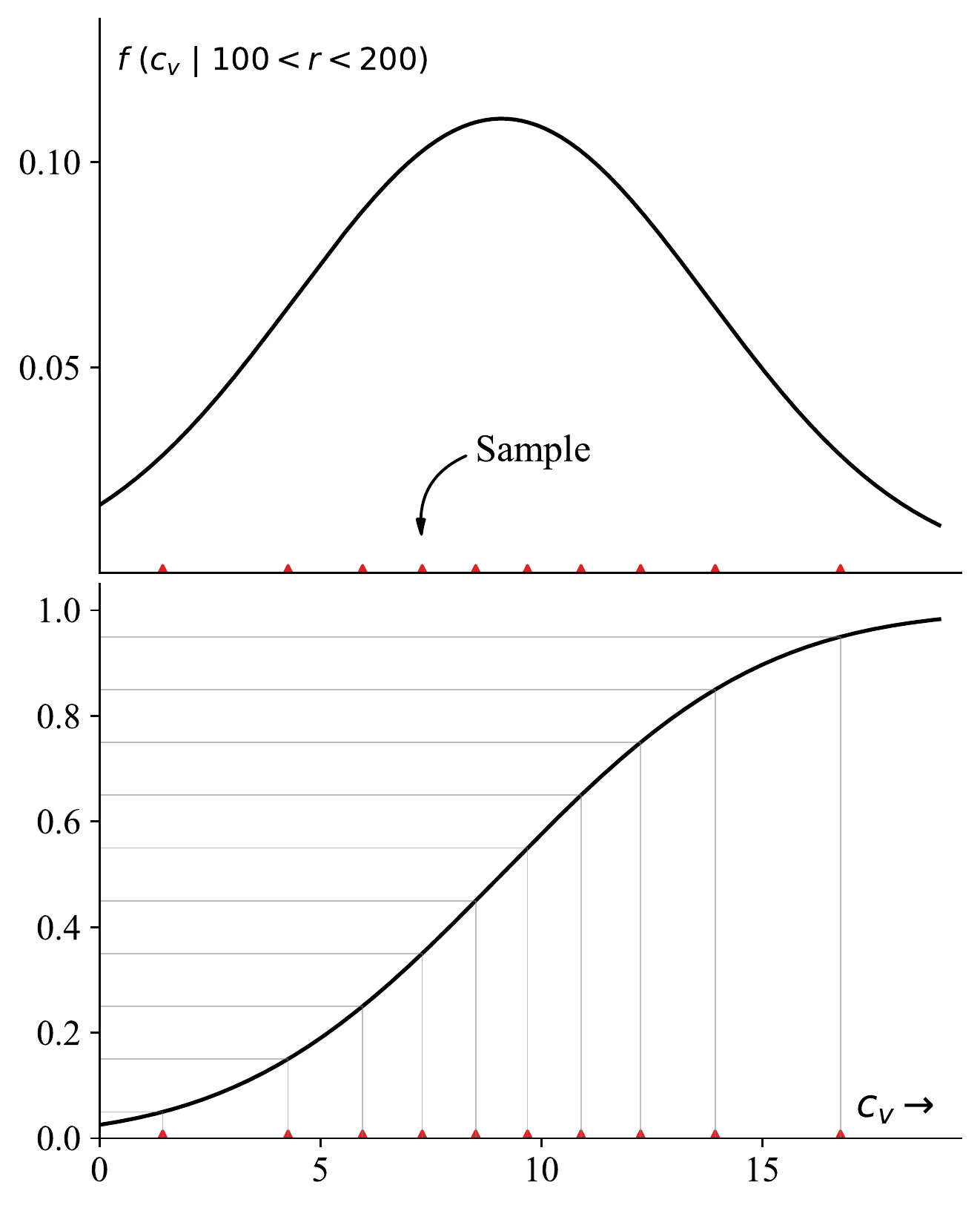}
    \caption{Pdf and cdf of scaling factor conditioned on a given range of distances. The ten red scatter points indicate samples drawn from the 10-quantile splitting procedure. We separate the pdf into 10 continuous intervals with equal probabilities from which we derive samples as the mean position of these intervals.}
    \label{fig:pdf_cdf}
\end{figure}

The top part of Fig.~\ref{fig:pdf_cdf} shows the resulting pdf when applying Eq. (\ref{eq:app_cond_dist_3}) to sources in Sco-Cen, where we assume a distance range of $r \in [100, 200]$. Here we can see an immediate caveat of our simple constant model uncertainty assumption; the resulting distribution has infinite support, thus, non-zero probability density for $f(y<0 \mid r)$. Although physically meaningless, the total probability of such events is small and, as seen below, does not drastically influence the final sample set.

We consider sampling strategies to obtain a set of scaling factors to use in the clustering process. Random sampling can generate almost identical realizations, so the possible solution space might not be covered evenly. Since we need to perform a separate clustering, run for each sample drawn, keeping the number as small as possible is essential. To cover the space evenly while considering the underlying probability distribution, we select a set of 10 samples that represent 10 quantiles of the pdf. We separate the pdf into 10 continuous intervals with equal probabilities from which we derive samples as the mean position of these intervals.

To compute the quantiles, we determine the cdf by solving the integral over the conditional pdf in Eq. (\ref{eq:app_cond_dist_final}); using functions $h(y)$ and $l(y)$ defined in Eq. (\ref{eq:h_l_definition}) the cdf becomes:
\begin{align}
    \begin{aligned}\label{eq:cdf}
    F(y \mid r_0 \le r \le r_1) &= \int f(y \mid r_0 \le r \le r_1) \,dy \\
    &= \frac{1}{2 \Delta g} [h(y) + l(y)] + C
    \end{aligned}
\end{align}
To obtain a proper cdf from Eq. (\ref{eq:cdf}), we set the constant of integration $C$ to $1/2$. Thus, the cdf becomes:
\begin{equation}
    F(y \mid r_0 \le r \le r_1) = \frac{1}{2} \left(1 + \frac{1}{\Delta g} [h(y) + l(y)] \right)
\end{equation}

The cdf defined in Eq. (\ref{eq:cdf}) for $r \in [100, 200]$ is shown in the bottom part of Fig.~\ref{fig:pdf_cdf}. The ten red scatter points\footnote{The velocity scaling values are:\\ $c_v = \{ 1.43,  4.26,  5.95,  7.3 ,  8.51,  9.68, 10.89, 12.24, 13.93, 16.77 \}$.} indicate samples drawn from the 10-quantile splitting procedure where horizontal lines indicate equal probability intervals. To invert the cdf and obtain scaling fraction samples from $F^{-1}(y \mid r_0 \le r \le r_1)$ we used a numerical approximation\footnote{We made use of the open source library \texttt{pynverse v0.1.4.4} to calculate the numerical inverse of the cdf.}.

\section{Consensus clustering}\label{app:consensus_clustering}
\begin{figure}
    \centering
    \includegraphics[width=0.93\columnwidth]{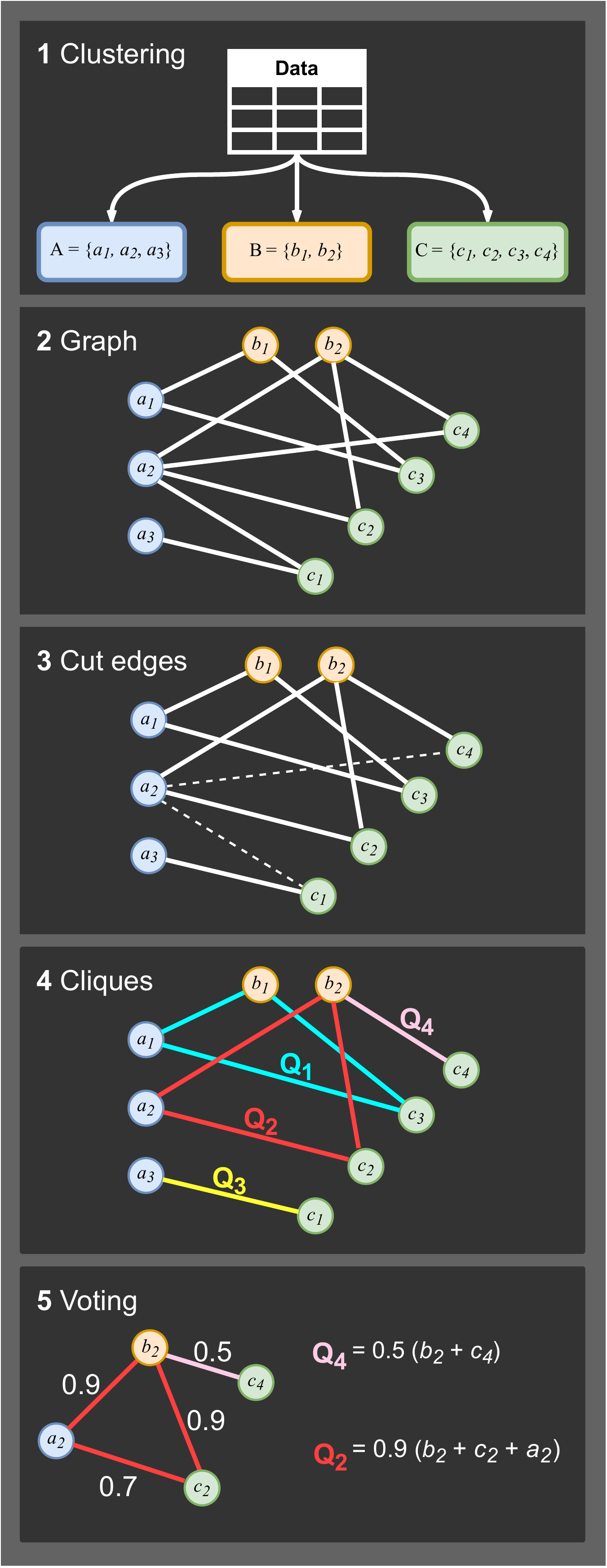}
    \caption{Consensus clustering pipeline for a simple example using an ensemble of three clustering solutions $A$, $B$, and $C$. (2) Clusters from the ensemble are linked in a graph based on overlapping points. (3) Edges between clusters are removed if the overlap between their members is insufficient to assume a common cluster solution. (4) Cliques are extracted representing stable cluster solutions, i.e., consensus clusters. (5) A voting strategy determines the assignment of individual sources to cliques.}
    \label{fig:consensus_pipeline}
\end{figure}

In this section we discuss algorithmic means to identify stable cluster solutions from an ensemble of clustering results. To identify robust solutions, we represent all clusters across the clustering ensemble in a graph. Each cluster of a single solution from the ensemble is represented by one node. We connect two nodes via an edge if two respective clusters share at least one common point. Figure~\ref{fig:consensus_pipeline} highlightes this step in the first two frames. The ensemble is comprised of three clustering solutions. Its individual runs $A$, $B$, and $C$ contain three, two, and four clusters, respectively. Each source is classified into a single group for a given run, resulting in disjoint sets. Thus, edges connect clusters from different runs.

Edges in the graph are weighted by the corresponding Jaccard similarity, which measures their common overlap. The Jaccard similarity is defined as the ratio of the intersection of two sets over their union. Typically, a value of $0.5$ or greater indicates high similarity between two sets. Since this linkage is fundamental for determining the consensus result, it should avoid connecting dissimilar clusters. Thus, we remove edges with a weight below $0.5$ (see Fig.~\ref{fig:consensus_pipeline}, step 3). This threshold is quite conservative, as it can separate even similar cluster solutions, e.g., if one cluster is a subset of the other. To avoid over-pruning the graph, we relax the cut criterion in the following way. Two clusters $a$ and $b$ with respective sizes $n_a$ and $n_b$, where $n_a > n_b$, are linked if the $n_b$ densest points of $a$ amount to a Jaccard similarity of $0.5$ or greater with points from cluster $b$. This criterion guarantees connectivity between cluster extractions at different isosurface thresholds while separating ties to clusters that randomly fragment into multiple subclusters. 

Robust clusters that exist throughout the ensemble will have strong connections to their counterparts from different runs while having none or very weak connections to other nodes in the graph. Thus, a robust cluster solution builds a strong clique in this graph\footnote{A clique in a graph is a subset of nodes that are all connected with each other. Thus, every pair of nodes in a clique is joined  by an edge.}. In contrast, unstable clusters will have many weak connections to many other clusters from different runs but none or very few strong ones; see Fig.~\ref{fig:consensus_pipeline}, panel 4. 

To extract all stable cluster solutions, we aim to identify all strongly connected cliques in the graph, i.e., the consensus result. Since individual sources and clusters can be part of several cliques we employ a voting strategy to determine the final data partitioning~\citep{Vega:2011}. Each clique is represented as a vector of length $N$, where $N$ is the number of sources in the data set.
The $N$ values correspond to the sum of the individual clusters represented as 0--1 vectors, where all entries are 1 for sources inside a cluster and 0 elsewhere. Each source is then associated to a single clique by maximizing the respective entry at the source's position (in the vector). Thus, the larger a clique, the more likely it wins a vote. To favor robust cluster solutions, we multiply each vector by the median of its connection strengths. This number is maximized if the cluster is unchanged throughout different runs. This step is summarized in panel 5 in Fig.~\ref{fig:consensus_pipeline}.

\section{Testing the independence assumption of resampled Gaia data}\label{app:resampling}
When testing for multimodality, \texttt{SigMA} simulates multiple realizations of the Gaia data set to see how robust density dips between pairs of neighboring peaks are. In Sect.~\ref{sec:uncertainty}, we discuss how these realizations are used in a global hypothesis test that determines if samples of pairwise adjacent density enhancements are part of a single underlying mode or not. This global hypothesis test combines individual tests from each realization and evaluates the global null hypothesis that no $p$-value is significant. 

However, the choice of combining different $p$-values is affected by the correlation structure between distinct tests. Typically, statistical tests require independence across tests to guarantee proper levels of specificity and sensitivity. In the case of the commonly used Fisher's combination test~\citep{fisher:1934}, positively correlated $p$-values increase the chance of type I errors (rejects a null hypothesis that is actually true). To investigate the independence assumption of $p$-values across resampled data sets we consider the effects that resampling has on the modality test procedure. 

The modality test is fully described by the dimensionality of the data and most importantly by the $k$-distances of modal and saddle points. Resampling the data set changes the latter. The sampling of new data points is done with Gaussians centered on mean astrometric observables with heteroscedastic error covariance matrices available in the Gaia database. We aim to use the covariance matrix in relation to the nearest neighbor distance to estimate the interdependence of $p$-values. 

As the entries of the covariance matrix shrink to zero, the resampled data points converge to the original data points until they eventually coincide. At the same time, the $p$-values across different data sets approach the same value leading to a perfect correlation between them. On the other hand, if the standard deviation along its principle axis (or any other direction for that matter) extends far beyond a point nearest neighbor or even its $k$-distance, the resampled data differs substantially from the original one, decorrelating the $p$-values. 

\begin{figure}[htbp]
    \centering
    \includegraphics[width=0.98\columnwidth]{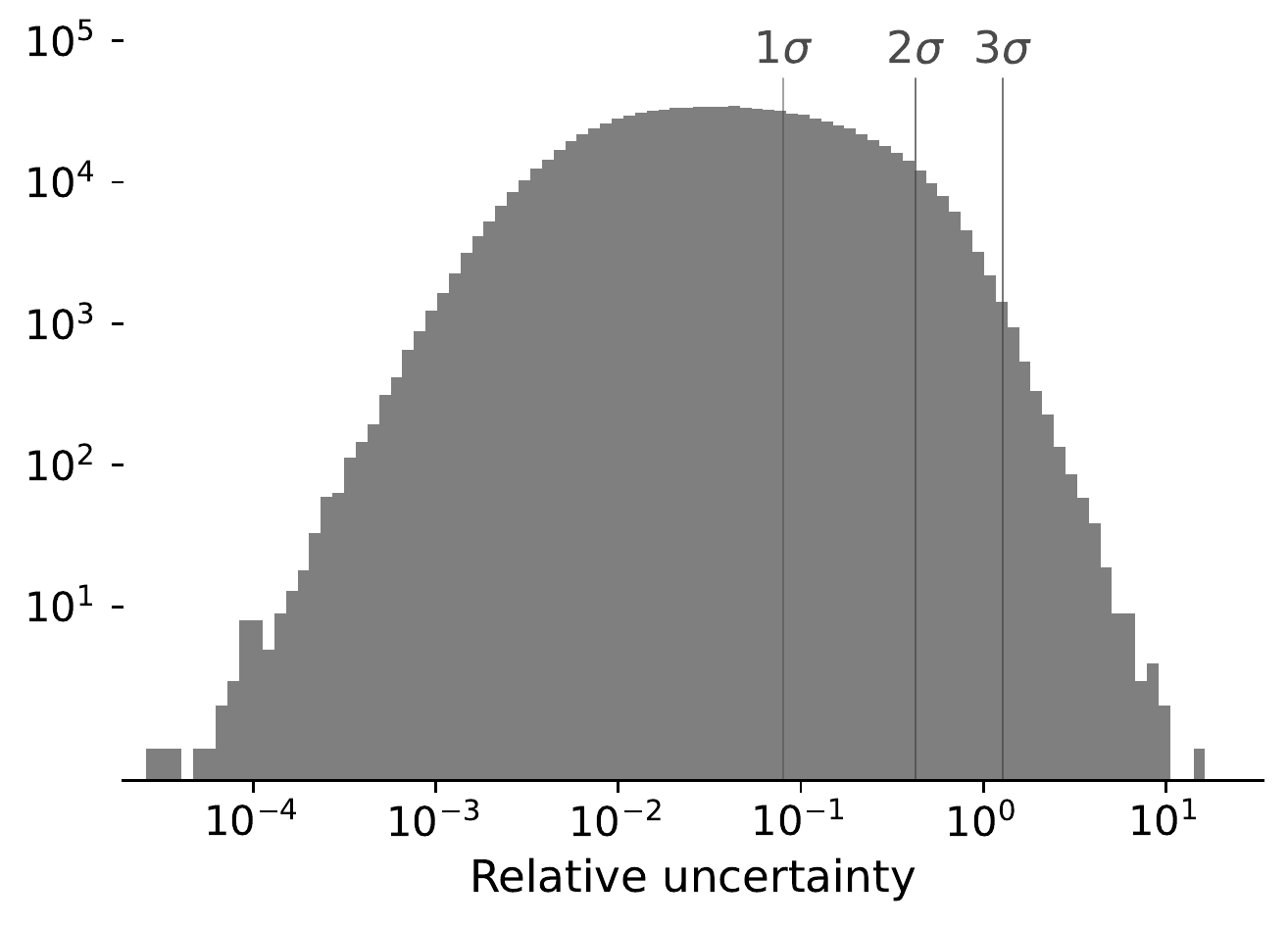}
    \caption{Log-log histogram showing the relative uncertainty of data points, i.e., the ratio of positional uncertainty given by the error covariance matrix over the nearest neighbor distance. The majority of data points are far below unity. The concentration of relative uncertainty values at zero indicates that $k$-distances across resampled data set are strongly correlated. The $1 \sigma,\,2 \sigma,\,3 \sigma$ lines indicate the percentiles containing 68, 95, and 99.7\% of the respective distribution.}  
    \label{fig:error_vs_nn}
\end{figure}

Figure~\ref{fig:error_vs_nn} illustrates this relationship for data in the Sco-Cen box. The histogram shows the relative uncertainty of data points, i.e., the ratio of positional uncertainty given by the error covariance matrix over the nearest neighbor distance. The distances are computed in the space of observed astrometric quantities where the uncertainties are assumed Gaussian. The majority of data points are far below unity, i.e., errors are small relative to their absolute values. The majority (68\%, i.e., $1 \sigma$) has a relative uncertainty below $0.08$ (see the 68--95--99.7\% percentiles in Fig.~\ref{fig:error_vs_nn} marked as $1 \sigma,\,2 \sigma,\,3 \sigma$, respectively). The concentration of relative uncertainty values at zero indicates that $k$-distances across resampled data set are strongly correlated. Thus, we cannot assume independence of $p$-values across different samples.

\section{Distribution of point-wise densities}\label{app:dist_density}
Instead of directly recovering clusters from phase space data, \texttt{SigMA} extracts modal regions which are a mixture of field stars and cluster members. To separate signal from background, we employ a density-based classifier that selects cluster members as an over-density over background. To automatically determine a density threshold we consider the distribution of field stars and cluster members in univariate density. 

As discussed in Sect.~\ref{noise-removal}, we approximately can the treat the field star content as a uniform distribution locally around a cluster in phase space. A uniform distribution in phase space translates to a Gaussian distribution in 1D density space (see Sect.~\ref{noise-removal} for a more detailed discussion). 

\begin{figure}[htbp]
    \centering
    \includegraphics[width=0.98\columnwidth]{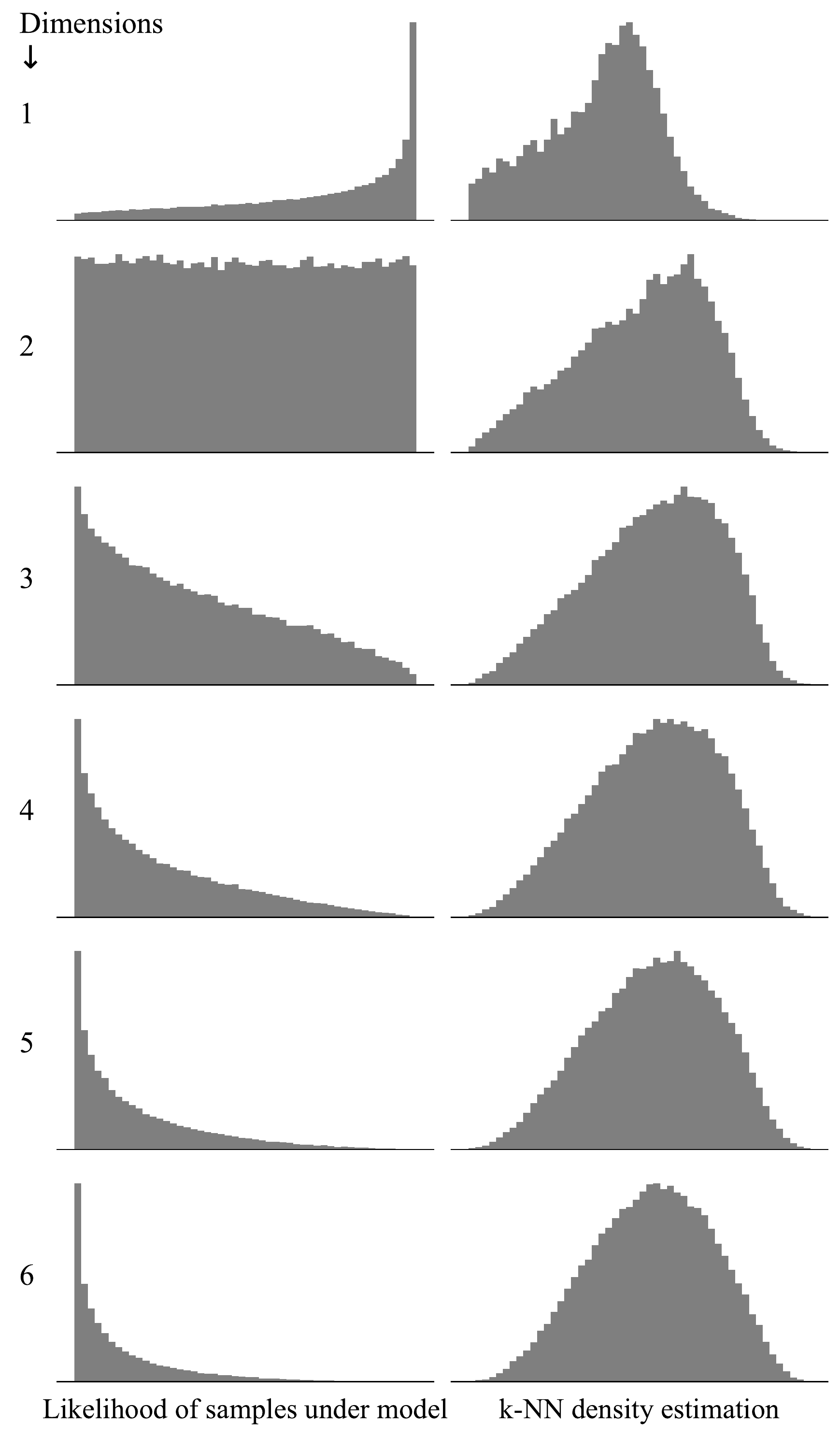}
    \caption{Distribution of point-wise density estimates of samples drawn from an $N$-dimensional Gaussian. The left column shows the likelihood of individual samples. It provides an assessment of the local density under the true model. The right column shows estimated point-wise densities by employing the $k$-NN technique. By increasing the dimensionality $N$ from one to six (top to bottom row) the distribution of point-wise densities approaches a Gaussian.}
    \label{fig:pdf_dist}
\end{figure}

Cluster members are commonly modeled as multivariate Gaussians \citep[e.g.][]{Gagne2014, Sarro2014, Crundall2019, Riedel2017}. Although observational findings point to more complex morphologies \citep[e.g.][]{MeingastAlves2019, Roeser2019, Jerabkova2021, Meingast2019, Kounkel2019, Cantat-Gaudin2019a, Wang2021, Coronado2022}, the Gaussianity assumption provides a good starting point to consider the point-wise univariate density distribution. Figure~\ref{fig:pdf_dist} shows multiple distributions of point-wise density estimates of $100,000$ samples drawn from an $N$-dimensional Gaussian\footnote{The Gaussian is chosen to have zero mean and identity covariance matrix}. The left column shows the likelihood of individual samples. It provides an assessment of the local density under the true model. The number of samples with a relatively high likelihood decreases exponentially as the dimension $N$ increases from one (top row) to six (bottom row).   

Neighborhood queries in high dimension are plagued by the curse of dimensionality. As the dimensionality grows, points are increasingly isolated making neighborhoods no longer local. This effect can already be seen for moderate dimensions in the right column of Fig.~\ref{fig:pdf_dist}. It shows point-wise density estimates obtained via the $k$-NN technique. Around the fourth dimension, the distribution of $k$-distances starts to converge to a normal distribution, which incorrectly suggests an underlying uniform distribution in $N$-dimensional input space. 

We compute point-wise density estimations in five and six dimensions. Although we cannot specifically write down a generative model for stellar groups in phase space, we can assume that neighborhood queries are strongly affected by the given dimensionality. Thus, we model both background and signal contributions as Gaussians in univariate density space.

\section{Parameter optimization}\label{app:grid_search}
The parameter choices of our proposed \texttt{SigMA} analysis pipeline are tuned to Gaia data (see Sect.~\ref{sect:parameter_selection}). In contrast, DBSCAN and HDBSCAN, which we use to compare and test our clustering technique, are general clustering techniques whose parameters we have to set. Instead of using subjective, error-prone prior knowledge to determine suitable parametrizations, we search the space of possible clustering results for the best result. This strategy provides a measure of the peak performance (in case of comprehensive/absolute prior knowledge) a clustering algorithm can achieve. Thus, a comparison against the best results allows for a discussion on methodological advantages and disadvantages rather than reflecting poor parameter selection. 

Compared to the original data set (see Sect.~\ref{sec:open_cluster_sample}), we generate 10 different realizations of the ``compact cluster sample'' to estimate an average performance across random configurations (see. Sect.~\ref{sec:dense_clusters}). Optimal parametrization can fluctuate between different data realizations in the compact cluster sample case. In reality however, the compact cluster sample is a single cluster region in which a unique parametrization should be applied to. Thus, after the parameter search on the 10 individual data samples, we take the median of the resulting parameters and re-run the algorithm on every sample to obtain a clustering score to report in Table~\ref{tab:results_simulation_oc_and_dense}.

To search the space of possible model configurations we employ a grid search in which we evaluate the clustering algorithm on a regular grid in parameters space. Compared to other parameter tuning methods such as random search or Bayesian optimization, a grid search has significant benefits in this scenario. First, a large portion of the parameter space is discrete. Thus, there is a finite grid step size that can cover the entire parameter space in these sub-spaces. Second, the parameter spaces are low dimensional (maximally four-dimensional) allowing densely spaced samples in each parameter axis. Third, compared to random search and Bayesian optimization, a grid search is deterministic and, thus, provides reproducibility. Further, as the grid is predetermined (compared to Bayesian optimization) it allows to fully parallelize the computation to guarantee dense sampling in reasonable times.

In the following subsections, we describe algorithm-specific parameter choices of the employed grid search. For each algorithm, we also search optimal values for the velocity scaling factor. We adopt values discussed in Appendix~\ref{app:scaling_factors}.

\subsection*{DBSCAN}
DBSCAN has two main parameters, \texttt{epsilon} and \texttt{min\_samples}. The parameter \texttt{epsilon} describes a neighborhood radius; in particular, it is the maximum distance within which two samples are considered to be neighbors. The parameter \texttt{min\_samples} denotes the minimum number of samples required in an \texttt{epsilon}-neighborhood to be considered a cluster (or specifically a core point which seeds a cluster).

Since we normalize the velocities to the spatial sub-space XYZ we can treat \texttt{epsilon} as a distance in parsecs. Together with the given minimum number of points it defines a minimum density that is needed to be considered a cluster. 
We search for optimal results within a range of \texttt{epsilon} $\in [2, 25]$\,pc with a step size of $0.5$\,pc. At the same time we vary minimum number of samples in the following range: \texttt{min\_samples} $\in [4, 40]$ with a step size of 2.

We find optimal parameters for the open cluster sample to be (\texttt{epsilon}, \texttt{min\_samples}, $c_v$) = (9.5, 6, 5.95).
For the compact cluster sample we find an optimal solution with (\texttt{epsilon}, \texttt{min\_samples}, $c_v$) = (9.0, 20, 5.95)

\subsection*{HDBSCAN}
HDBSCAN has three main parameters, \texttt{min\_cluster\_size}, \texttt{min\_samples}, and \texttt{cluster\_selection\_method}. Intuitively, \texttt{min\_cluster\_size} determines the smallest cluster sizes that HDBSCAN considers. Which points are still associated to a cluster is determined by \texttt{min\_samples}. By increasing \texttt{min\_samples} clusters are progressively more forced into denser areas leaving more points to be declared as noise. The parameter \texttt{cluster\_selection\_method} determines how clusters are selected from the cluster tree hierarchy, see Appendix~\ref{app:methods} for a detailed discussion.

We search for optimal results within a range of \texttt{min\_cluster\_size} $\in [20, 100]$ with a step size of $2$. At the same time, we vary the minimum number of samples \texttt{min\_samples} in the same range and step size while requiring \texttt{min\_cluster\_size} $\ge$ \texttt{min\_samples}\footnote{This is an intrinsic requirement of the algorithm.}. The selection method parameter can take the following two values: \texttt{cluster\_selection\_method} $\in$ [``leaf'', ``eom'']. 

We find optimal parameters for the open cluster sample to be (\texttt{min\_cluster\_size}, \texttt{min\_samples}, \texttt{cluster\_selection\_method}, $c_v$) = (60, 60, ``eom'', 8.51).
For the compact cluster sample, we find an optimal solution with (\texttt{min\_cluster\_size}, \texttt{min\_samples}, \texttt{cluster\_selection\_method}, $c_v$) = (24, 18, ``eom'', 5.95).

\section{Projected velocities} \label{app:vtan}

\begin{figure}[!ht]
    \centering
    \includegraphics[width=1\columnwidth]{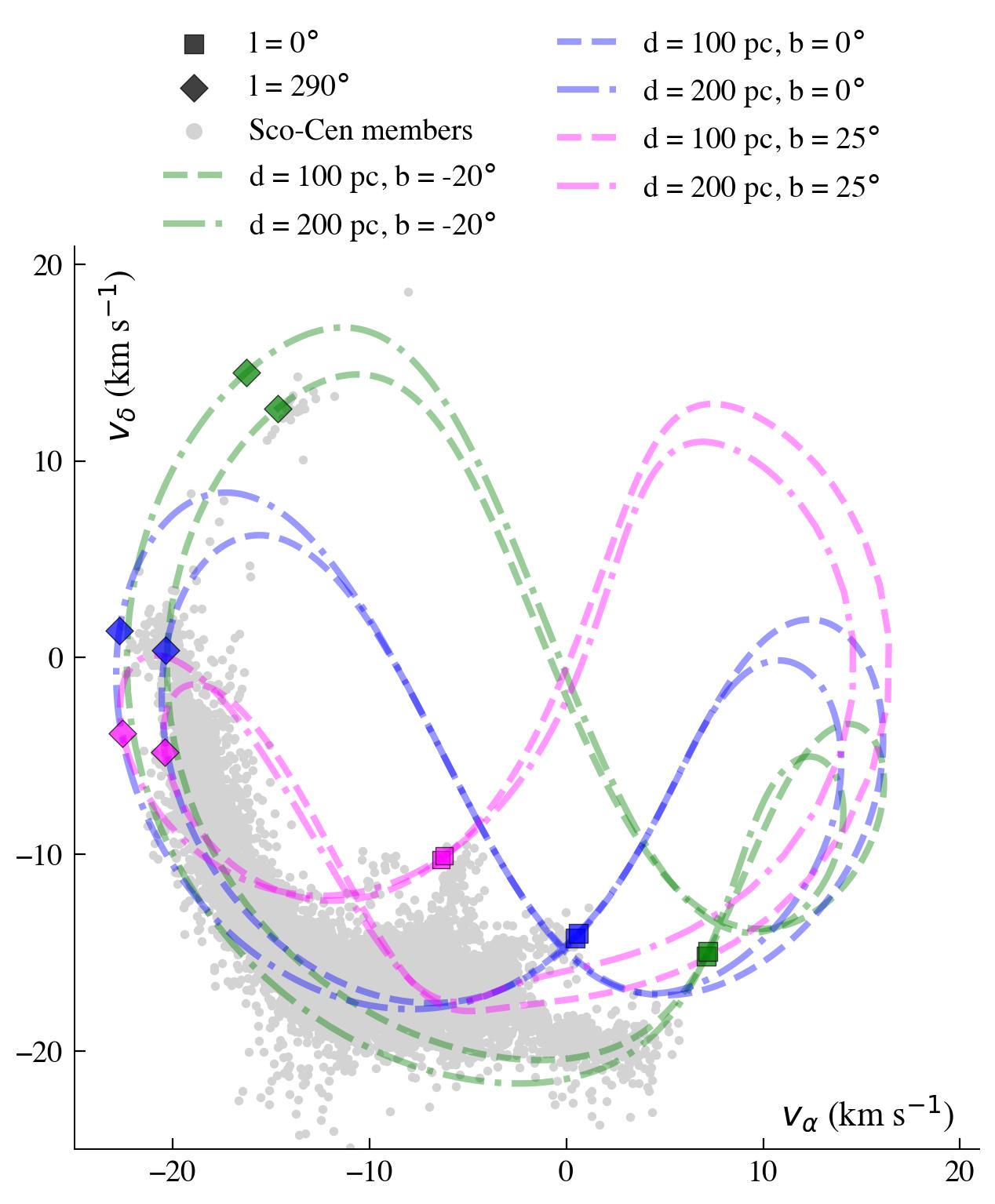}
    \caption{Tangential velocities in the $v_\alpha$/$v_\delta$ plane, showing the theoretical locations of sources with circular Galactic orbits and LSR velocities. Shown are six different cases, while each of the lines represents sources at all $l$ positions. The six cases are for two different distances (100\,pc, dashed lines; 200\,pc, dash-dotted lines), and for three different $b$ positions ($b=-20^\circ$, green; $b=0^\circ$, blue; $b=25^\circ$, magenta). The indicated longitude positions at $l=0^\circ$ (box symbols) and $l=290^\circ$ (diamond symbols) roughly mark the eastern and western borders of Sco-Cen. The \texttt{SigMA} selected Sco-Cen members are shown as gray dots (without stability cut). See also Fig.~\ref{fig:vtan} for a separation of the clusters and for the $v_{\alpha,\mathrm{LSR}}$/$v_{\delta,\mathrm{LSR}}$ plane.}  
    \label{fig:vtan_theoretical}
\end{figure}

The reflex motion of the Sun influences how the observed tangential velocities are distributed in $v_\alpha$/$v_\delta$ space. Figure~\ref{fig:vtan_theoretical} shows the theoretical positions of objects if they follow a circular orbit around the Galactic center at given positions within the Galactic potential. The orbits are estimated within a Milky Way potential including a disk, bulge, and halo component, using the python package \texttt{galpy} by \citet{Bovy2015} (\texttt{galpy.potential.MWPotential2014}; \texttt{galpy.potential.vcirc}) and assuming the local standard of rest (LSR) velocity from \citet{Schoenrich2010}. The projected motions are given for all Galactic longitude ($l$) positions at two distances ($d$) of 100\,pc and 200\,pc and at three Galactic latitudes ($b$) of $-20^\circ$, $0^\circ$, and $25^\circ$. These $d$ and $b$ ranges encompass the Sco-Cen region, which reaches from about $l = 0^\circ$ to $290^\circ$. The members of Sco-Cen within the selected \texttt{SigMA} clusters are plotted as gray dots in Fig.~\ref{fig:vtan_theoretical}. 

Overall, the young stellar groups in Sco-Cen seem to roughly follow expected motions in our Galaxy assuming they follow the LSR velocity.  
The figure additionally highlights the issues that come with the projected tangential velocity plane $v_\alpha$/$v_\delta$, which is a function of position in the sky and the Sun's motions. Very nearby stellar groups, like in nearby young local associations, could cover large areas in the sky and consequently also in the tangential velocity plane, while being at the same time confined in 3D velocity space (UVW).
This effect can be reduced if computing the tangential velocities relative to the LSR $v_{\alpha,\mathrm{LSR}}$/$v_{\delta,\mathrm{LSR}}$, which eliminates the influence of the reflex motion of the Sun (see also Sect.~\ref{sec:data} and Appendix~\ref{app:box}). A comparison of the two velocity spaces ($v_\alpha$/$v_\delta$ and $v_{\alpha,\mathrm{LSR}}$/$v_{\delta,\mathrm{LSR}}$) is shown in Fig.~\ref{fig:vtan}, displaying the \texttt{SigMA} selected members of the Sco-Cen clusters.

\section{The Gaia DR3 HRD} \label{app:hrd}

In this section we first describe our procedure to estimate the fraction of possible contaminants form older stellar populations (or field stars) in the \texttt{SigMA} selected Sco-Cen sample using the Gaia HRD, and second we estimate the fraction of sub-stellar objects (brown dwarf candidates).

\subsection{Estimating the contamination from older sources} \label{ap:old-star-cont}

The HRD in Fig.~\ref{fig:hrd} in Sect.~\ref{sec:results} shows a color-absolute-magnitude diagram (equivalent to an Hertzsprung-Russel diagram, HRD) using the magnitudes from the Gaia DR3 passbands $G$ versus $G_\mathrm{BP}-G_\mathrm{RP}$. Not all sources have detections in all three passbands; within the SigMA selected Sco-Cen members there are 12,724 (97\%) sources which have an entry in all three passbands. The absolute magnitude $G_\mathrm{abs}$ is calculated with the distance modulus using the inverse of the parallax as distance.
To estimate how many \texttt{SigMA} selected Sco-Cen members are consistent with the expected ages below $\sim$20\,Myr, and which sources could be contaminants from older populations, we first need to apply photometric quality criteria. Inferior photometric quality mostly affects the fainter low-mass sources and shifts them to the left in the Gaia HRD. We use the magnitude errors and the photometric flux excess factor $C$, which are defined as follows:
\begin{equation}
    \label{eq:mag-err}
    \begin{aligned}
        & G_\mathrm{err} = 1.0857/\mathtt{phot\_g\_mean\_flux\_over\_error} \\
        & G_\mathrm{RP,err} = 1.0857/\mathtt{phot\_rp\_mean\_flux\_over\_error}  \\
        & G_\mathrm{BP,err} = 1.0857/\mathtt{phot\_bp\_mean\_flux\_over\_error} \\
        & C = \mathtt{phot\_bp\_rp\_excess\_factor} =  (I_\mathrm{BP} + I_\mathrm{RP})/I_G \\
        & C^* = \mathrm{corrected} \,\,\, C
    \end{aligned}
\end{equation}
The flux excess factor $C$ \citep{Evans2018, Riello2021} gives the flux excess in the $G_\mathrm{BP}-G_\mathrm{RP}$ color relative to the $G$ band flux. It is recommended to use the corrected $C$ (denoted as $C^*$) as given in \citet{Riello2021}, which reduces the color dependence.
Using these parameters, we apply the following quality criteria to the photometry.
\begin{equation}
    \label{eq:phot-cut}
    \begin{aligned}
        & G_\mathrm{err} < 0.006, \,\, 
        G_\mathrm{RP,err} < 0.02, \,\, 
        G_\mathrm{BP,err} < 0.1, \,\,
        C^* < 0.3  
    \end{aligned}
\end{equation}
These cuts reduce the number of Sco-Cen members from 12,724 with complete phototmetric information to 11,132 (leaving 87\%), mainly reducing the number of the fainter low-mass stars. 
If not applying any quality criteria, a significant number of sources would be shifted toward older ages, only because of unreliable Gaia colors. 

Figure~\ref{fig:hrd} shows two isochrones. The dashed line shows a 25\,Myr isochrone from \citet{Baraffe2015} (BHAC15\footnote{\url{http://perso.ens-lyon.fr/isabelle.baraffe/BHAC15dir/}}) for Gaia DR3/EDR3 passbands.  
Additionally, the solid line shows a 25\,Myr isochrone from PARSEC\footnote{\url{http://stev.oapd.inaf.it/cgi-bin/cmd}; assuming solar metallicity (metal fraction $z = 0.0152$) and no extinction.} for Gaia DR3/EDR3 passbands \citep[e.g.,][]{Bressan2012, Chen2014parsec, Chen2015parsec, Marigo2017parsec, Riello2021}, including the upper--main-sequence (UMS), which is missing in the BHAC15 models. We use both models, since BHAC15 models deliver a better representation for low-mass stars. 

To get a measure for the contamination from older sources (older than the expected $\sim20$\,Myr), we select sources to the left of the two 25\,Myr isochrones, allowing for random scatter around the 20\,Myr isochrone (in particular young stars often show higher variability than main-sequence stars). 
Additionally, we do not consider sources at the UMS, since there the trend reverses (younger sources are to the left of the UMS). Hence, we apply a cut at $G_\mathrm{abs}>3$\,mag, only including fainter sources to test the contamination from older sources. 

The combined conditions deliver 737 candidate contaminants (and 10,395 young Sco-Cen candidates) out of 11,132 sources with applied photometric quality criteria, hence about 6--7\%, which are possible contaminants from older populations within the \texttt{SigMA} clusters. Considering the chosen borders, we like to stress that this separation can only be seen as a  rough estimate. In particular, we did not consider any possible contaminants in the UMS regime, where it is more difficult to distinguish young members from older stellar populations. Moreover, the chosen isochrone models have intrinsic uncertainties, and any change in metallicity or extinction is ignored in our test. 
Finally, we only examine sources in the HRD in Fig.~\ref{fig:hrd} with the applied photometric quality criteria from Equ.~(\ref{eq:phot-cut}). Hence, we can not make any statement for sources with inferior photometric quality, which often also suffer from higher astrometric uncertainties. Consequently, such sources could also have a higher probability of having a wrong cluster membership, solely due to the generally larger scatter in various dimensions.

To better understand the influence of the quality criteria, we repeat the selection of old star contamination with different photometric quality cuts. First, we consider the case of applying no cuts, using the 12,724 source with entries in all three Gaia bands, delivering a contamination fraction from older sources of about 12\%. Next we apply somewhat looser and also stricter quality criteria than given in Equ.~(\ref{eq:phot-cut}) as follows, first showing the looser cuts (\ref{eq:phot-cut-loose}), and next the stricter cuts (\ref{eq:phot-cut-strict}):
\begin{equation}
    \label{eq:phot-cut-loose}
    \begin{aligned}
        & G_\mathrm{err} < 0.01, \,
        G_\mathrm{RP,err} < 0.045, \,
        G_\mathrm{BP,err} < 0.25, \,
        C^* < 0.5  
    \end{aligned}
\end{equation}
\begin{equation}
    \label{eq:phot-cut-strict}
    \begin{aligned}
        & G_\mathrm{err} < 0.004, \,
        G_\mathrm{RP,err} < 0.015, \,
        G_\mathrm{BP,err} < 0.05, \,
        C^* < 0.3
    \end{aligned}
\end{equation}
With the looser cuts we get a contamination fraction of about 10\% and with the stricter cuts about 4\%. It gets clear that the fraction of sources to the left of the chosen isochrones decreases significantly (from about 12\% to 4\%) when using superior photometry, which indicates that many sources indeed get erroneously shifted to older ages when not considering the influence of photometric errors. 
In conclusion, we estimate the contamination fraction from older populations or field stars to be between 4--10\%.

Additionally, we could investigate the influence of the membership stability as delivered by the \texttt{SigMA} algorithm (Sect.\,\ref{sec:scale-space}). Figure~\ref{fig:stability} shows the influence of different stability cuts on the ``old star contamination fraction'' when using the quality criteria from Equ.\,(\ref{eq:phot-cut}). It can be seen that for low stability ($\lesssim$11\%) there is also a significant increase in the contamination fraction. When only using sources with \mbox{stability\,>\,11\%}, we would get a ``old star contamination fraction'' in the range of about 2--9\% for the cases of no photometric cuts to strict photometric cuts. 
Hence, 2--4\% can be considered as the lower limit for contamination from older stellar populations (or field stars), while about 10\% is likely the upper limit. 
In conclusion, the majority of the \texttt{SigMA} selected Sco-Cen members (likely more than 90\%) are sources with young Sco-Cen ages and therefore likely not contaminants from interloping older populations or field stars.

\subsection{Estimating the fraction of sub-stellar sources} \label{ap:substellar}

To get an estimate of sub-stellar sources in our sample (brown dwarf candidates), we use a 0.08\,M$_\odot$ iso-mass line in Fig.~\ref{fig:hrd} (right panel), which is extracted from BHAC15 models using ages from $10^4$ to $10^{10}$\,yr, to get a wide range. 
We select sources below 0.08\,M$_\odot$, which is defined as the approximate hydrogen-burning limit \citep[e.g.,][]{Baraffe1998, Burrows2001, Freytag2010, Freytag2012, Dieterich2014}.
The uncertainties at the low-mass regime make this selection only a rough estimate, in combination with the uncertainties of the stellar models (e.g., unknown metallicity, neglected extinction, different models give different results). Additionally, all the uncertainties mentioned above in Sect.~\ref{ap:old-star-cont} (e.g., chosen error or stability cuts) should be considered. 

Using a cut 0.08\,M$_\odot$ and the quality criteria from Equ.\,(\ref{eq:phot-cut}) we find that there are 1944 out of 10,395 (18.7\%) potential substellar sources, when considering only the younger sources from the middle panel in Fig.~\ref{fig:hrd}. This selection indicates a fraction of sub-stellar objects of about 19\% within the \texttt{SigMA} clusters.
If applying less strict error cuts (no cuts or Equ.\,(\ref{eq:phot-cut-loose})), the fraction stays at about 19\%, and if applying more strict error cuts (Equ.\,(\ref{eq:phot-cut-strict})) the fraction decreases to about 12\%. This is expected since more conservative photometric error cuts affect in particular faint sources. Changing the stability criteria does not influence these different fractions significantly, since sources both in the stellar and substellar regime seem to be affected almost equally. 
Concluding, there are about 19\% of brown dwarf candidates in the \texttt{SigMA} selected Sco-Cen sample.

\section{Auxiliary tables and figures} \label{app:tab}

In Table~\ref{tab:member-catalog} we give an overview of the contents of the source catalog containing all Sco-Cen members as selected in this work, including labels for cluster membership. The full version of the table is available online as a machine-readable version.  

We provide three additional tables, giving an overview of the literature comparisons between the \texttt{SigMA} clusters and other Sco-Cen samples. 
In Table~\ref{tab:damiani} we compare to \citet{Damiani2019} and \citet{Zerjal2021arxiv}, 
in Table~\ref{tab:squicc} we compare to \citet{Squicciarini2021} and \citet{MiretRoig2022b},
and in Table~\ref{tab:kerr} to \citet{Kerr2021}. 
More details on the comparisons can be found in Sect.~\ref{sec:compare}. 

\begin{table*}[!ht] 
\begin{center}
\begin{small}
\caption{Catalog of the 13,103 Sco-Cen members labeled for cluster membership as identified with \texttt{SigMA}. }
\renewcommand{\arraystretch}{1.2}
\resizebox{0.95\textwidth}{!}{%
\begin{tabular}{lll}
\hline \hline
\multicolumn{1}{l}{Column name} &
\multicolumn{1}{l}{Unit} &
\multicolumn{1}{l}{Column description} \\
\hline

\texttt{dr3\_source\_id} &  & The source ID from Gaia DR3.  \\
\texttt{SigMA} &  & \texttt{SigMA} membership label for each source, as defined in Table~\ref{tab:averages-pos}.   \\
\texttt{stability} &  & Membership stability of each source between 0--100\%   \\

\texttt{distance} & pc  & Distance derived from inverse of the parallax   \\
\texttt{e\_d\_upper} & pc & Upper 1\,$\sigma$ distance uncertainty determined from the 68.3 percentile of the sampled $d$ distribution   \\
\texttt{e\_d\_lower} & pc & Lower 1\,$\sigma$ distance uncertainty determined from the 68.3 percentile of the sampled $d$ distribution   \\

\texttt{X} & pc  & Heliocentric Galactic Cartesian coordinate, X-axis grows positive toward the Galactic center   \\
\texttt{Y} & pc  & Heliocentric Galactic Cartesian coordinate, Y-axis grows positive in direction of Galactic rotation   \\
\texttt{Z} & pc  & Heliocentric Galactic Cartesian coordinate, Z-axis grows positive toward the Galactic North-pole   \\
\texttt{e\_X\_upper} & pc  & Upper 1\,$\sigma$ X uncertainty determined from the 68.3 percentile of the sampled X distribution   \\
\texttt{e\_Y\_upper} & pc  & Upper 1\,$\sigma$ Y uncertainty determined from the 68.3 percentile of the sampled Y distribution   \\
\texttt{e\_Z\_upper} & pc  & Upper 1\,$\sigma$ Z uncertainty determined from the 68.3 percentile of the sampled Z distribution   \\
\texttt{e\_X\_lower} & pc  & Lower 1\,$\sigma$ X uncertainty determined from the 68.3 percentile of the sampled X distribution   \\
\texttt{e\_Y\_lower} & pc  & Lower 1\,$\sigma$ Y uncertainty determined from the 68.3 percentile of the sampled Y distribution   \\
\texttt{e\_Z\_lower} & pc  & Lower 1\,$\sigma$ Z uncertainty determined from the 68.3 percentile of the sampled Z distribution   \\

\texttt{v\_alpha} & km\,s$^{-1}$  & Tangential velocity in the direction of $\alpha$   \\
\texttt{v\_delta} & km\,s$^{-1}$  & Tangential velocity in the direction of $\delta$   \\
\texttt{e\_v\_alpha\_upper} & km\,s$^{-1}$  & Upper 1\,$\sigma$ $v_\alpha$ uncertainty determined from the 68.3 percentile of the sampled $v_\alpha$ distribution   \\
\texttt{e\_v\_delta\_upper} & km\,s$^{-1}$  & Upper 1\,$\sigma$ $v_\delta$ uncertainty determined from the 68.3 percentile of the sampled $v_\delta$ distribution    \\
\texttt{e\_v\_alpha\_lower} & km\,s$^{-1}$  & Lower 1\,$\sigma$ $v_\alpha$ uncertainty determined from the 68.3 percentile of the sampled $v_\alpha$ distribution    \\
\texttt{e\_v\_delta\_lower} & km\,s$^{-1}$  & Lower 1\,$\sigma$ $v_\delta$ uncertainty determined from the 68.3 percentile of the sampled $v_\delta$ distribution   \\

\texttt{v\_alpha\_LSR} & km\,s$^{-1}$  & Tangential velocity in the direction of $\alpha$ and relative to the LSR   \\
\texttt{v\_delta\_LSR} & km\,s$^{-1}$  & Tangential velocity in the direction of $\delta$ and relative to the LSR   \\
\texttt{v\_ERV} & km\,s$^{-1}$  & Estimated radial velocity, given as $\hat{v}_r$ in Sect.~\ref{sec:bulkvelocity}  \\

\hline
\end{tabular}
} 
\renewcommand{\arraystretch}{1}
\label{tab:member-catalog}
\tablefoot{The full machine readable version of the catalog is given online, while a column overview is given here. We list all relevant derived parameters. Original Gaia DR3 parameters can be queried from the Gaia Archive by using the Gaia \texttt{dr3\_source\_id}.
}
\end{small}
\end{center}
\end{table*}

\begin{table*}[!ht] 
\begin{center}
\begin{small}
\caption{Comparing the \texttt{SigMA} clusters with stellar group selections from \citet{Damiani2019} and from \citet{Zerjal2021arxiv}. Only those \texttt{SigMA} clusters which have cross-matches with either of the two literature samples are given here. \vspace{-2mm}}
\renewcommand{\arraystretch}{1.2}
\resizebox{1\textwidth}{!}{%
\begin{tabular}{llcllll}
\hline \hline
\multicolumn{1}{l}{\texttt{SigMA}} &
\multicolumn{1}{l}{Name (\texttt{SigMA})} &
\multicolumn{1}{c}{Nr\tablefootmark{a}} &
\multicolumn{1}{l}{Matches with DDP19\tablefootmark{b}} &
\multicolumn{1}{l}{Matches with ZIC21\tablefootmark{c}} \\
\hline
%
1 & rho Oph & 535 & US-f(3)US-n(316)US-D2(68)N(20) & C-USco(405)E-USco-multi(13) \\
2 & nu Sco & 150 & US-n(62)US-D2(65)N(1) & C-USco(127)E-USco-multi(1) \\
3 & delta Sco & 691 & US-f(33)US-n(74)D1(88)D2a(8)D2b(2)US-D2(387)N(6) & G-UCL-East(3)C-USco(489)E-USco-multi(24) \\
4 & beta Sco & 285 & US-f(58)US-n(17)US-D2(152)N(8) & C-USco(189)E-USco-multi(19) \\
5 & sigma Sco & 544 & US-f(180)US-n(3)D1(14)D2a(3)US-D2(227)N(7) & G-UCL-East(2)C-USco(105)E-USco-multi(184) \\
6 & Antares & 502 & US-f(67)US-n(40)D1(19)D2a(5)US-D2(249)N(29) & C-USco(78)E-USco-multi(252) \\
7 & rho Sco & 240 & US-f(14)US-n(2)D1(159)D2a(3)US-D2(10)N(15) & G-UCL-East(48)C-USco(1)E-USco-multi(7) \\
8 & Sco-Body & 373 & D1(2)D2a(221)US-D2(34)N(7) & E-USco-multi(291) \\
9 & US-fg & 276 & D1(188)D2b(1)US-D2(7)N(12) & G-UCL-East(16)E-USco-multi(29) \\
\hline
10 & V1062-Sco & 1029 & UCL-1(554)D1(11)D2a(222)D2b(4)N(10) & D-UCL-V1062-Sco(499)F-UCL-V1062-Sco(228)G-UCL-East \\
11 & mu Sco & 54 & UCL-1(36)D1(2)D2a(5) & D-UCL-V1062-Sco(7)F-UCL-V1062-Sco(30) \\
12 & Libra-S & 71 & D1(1)D2a(8)D2b(13)US-D2(32) & E-USco-multi(4) \\
13 & Lup\,1-4 & 226 & LupIII(67)D2a(47)D2b(65)N(4) & G-UCL-East(6)T-UCL-West(1)E-USco-multi(109) \\
14 & eta Lup & 769 & UCL-3(3)D1(549)D2a(43)D2b(10)US-D2(1)N(14) & G-UCL-East(419)E-USco-multi(15) \\
15 & phi Lup & 1114 & UCL-3(48)D1(627)D2a(62)D2b(148)N(38) & G-UCL-East(652)T-UCL-West(28)E-USco-multi(17) \\
16 & Norma-N & 42 & D1(1)N(6) &  \\
17 & e Lup & 516 & D1(319)D2a(18)D2b(80)N(5) & G-UCL-East(349)T-UCL-West(15) \\
18 & UPK\,606 & 131 & UCL-2(50)D1(2)D2b(57)N(1) & G-UCL-East(54)T-UCL-West(9) \\
19 & rho Lup & 246 & D1(17)D2b(189)N(2) & A-LCC-North (45)G-UCL-East(10)T-UCL-West(110) \\
20 & nu Cen & 1737 & UCL-2(2)D1(54)D2a(12)D2b(1270)US-D2(3)N(50) & A-LCC-North (70)G-UCL-East(116)T-UCL-West(790)E-US \\
\hline
21 & Cen-Far & 99 & D2b(41)N(1) &  \\
22 & sig Cen & 1805 & LCC-1(1)D1(45)D2b(1384)N(43) & A-LCC-North (1077)U-LCC-South(56)T-UCL-West(66) \\
23 & Acrux & 394 & LCC-1(89)D1(11)D2b(242)N(4) & A-LCC-North (25)U-LCC-South(316) \\
24 & Musca-fg & 95 & D2b(35)N(2) & U-LCC-South(76) \\
25 & eps Cham & 39 &  & U-LCC-South(25) \\
26 & eta Cham & 30 &  & U-LCC-South(3) \\
\hline
27 & B59 & 32 & D2a(1)N(20) & E-USco-multi(2) \\
28 & Pipe-North & 42 &  & E-USco-multi(38) \\
29 & tet Oph & 98 & D2a(37)US-D2(6)N(2) & E-USco-multi(87) \\
\hline
30 & CrA-Main & 96 &  & E-USco-multi(2) \\
32 & Sco-Sting & 132 & D1(22)D2a(1) &  \\
\hline
35 & L134/L183 & 24 &  & E-USco-multi(16) \\

\hline
\end{tabular}
} 
\renewcommand{\arraystretch}{1}
\label{tab:damiani}
\tablefoot{
\tablefoottext{a}{Number of sources from this work, for a direct comparison with the number of cross-matches given in brackets in Cols.~4--5.}
\tablefoottext{b}{The \citetalias{Damiani2019} group shortcuts are given for eight compact clusterings (UCL-1, UCL-2, UCL-3, Lupus 3, LCC-1, US-far, US-near), four diffuse populations  (D1, D2a, D2b, US-D2), and sources that have not been assigned to any of these groups (N). The number in brackets gives the number of matches with the respective \texttt{SigMA} cluster. See details in Sect.\,\ref{sec:damiani}.}
\tablefoottext{c}{\citetalias{Zerjal2021arxiv} report eight sub-groups in Sco-Cen (C, E, D, F, G, T, A, U) and two additional older groups (H, I; IC\,2602 and Platais\,8), while there are no matches of H or I with the 37 \texttt{SigMA} clusters. Again, the number of matches is given in brackets. The groups contain the following numbers of sources in \citetalias{Zerjal2021arxiv}: C-USco 1432, E-USco-multi 1483, D-UCL-V1062-Sco 506, F-UCL-V1062-Sco 273, G-UCL-East 1713, T-UCL-West 1057, A-LCC-North 1234, and U-LCC-South 487. See further details in Sect.\,\ref{sec:zerjal}.}
}
\end{small}
\end{center}
\end{table*}

\begin{table*}[!ht] 
\begin{center}
\begin{small}
\caption{Comparing the \texttt{SigMA} clusters with \citet{Kerr2021} clusters toward Sco-Cen. Only those \texttt{SigMA} groups which have matches with \citetalias{Kerr2021} are given here, while Oph-NorthFar is the only \texttt{SigMA} cluster without matches.}
\renewcommand{\arraystretch}{1.2}
\resizebox{0.9\textwidth}{!}{%
\begin{tabular}{llcllll}
\hline \hline
\multicolumn{1}{l}{\texttt{SigMA}} &
\multicolumn{1}{l}{Name (\texttt{SigMA})} &
\multicolumn{1}{c}{Nr\tablefootmark{a}} &
\multicolumn{1}{l}{TLC\tablefootmark{b}} &
\multicolumn{1}{l}{EOM\tablefootmark{c}} &
\multicolumn{1}{l}{LEAF\tablefootmark{d}} &
\multicolumn{1}{l}{Name (KRK21)\tablefootmark{e}} \\
\hline

1 & rho Oph/L1688 & 535 & 22(308) & 17(272) & I(109) & US-I-rho Oph \\
2 & nu Sco & 150 & 22(91) & 17(90) & E(54) & US-E \\
3 & delta Sco & 691 & 22(414) & 17(378) & F(1)H(102)I(1) & US-H \\
4 & beta Sco & 285 & 22(167) & 17(137) & G(29) & US-G \\
5 & sigma Sco & 544 & 22(296) & 17(248) & A(1)C(17)D(22) & US-C/D \\
6 & Antares & 502 & 22(292) & 17(239) & A(1)B(11)C(1)F(24) & US-B/F \\
7 & rho Sco & 240 & 22(128) & 17(64) & A(9) & US-A \\
8 & Scorpio-Body & 373 & 22(193) & 16(12)17(45) &  & EOM-16/US \\
9 & US-foreground & 276 & 22(111) & 13(30) &  & EOM-13 \\
\hline
10 & V1062-Sco & 1029 & 22(503) & 14(20)15(347) &  & LowerSco/EOM-14 \\
11 & mu Sco & 54 & 22(28) & 15(23) &  & LowerSco \\
12 & Libra-South & 71 & 22(23) &  &  &  \\
13 & Lupus-1-4 & 226 & 22(143) & 12(102) & A(46)B(14) & Lupus-IV/III \\
14 & eta Lup & 769 & 22(411) & 9(15)22(102)23(6) &  & EOM-9/22/23 \\
15 & phi Lup & 1114 & 22(391) & 17(4)19(10)23(6) &  & EOM-19/23 \\
16 & Norma-North & 42 & 22(4) &  &  &  \\
17 & e Lup & 516 & 22(257) & 11(1)20(76)21(8) &  & EOM-20 \\
18 & UPK606 & 131 & 21(1)22(53) & 11(32) &  & UPK606 \\
19 & rho Lup & 246 & 22(123) & 21(2)25(10) &  & EOM-25 \\
20 & nu Cen & 1737 & 22(573) & 11(3)21(2)24(108)26(14)27(2) &  & EOM-24 \\
\hline
21 & Centaurus-Far & 99 & 21(36) & 3(30) &  & Cen-South \\
22 & sig Cen & 1805 & 22(987) & 25(1)26(26)27(421) & C(4)D(12)E(48) & EOM-26/LCC-D/E \\
23 & Acrux & 394 & 22(258) & 27(208) & B(1)C(96) & LCC-C-Crux S \\
24 & Musca-foreground & 95 & 22(65) & 27(46) & B(16) & LCC-B \\
25 & eps Cham & 39 & 22(23) & 27(20) & A(17) & LCC-A-eps Cha \\
26 & eta Cham & 30 & 22(18) & 18(17) &  & eta Cha \\
\hline
27 & B59 & 32 & 22(14) & 6(13) &  & Pipe \\
28 & Pipe-North & 42 & 22(19) &  &  &  \\
29 & tet Oph & 98 & 22(49) & 10(28) &  & Theia67 \\
\hline
30 & Corona Australis & 96 & 22(53) & 8(52) &  & CrA \\
31 & CrA-North & 351 & 22(207) & 7(1)8(195) &  & CrA \\
32 & Scorpio-Sting & 132 & 22(62) & 7(11) &  & EOM-7 \\
\hline
33 & Chamaeleon-1 & 192 & 21(101) & 1(101) &  & Cha-1 \\
34 & Chamaeleon-2 & 54 & 21(30) & 2(30) &  & Cha-2 \\
\hline
35 & L134/L183 & 24 & 22(6) &  &  &  \\
36 & Oph Southeast & 61 & 4(20) &  &  & Oph Southeast \\

\hline
\end{tabular}
} 
\renewcommand{\arraystretch}{1}
\label{tab:kerr}
\tablefoot{
\tablefoottext{a}{Number of sources from this work, for a direct comparison with the number of cross-matches as given in brackets in Cols.~4--6.}
\tablefoottext{b}{Col.\,4 lists the TLC group labels if there are matches with \texttt{SigMA}, with the number of cross-matches in brackets. There have been only matches with the TLC groups 4, 21, and 22.}
\tablefoottext{c}{Col.\,5 lists the EOM group labels if there are matches with \texttt{SigMA}, with the number of cross-matches in brackets, while each EOM represents a sub-clustering within a lower level TLC group.}
\tablefoottext{d}{The letters in Col.\,6 correspond to LEAF sub-groups, with the number of cross-matches in brackets, wile a LEAF group represents a sub-clustering within the lower level EOM group.}
\tablefoottext{e}{Group names from \citetalias{Kerr2021}, if there is a significant overlap with \texttt{SigMA}. Only the (sub)group with the most significant number of cross-matches is given (in few cases more than one), as apparent from the numbers in brackets in Cols.~5 \& 6.}
See more details in Sect.\,\ref{sec:kerr}.
}
\end{small}
\end{center}
\end{table*}

\begin{table*}[!t] 
\begin{center}
\begin{small}
\caption{Comparing the \texttt{SigMA} clusters with stellar group selections from \citet{Squicciarini2021} and from \citet{MiretRoig2022b}. Only those SigMA groups which have cross-matches with either of the two literature samples are given here. \vspace{-3mm}}
\renewcommand{\arraystretch}{1.2}
\resizebox{1\textwidth}{!}{%
\begin{tabular}{llcllll}
\hline \hline
\multicolumn{1}{l}{\texttt{SigMA}} &
\multicolumn{1}{l}{Name (\texttt{SigMA})} &
\multicolumn{1}{c}{Nr\tablefootmark{a}} &
\multicolumn{1}{l}{Matches with SGB21\tablefootmark{b}} &
\multicolumn{1}{l}{Matches with MR22\tablefootmark{c}} \\
\hline

1 & rho Oph & 535 & G1(428)G2(2)G3(1)G4(1)G8(2)D(51) & $\alpha$Sco(3)$\delta$Sco(9)$\nu$Sco(1)$\sigma$Sco(67)$\rho$Oph(370) \\
2 & nu Sco & 150 & G1(1)G2(110)G6(10)D(23) & $\beta$Sco(2)$\delta$Sco(2)$\nu$Sco(110)$\sigma$Sco(22) \\
3 & delta Sco & 691 & G1(19)G2(2)G3(390)G4(9)G5(50)G6(2)G7(1)G8(2)D(136) & $\alpha$Sco(27)$\beta$Sco(8)$\delta$Sco(410)$\nu$Sco(29)$\pi$Sco(38)$\sigma$Sco(90) \\
4 & beta Sco & 285 & G1(1)G4(141)G5(11)G6(46)D(57) & $\alpha$Sco(2)$\beta$Sco(169)$\sigma$Sco(70) \\
5 & sigma Sco & 544 & G1(2)G4(2)G5(104)G8(3)D(377) & $\alpha$Sco(268)$\delta$Sco(3)$\pi$Sco(7)$\sigma$Sco(163) \\
6 & Antares & 502 & G1(13)G3(5)G7(44)G8(33)D(313) & $\alpha$Sco(290)$\pi$Sco(29)$\sigma$Sco(52)$\rho$Oph(37) \\
7 & rho Sco & 240 & D(168) & $\alpha$Sco(3)$\pi$Sco(180) \\
8 & Sco-Body & 373 &  & $\alpha$Sco(2)$\sigma$Sco(2) \\
9 & US-fg & 276 & D(13) & $\alpha$Sco(1)$\pi$Sco(194) \\
\hline
12 & Libra-S & 71 &  & $\sigma$Sco(1) \\
14 & eta Lup & 769 &  & $\pi$Sco(12) \\
15 & phi Lup & 1114 & D(2) & $\pi$Sco(3) \\
20 & nu Cen & 1737 &  & $\sigma$Sco(1) \\

\hline
\end{tabular}
} 
\renewcommand{\arraystretch}{1}
\label{tab:squicc}
\tablefoot{
\tablefoottext{a}{Number of sources from this work, for a direct comparison with the number of matches as given in brackets in Cols.~4--5.}
\tablefoottext{b}{The \citetalias{Squicciarini2021} groups (G) are numbered from 1 to 8, and their diffuse population is given with D. The number of cross-matches is given in brackets. The eight groups in  \citetalias{Squicciarini2021} are associated with the brightest star in each group as follows: G1--$i$\,Sco; G2--$\nu$\,Sco\,B; G3--$b$\,Sco; G4--HD\,144273; G5--HIP\,77900; G6--HIP\,78968; G7--HIP\,79910; G8--HD\,146467. See details in Sect.\,\ref{sec:squicc}.}
\tablefoottext{c}{Comparison to the seven groups in US as identified by \citetalias{MiretRoig2022b}. Again, the number of cross-matches is given in brackets. See details in Sect.\,\ref{sec:miret}.}
}
\end{small}
\end{center}
\end{table*}

\begin{figure*}[!h]
    \centering
    \includegraphics[width=0.96\textwidth]{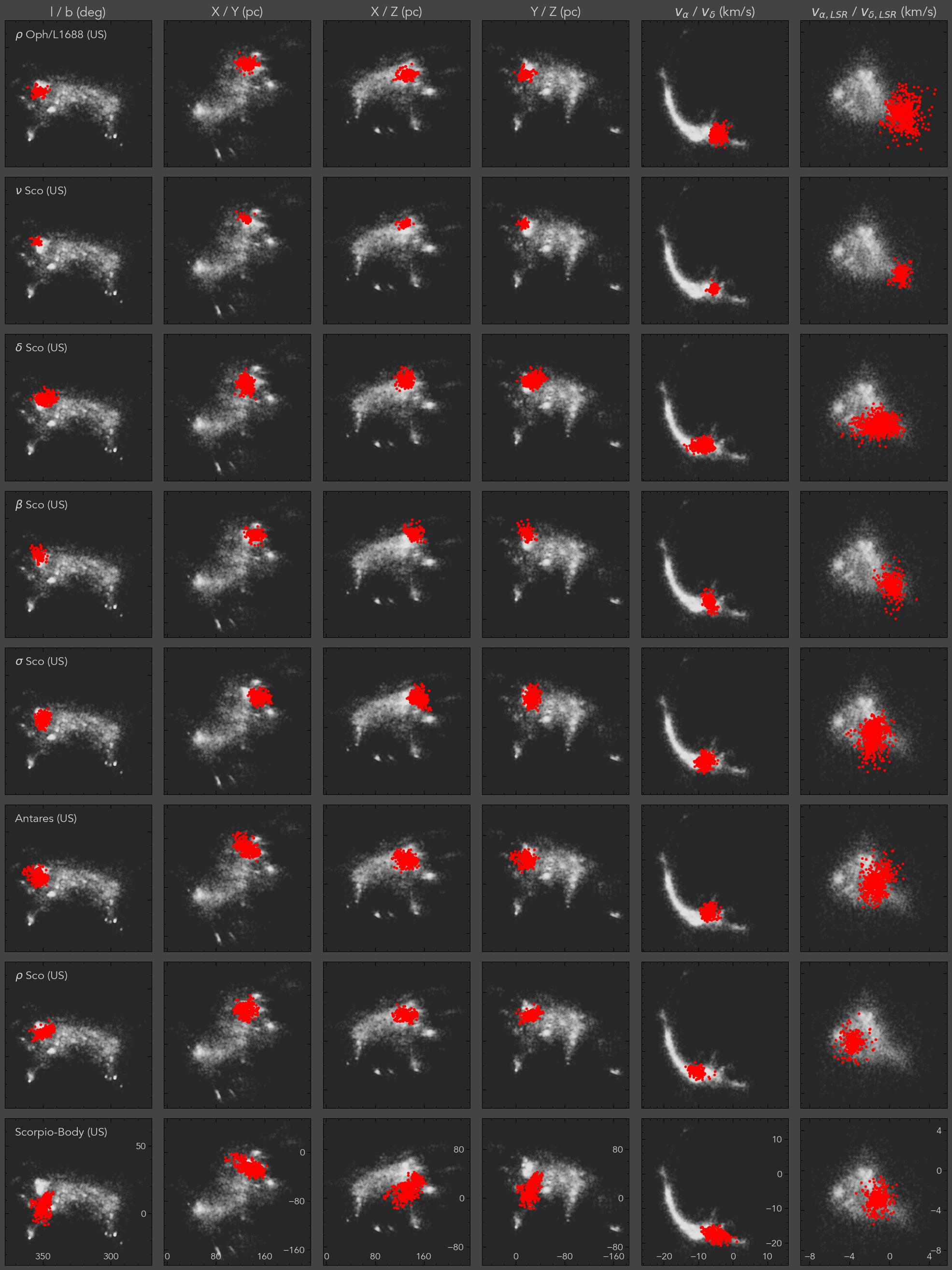}
    \caption{Six parameter spaces, highlighting the individual clusters in red. Shown are the clusters \texttt{SigMA} 1--8 (part of US). The gray background sources are all \texttt{SigMA}-selected Sco-Cen members. Cluster names are given in the left panels of each row. The used xy-axes are given as title at the top of each column, and tick labels are only given in the bottom row. 
    See also Figs.~\ref{fig:lb-map}--\ref{fig:vtan} and text for more details.
    }
    \label{fig:parameter-multis-1}
\end{figure*}
\begin{figure*}[!h]
    \centering
    \includegraphics[width=0.96\textwidth]{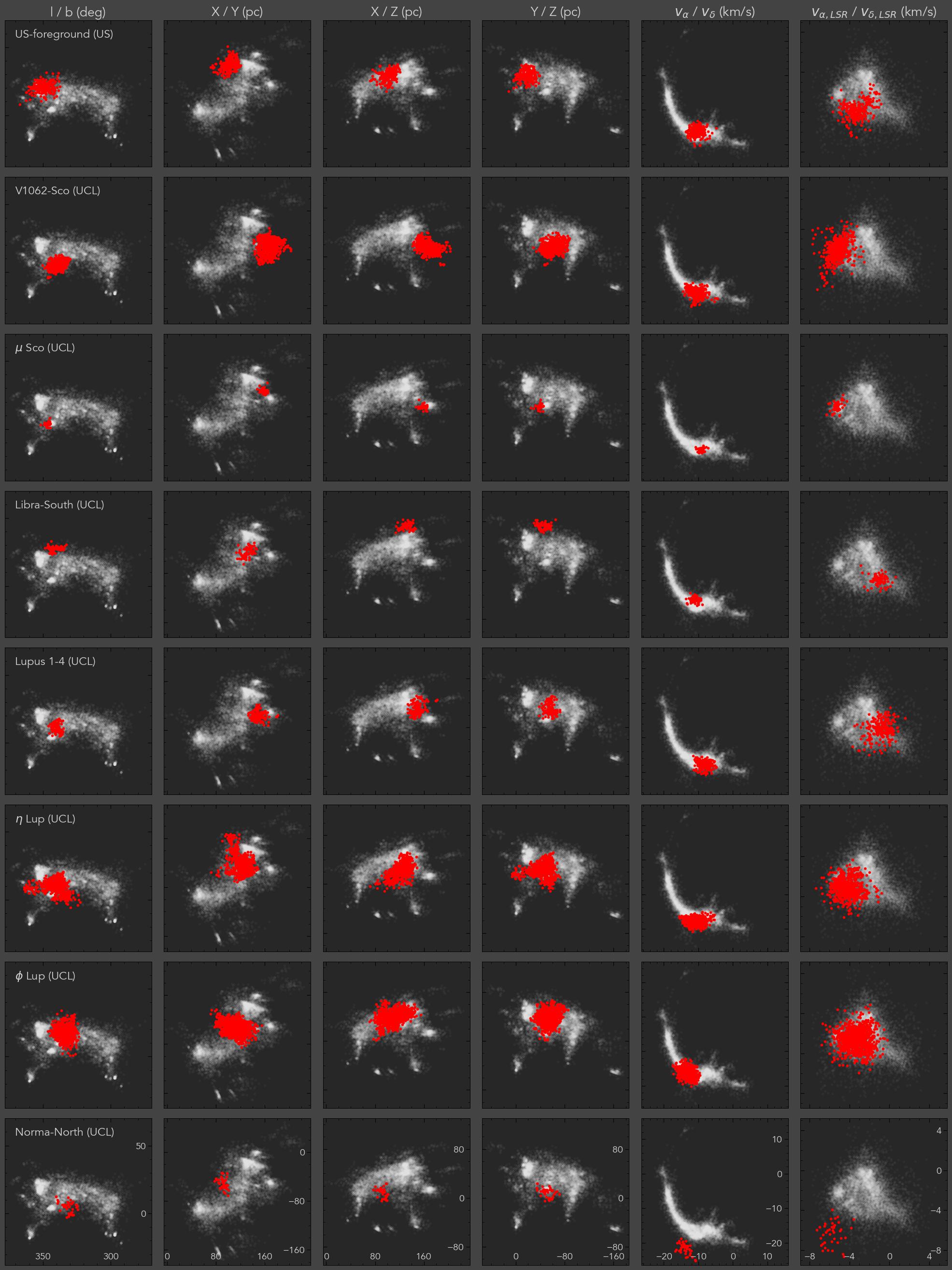}
    \caption{Same as Fig.\ref{fig:parameter-multis-1}, but for the clusters \texttt{SigMA} 9--16 (part of US and UCL).}
    \label{fig:parameter-multis-2}
\end{figure*}
\begin{figure*}[!h]
    \centering
    \includegraphics[width=0.96\textwidth]{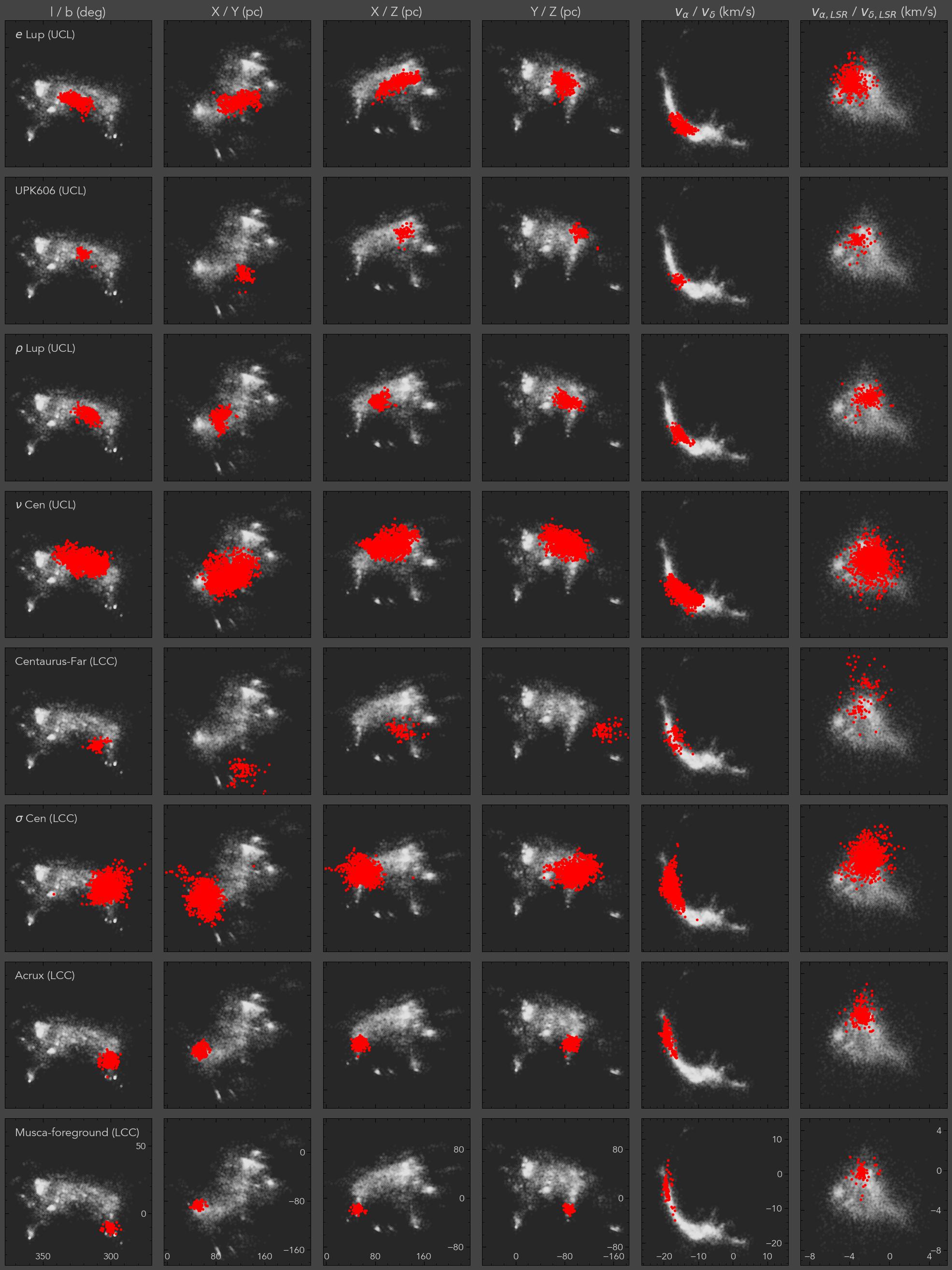}
    \caption{Same as Fig.\ref{fig:parameter-multis-1}, but for the clusters \texttt{SigMA} 17--24 (part of UCL and LCC).}
    \label{fig:parameter-multis-3}
\end{figure*}
\begin{figure*}[!h]
    \centering
    \includegraphics[width=0.96\textwidth]{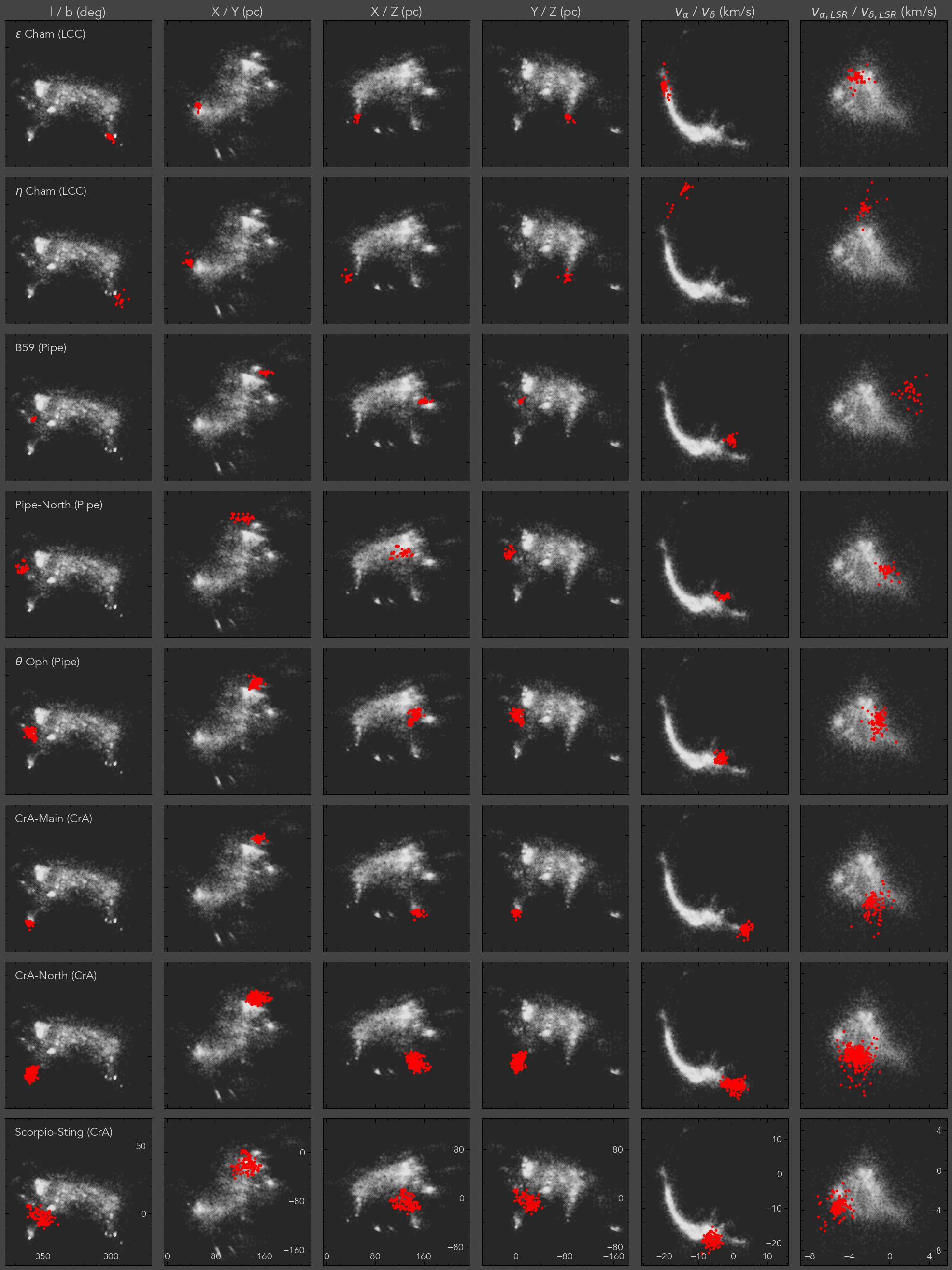}
    \caption{Same as Fig.\ref{fig:parameter-multis-1}, but for the clusters \texttt{SigMA} 25--32 (part of LCC, Pipe, and CrA).}
    \label{fig:parameter-multis-4}
\end{figure*}
\begin{figure*}[!h]
    \centering
    \includegraphics[width=0.96\textwidth]{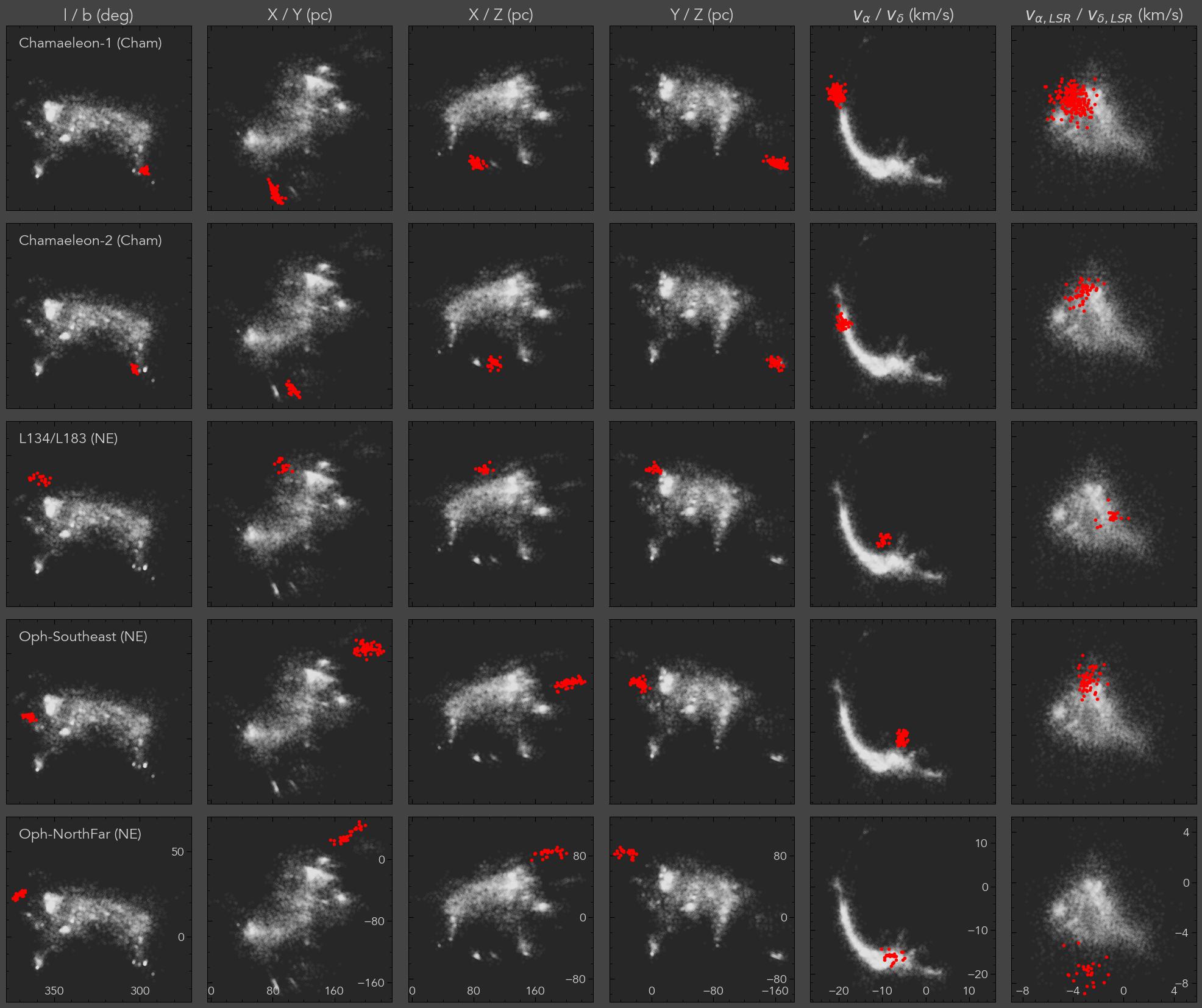}
    \caption{Same as Fig.\ref{fig:parameter-multis-1}, but for the clusters \texttt{SigMA} 33--37 (part of Cham and NE).}
    \label{fig:parameter-multis-5}
\end{figure*}

Finally, we provide additional figures of the position and velocity spaces of the individual 37 \texttt{SigMA} clusters. 
This allows a better appreciation of the individual cluster source distribution in each parameter space. The Figs.\,\ref{fig:parameter-multis-1}--\ref{fig:parameter-multis-5} are constructed as follows.
The cluster names are given in the left panel ($l,b$ panel) of each row.
Each column shows one of the six different parameter spaces for one cluster. The parameter spaces are the same as in Figs.~\ref{fig:lb-map}--\ref{fig:vtan}, namely $l$ vs $b$\,(deg), $X$ vs $Y$\,(pc), $X$ vs $Z$\,(pc), $Y$ vs $Z$\,(pc), $v_\alpha$ vs $v_\delta$\,(km\,s$^{-1}$), and $v_{\alpha,LSR}$ vs $v_{\delta,LSR}$\,(km\,s$^{-1}$). These axes labels are given at the top of each column. We like to highlight that Col.\,5 (tangential velocities) shows a larger velocity range than Col.\,6 (tangential velocities relative to the LSR), where the clusters actually occupy a smaller velocity space, hence show a smaller velocity dispersion.
All \texttt{SigMA} selected Sco-Cen members are plotted in gray in all panels and the given cluster is over-plotted with red dots. 
See also Figs.\,\ref{fig:lb-map}--\ref{fig:vtan} for an alternative view of the 37 \texttt{SigMA} clusters, and the interactive \href{https://homepage.univie.ac.at/josefa.elisabeth.grossschedl/wp-content/uploads/2022/11/scocen_interactive_2D.html}{2D} and \href{https://homepage.univie.ac.at/josefa.elisabeth.grossschedl/wp-content/uploads/2022/11/scocen_interactive_3D-1.html}{3D}  versions.

\end{appendix}
\end{document}